\documentclass{SCGEcontent}
\usepackage{titletoc}
\begin{document}

\begin{picture}(0,0){\rm
\put(0,-20){\makebox[160truemm][l]{\bf {\sanhao\raisebox{2pt}{.}}
Invited Review  {\sanhao\raisebox{1.5pt}{.}}}}}
\put(0,-34){\jiuwuhao {\textcolor[rgb]{0.5,0.5,0.5}{\sf Special Topic: the Next Detectors for Gravitational Wave Astronomy
}}}
\end{picture}

\def\bm{\boldsymbol}

\def\dl{\displaystyle}
\def\du{\end{document}}
\def\d{{\rm d}}
\def\e{{\rm e}}
\def\r{{\bm r}}
\def\P{{\bm P}}
\def\A{{\bm A}}
\def\k{{\bm k}}
\def\Q{{\bm Q}}
\def\pi{{\uppi}}
\def\cp#1{\mathbf{#1}}

\Year{X} %
\Month{X} %
\Vol{X} 
\No{X} 
\BeginPage{1} 
\EndPage{11} 
\AuthorMark{{\rm Blair D}, et al.}  
\AuthorMarkCite{{\rm Zhang Y, Liu Y X, Hou Z F}, et al.} 
\DOI{10.1007/s11433-015-5748-6} 
\ArtNo{XXXXXX}

\title{Gravitational wave astronomy: the current status\footnotemark[2]\footnotetext[2]{Sect. 1 is contributed by BLAIR David, JU Li, ZHAO ChunNong, WEN LinQing, CHU Qi, FANG Qi,  CAI RongGen, GAO JiangRui, LIN XueChun, LIU Dong, WU Ling-An, ZHU ZongHong
(corresponding author, BLAIR David, email: david.blair@uwa.edu.au);
sect. 2 is contributed by REITZE David H., ARAI Koji, ZHANG Fan (corresponding authors, REITZE David H., email: reitze@ligo.caltech.edu; ZHANG Fan, email: fnzhang@bnu.edu.cn); sect. 3 is contributed by FLAMINIO Raffaele (email:
raffaele.flaminio@nao.ac.jp); sect. 4 is contributed by ZHU XingJiang, WEN LinQing, HOBBS George, MANCHESTER Richard N., SHANNON Ryan M. (corresponding authors, ZHU XingJiang, email: xingjiang.zhu@uwa.edu.au; WEN LinQing, email:
linqing.wen@uwa.edu.au); sect. 5 is contributed by BACCIGALUPI Carlo (email: bacci@sissa.it); sect. 6 is contributed by GAO Wei, XU Peng, BIAN Xing, CAO ZhouJian, CHANG ZiJing, DONG Peng, GONG XueFei, HUANG ShuangLin, JU Peng,
LUO ZiRen, QIANG Li'E, TANG WenLin, WAN XiaoYun, WANG Yue, XU ShengNian, ZANG YunLong, ZHANG HaiPeng, LAU Yun-Kau (corresponding author, LAU Yun-Kau, email: lau@amss.ac.cn); sect. 7 is contributed by NI Wei-Tou (email: weitou@gmail.com)}}

\author[1]{BLAIR David}{}
\author[1]{JU Li}{}
\author[1]{ZHAO ChunNong}{}
\author[1]{WEN LinQing}{}
\author[1]{CHU Qi}{}
\author[1]{FANG Qi}{}
\author[2]{\vspace*{1.1mm}\\CAI RongGen}{}

\author[3]{GAO JiangRui}{}
\author[4]{LIN XueChun}{}
\author[5]{LIU Dong}{}
\author[6]{WU Ling-An}{}
\author[7]{ZHU ZongHong}{}

\author[8,9]{\vspace*{1.1mm}\\REITZE David H.}{}
\author[8]{ARAI Koji}{}
\author[7,10]{ZHANG Fan}{}
\author[11]{FLAMINIO Raffaele}{}
\author[1,12]{ZHU XingJiang}{}
\author[12]{\vspace*{1.1mm}\\HOBBS George}{}
\author[12]{MANCHESTER Richard N.}{}
\author[12,13]{SHANNON Ryan M.}{}
\author[14]{\vspace*{1.1mm}\\BACCIGALUPI Carlo}{}
\author[15,16]{GAO Wei}{}
\author[15,17]{XU Peng}{}
\author[15,16]{BIAN Xing}{}
\author[15,18]{CAO ZhouJian}{}
\author[19]{\vspace*{1.1mm}\\CHANG ZiJing}{}
\author[15,17]{DONG Peng}{}
\author[15]{GONG XueFei}{}
\author[20]{HUANG ShuangLin}{}
\author[21]{JU Peng}{}
\author[22,23]{\vspace*{1.1mm}\\LUO ZiRen}{}
\author[21]{QIANG Li'E }{}
\author[24]{TANG WenLin}{}
\author[25]{WAN XiaoYun}{}
\author[19]{WANG Yue}{}
\author[15]{\vspace*{1.1mm}\\XU ShengNian}{}
\author[15,16]{ZANG YunLong}{}
\author[20]{ZHANG HaiPeng}{}
\author[15,17,19]{LAU Yun-Kau}{}
\author[26]{NI Wei-Tou}{}

\address[{\rm1}]{School of Physics, The University of Western Australia, Crawley WA 6009, Australia;}
\address[{\rm2}]{State Key Laboratory of Theoretical Physics, Institute of Theoretical Physics, Chinese Academy of Sciences, Beijing 100190, China;}
\address[{\rm3}]{The School of Physics \& Electronic Engineering, Shanxi University, Taiyuan 030006, China;}
\address[{\rm4}]{Laboratory of All-Solid-State Light Sources, Institute of Semiconductors, Chinese Academy of Science, Beijing 100083, China; }
\address[{\rm5}]{State Key Laboratroy of Modern Optical Instrumentation, Department of Optical Engineering, Zhejiang University, Hangzhou 310027, China;}
\address[{\rm6}]{Laboratory of Optical Physics, Institute of Physics, Chinese Academy of Sciences, Beijing 100190, China; }
\address[{\rm7}]{Gravitational Wave and Cosmology Laboratory, Department of Astronomy, Beijing Normal University, Beijing 100875, China;}

\address[{\rm8}]{LIGO Laboratory, California Institute of Technology, MC 100-36, Pasadena, California 91125, USA;}
\address[{\rm9}]{Department of Physics, University of Florida, P.O. Box 118440, Gainsville, Florida 32611, USA;}

\address[{\rm10}]{~~Department of Physics, West Virginia University, PO Box 6315, Morgantown, WV 26506, USA;}
\address[{\rm11}]{~~National Astronomical Observatory of Japan, 2-21-1 Osawa, Mitaka, 181-8588 Tokyo, Japan;}
\address[{\rm12}]{~~CSIRO Astronomy and Space Science, PO Box 76, Epping NSW 1710, Australia;}
\address[{\rm13}]{~~International Centre for Radio Astronomy Research, Curtin University, Bentley WA 6102, Australia;}
\address[{\rm14}]{~~SISSA, Astrophysics Sector, via Bonomea 265, 34136, Trieste, Italy;}
\address[{\rm15}]{~~Institute of Applied Mathematics, Academy of Mathematics and Systems Science, Chinese Academy of Sciences, Beijing 100190, China;}
\address[{\rm16}]{~~University of Chinese Academy of Sciences, Beijing 100049, China;}
\address[{\rm17}]{~~Morningside Center of Mathematics, Chinese Academy of Sciences, Beijing 100190, China;}
\address[{\rm18}]{~~State Key Laboratory of Scientific and Engineering Computing, Academy of Mathematics and Systems Science, \\Chinese Academy of Sciences, Beijing 100190, China;}
\address[{\rm19}]{~~Department of Mathematics, Henan University, Kaifeng 475001, China;}
\address[{\rm20}]{~~Department of Mathematics, Capital Normal University, Beijing 100089, China;}
\address[{\rm21}]{~~Department of Geophysics, College of the Geology Engineering and Geomatics, Chang'an University, Xi'an 710054, China;}
\address[{\rm22}]{~~QUEST Centre of Quantum Engineering and Space-Time Research, Leibniz Universit$\ddot{a}$t Hannover, 30167, Hannover, Germany;}
\address[{\rm23}]{~~Max-Planck-Institut f\"{u}r Gravitationsphysik (Albert Einstein Institut), D-30167 Hannover, Germany;}
\address[{\rm24}]{~~Aerospace Flight Dynamics Laboratory, Beijing Aerospace Control Center, Beijing 100094, China;}
\address[{\rm25}]{~~Qian Xuesen Laboratory of Launch Vehicle Technology, Beijing 100094, China;}
\address[{\rm26}]{~~Center for Gravitation and Cosmology, Department of Physics, Tsing Hua University, Hsinchu 30013, China}

\maketitle \vspace{-3.5mm}{\footnotesize\begin{center} Received September 24, 2015; accepted September 28, 2015
\end{center}}\vspace*{-5mm}

\newpage

\begin{center}
\rule{16.5cm}{0.4pt}
\parbox{16.5cm}
{\begin{abstract}
In the centenary year of Einstein's General Theory of Relativity, this paper reviews the current status of gravitational wave astronomy across a spectrum which stretches from attohertz to kilohertz frequencies. Sect. 1 of this paper reviews the historical development of gravitational wave astronomy from Einstein's first prediction to our current understanding the spectrum.  It is shown that detection of signals in the audio frequency spectrum can be expected very soon, and that a north-south pair of next generation detectors would provide large scientific benefits. Sect. 2 reviews the theory of gravitational waves and the principles of detection using laser interferometry. The state of the art Advanced LIGO detectors are then described.  These detectors have a high chance of detecting the first events in the near future. Sect. 3 reviews the KAGRA detector currently under development in Japan, which will be the first laser interferometer detector to use cryogenic test masses. Sect. 4 of this paper reviews gravitational wave detection in the nanohertz frequency band using the technique of pulsar timing. Sect. 5 reviews  the status of gravitational wave detection in the attohertz frequency band, detectable in the polarisation of the cosmic microwave background, and discusses the prospects for detection of primordial waves from the big bang. The techniques described in sects. 1--5 have already placed significant limits on the strength of gravitational wave sources. Sects. 6 and 7 review ambitious plans for future space based gravitational wave detectors in the millihertz frequency band. Sect. 6 presents a roadmap for development of space based gravitational wave detectors by China while sect. 7 discusses a key enabling technology for space interferometry known as time delay interferometry.
\end{abstract}}
\end{center}\vspace*{-0.6cm}

\begin{center}
\parbox{16.5cm}
{\bf\jiuhao gravitational waves, ground based detectors, pulsar timing, spaced based detectors, CMB}
\end{center}

\begin{center}
{}
\Cit{Blair D, Ju L, Zhao C N, et al. Gravitational wave astronomy: the current status. Sci China-Phys Mech Astron,
XX, X: XXXXXX, doi: 10.1007/s11433-015-5748-6}
\end{center}

\textwidth=178truemm \textheight=236truemm

\wuhao\vspace*{1.5mm}
\tableofcontents
\vspace*{3mm}
\begin{multicols}{2}

\renewcommand{\baselinestretch}{1.08} \baselineskip 12.2pt\parindent=10.8pt

\renewcommand{\thefootnote}


\section{Ground based gravitational wave astronomy and opportunities for Australia-China collaboration}

\emph{Gravitational wave (GW) detection has become a high priority for physics since advanced technology has enabled the creation of devices with sufficient sensitivity to detect predicted astronomical signals. This section first presents the history of detector developments to highlight the enormous achievements of the field. We summarise the case for building terrestrial detectors, for which it is expected that signals from coalescing pairs of neutron stars will be detectable in the short term.  We emphasise the importance of a worldwide array of detectors to create an optimum GW telescope.  The importance of developing the next generation of detectors is emphasised, and  the case for a north-south pair of 8 km arm length detectors  is presented. Such a  system could monitor a volume of the universe 60 times larger than the best current detectors, enabling precision testing of black hole and neutron star physics.}

\subsection{Gravitational wave astronomy}\label{sec:1}

The spectrum of electromagnetic waves was predicted in  the paper ``A dynamical theory of the electromagnetic field'' by Maxwell \cite{ref1}, published in 1865, exactly 150 years ago.
Maxwell's paper predicted waves travelling close to the speed of light, and implied that light was an electromagnetic wave. The harnessing of electromagnetic waves began with the first experiments of Hertz \cite{ref2} between 1886 and 1889. During the next century as technology developed, the whole spectrum, covering at least 33 decades of frequency, has been harnessed, from the highest energy cosmic ray photons with energies $\sim1$ J to Schumann resonances with photon energies $\sim10^{-33}$ J.

\begin{table*}[t]
\caption{The four main frequency bands for gravitational astronomy\hspace*{150mm}}\label{tab1_sources}
\footnotesize \begin{center}
\begin{tabular*}{\textwidth}
{p{1.8cm}p{3.2cm}p{3.5cm}p{3.61cm}p{3.5cm}}\toprule[0.65pt]  
\bf{Frequency band} & Cosmological

10$^{-16}$ Hz & Nanohertz band

 10$^{-9}$--10$^{-7}$ Hz & Millihertz band

  10$^{-4}$--10$^{-3}$ Hz & Audio band

  10--10$^4$ Hz \\
\hline
\bf{Signal source} & frozen relic waves from the big bang at ultralow frequency & waves from supermassive black holes at a frequency 1 cycle per 3 years & waves from massive black hole binaries at ~1 cycle per minute partially masked by galactic binary star systems & audio frequency waves, from coalescence of stellar mass neutron stars and black holes \\
\bf{Detection technique} & B-mode polarisation of the cosmic microwave background & correlated  pulse
arrival time variations of millisecond pulsar signals & drag free space interferometers of ~ 10$^6$ km baselines & high power ground based multi-kilometer baseline interferometers \\
\hline
\end{tabular*}
\end{center}
\end{table*}

The spectrum of GWs is a direct prediction of Einstein's General Theory of Relativity published in November 1915, exactly one hundred years ago. This theory predicts that GWs travel at the speed of light \cite{ref3}. The prediction of GWs was not included in the 1915 paper, but was published in 1916 \cite{ref4}, and then corrected in 1918 \cite{ref5}.

For decades there were arguments about the reality of GWs. The argument was first clarified by Feynman in 1957 when he showed that GWs can do work against frictional forces \cite{ref6}. In 1975  Hulse and  Taylor \cite{ref7} discovered the binary pulsar PSR1913+16. Taylor and co-workers monitored this system for 20 years, and demonstrated that the system lost energy exactly as predicted for GW emission, leading to coalescence in a short time compared with the Hubble time. Hulse and Taylor were awarded the Nobel Prize in 1993 for this dramatic proof of the existence of GWs. Their discovery  also demonstrates the existence of a strong, and well defined class of GW sources: coalescing compact binary systems. Such sources create a well defined  chirp signal in the audio frequency band when the two stars coalesce. In the Hubble volume of the universe such events are predicted to be occurring at a rate of $\sim5$ per minute.\cite{zhux-j}

 Weber first discussed the possibility of direct detection of GWs in the early 1960's with publication of a paper \cite{ref8} and a book \cite{ref9} in which he presented the rationale for GW detection. He was the first to show that the GW luminosity could reach $\rm ~c^5$/G, or $\sim10^{23}$ solar luminosities when two black holes coalesce. Thus GW emission could represent the most powerful source of radiation since the big bang.

As technology has developed the possibility of detection has increased, and today the detection of GWs has become a greater and greater priority for physics. Many techniques for GW detection are currently under development, as summarised in Table~\ref{tab1_sources}. The spectrum spans at least 20 decades of frequency, and could extend much higher although sources at higher frequency are speculative. Active programs are underway across the frequency bands indicated in Table~\ref{tab1_sources}, because each frequency band offers different sources. Figure~1 summarises the GW spectrum and the instrumental sensitivity targets.  Only the most promising sources are summarised in the figure. The diagonal line is a possible spectrum for stochastic waves created in the big bang, for which the energy density is equal in each decade.

The history of space and ground based detectors is shown schematically in Figure 2.
In the rest of this paper we will focus on audio frequency detection, which offers a rich range of sources \cite{JuRPP} and probably the best near term possibility of detection. The first  GW detectors, developed by Weber,  consisted of aluminium bars instrumented with piezo-electric crystals to detect small vibrations induced by GWs. Bars of this type achieved a GW strain sensitivity $h\sim10^{-15}$--$10^{-16}$ in a narrow bandwidth near 1 kHz. Despite some claims, there was a clear consensus that these detectors had insufficient sensitivity to detect expected astrophysical signals.

Cryogenic resonant mass GW detectors were developed from 1972--2000. They consisted of massive
metal bars cooled to cryogenic temperatures with superconducting transducers \cite{dgbBook}. These achieved at least 6 orders
of magnitude greater flux sensitivity than the original Weber-type detectors \cite{ref10,ref11}. In the 1990s after 2 decades of development an array of 5 detectors at Baton Rouge, Frascati, Legnaro, CERN and Perth, called the International Gravitational Events Collaboration undertook long term searches. These also failed to identify any signals although there were a few false alarms. The cryogenic bar detectors were mainly sensitive to broadband bursts of GW emission from events such as\linebreak
\vspace*{-4mm}

\begin{figure}[H]
\centering
\includegraphics[scale=1.05]{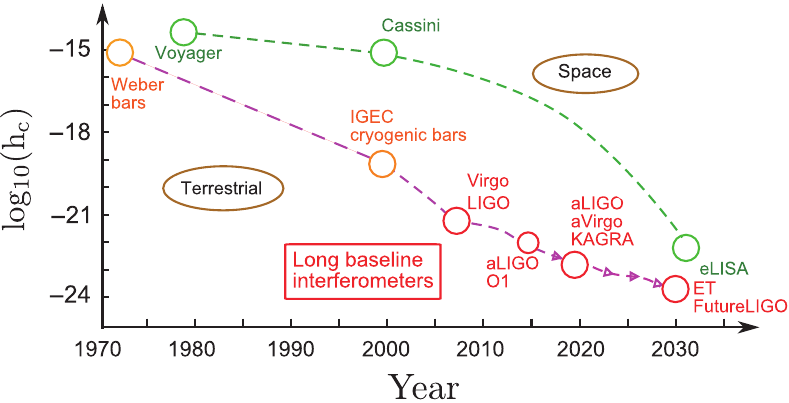}
\caption{(Color online) Progress in the development of GW detectors. The strain sensitivity figures are indicative only because  different data analysis methods can lead,
for example, to much higher sensitivity for searches for continuous wave sources. Ground based detectors have improved by seven order of magnitude over 45 years.
Space detectors which first used Doppler tracking of distant spacecraft (Voyager and Cassini) will become very sensitive when space interferometers are implemented.} 
\label{fig:figure2}
\end{figure}

\begin{figure}[H]
\centering
\includegraphics[scale=1]{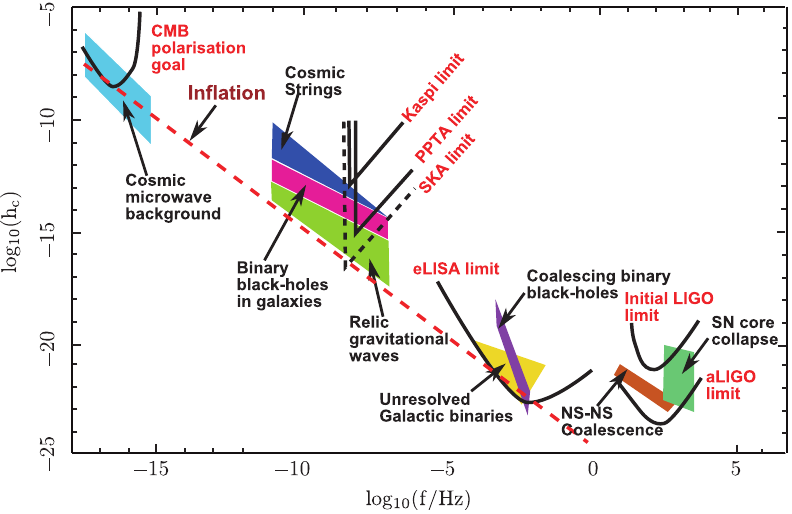}
\caption{(Color online) The strain amplitude, main signal sources and detector sensitivity for the four main GW detection bands. The sensitivity is only approximate because integration and correlation techniques allow different sensitivities according to data analysis techniques. Cosmic Microwave Background polarisations measurements have recently been shown to be limited by effects of galactic dust. For pulsar timing most sources within the yellow band have been ruled out by observations. The Square Kilometer Array (SKA) is predicted to allow deeper searches than the recently reported Parkes Pulsar Time Array (PPTA) searches. Advanced LIGO is beginning initial observations in late 2015.} 
\label{fig:figure1}
\end{figure}

\noindent  the collision of black holes or core collapse supernovae, but
sensitivity was limited to events occurring within the Milky Way galaxy.  There was a consensus that the event rate for such events was too low. Unfortunately sensitivity was insufficient to detect events in more distant galaxies such as the Virgo Cluster, for which the total event rate would be much larger.

Over the period that cryogenic bars were being prepared for operation, the technology for another class of detector---long baseline laser interferometers---was being developed at Glasgow, Munich, MIT, Caltech, Hannover, Paris, Pisa and Tokyo. These detectors were ideally suited to detect chirp signals from the coalescence of binary neutron stars such as the system discovered by Hulse and Taylor. Since the GW-induced relative displacement increases linearly with arm length, (for light storage times less than half a GW period) long baseline interferometers could in principle be much more sensitive than resonant bars, whose length is limited both by practical considerations and by the sound velocity in the bar. By using multiple reflections, interferometers could achieve an interaction time of up to half a cycle of GWs.

Following successful testing of small scale prototypes, funding was received to build the 300 m TAMA detector in Japan, the 600 m GEO detector near Hannover, Germany, the 3 km French-Italian Virgo detector near Pisa in Italy, and two 4 km LIGO detectors in the USA, one in Livingston, Louisiana, and the other near Hanford in Washington State. In their initial phase, the long baseline detectors were not expected to have a high chance of detection, but a second Advanced stage was foreshadowed, that would bring the detectors to a sensitivity sufficient to detect the coalescence of binary neutron stars at a reasonable rate.

\subsection{The event rate for ground based gravitational wave astronomy}\label{sec:2}
Neutron star coalescence signals have a well defined waveform, so that amplitude uncertainty in mainly due to the geometrical orientation of the system relative to the detector and the spectral sensitivity of the interferometer. Matched filtering allows optimal signal extraction. This enables standardisation of the neutron star inspiral range, assuming a pair of 1.4 solar mass neutron stars coalescing in a random orientation relative to the detector, and a signal to noise ratio (SNR) of 8. This is a useful calibration standard to allow comparison of detector sensitivities. However it is important to understand that the range differs according to the type of source. In particular the coalescence of a pair of 15 solar mass  black holes can be detectable to much greater distances.

The initial long baseline laser interferometers  achieved a neutron star coalescence range exceeding 16 Mpc, corresponding to about 1000 times the diameter of the Milky Way. This was the first time sensitivity had been achieved for a known class of source far beyond the Milky Way.

Initial LIGO and Virgo accumulated long observational runs (two years of operation with about 50\% multi-detector duty cycle). At the range achieved, which slightly exceeded the design sensitivity predicted in the 1990s, the neutron star binary coalescence rate was expected to be about one event per 20--100 years.

The successful development of sensitive laser interferometers required and motivated very important technological innovations, including: a) the development of supermirrors with extremely low optical losses, b) high purity fused silica with very low optical and acoustic losses, c) high performance vibration isolation and suspension systems, d) optical coatings with reduced acoustic losses, e) the development of the concept of optical power recycling which allowed high optical power interferometers to be created in which the laser power was built up by resonant amplification. Advanced interferometer techniques were also developed, including the concept of signal recycling and variations in the optical configuration in which the signal sidebands can be enhanced. In turn, all of the above technologies have opened whole new areas of quantum opto-mechanics research.

The Advanced LIGO GW detectors are an implementation of most of the above concepts, as discussed below. and reported in sect. 2 of this paper by Reitze et al.  In late 2015 Advanced LIGO will begin operation with a greatly increased range, of about 70 Mpc. Since the sensitivity is now extended to regions where the universe is moderately homogeneous, the expected event rate is proportional to the horizon volume of the universe. Compared to the initial LIGO horizon distance, the expected event rate at the time of writing should be increased by a factor of $(70/16)^3$, corresponding to a neutron star coalescence event rate of $\sim4$--0.8 per year. Advanced Virgo is expected to achieve comparable sensitivity one year later.

The plausibility of the expected Advanced LIGO event rate can be arrived at very simply, based on the observed population of compact binary neutron stars, assuming only that we are not preferred observers looking at an exceptional population of binary pulsars.  The event rate is determined statistically, based on the six observed short lived binary pulsar systems. From these 6 local systems in our galaxy, we can estimate the total population of potentially observable systems---ones which could be observed in the Milky Way given an arbitrarily large radio telescope. This number depends on the local distribution of observed pulsars, the size of the Milky Way and the distribution of systems in the Milky Way. A typical estimate is $\sim1000$.

Because pulsar signals are beamed, we observe only a fraction of all active pulsars. For example, quite soon precession will cause the Hulse-Taylor pulsar to become invisible due to mis-orientation. We can estimate the \textit{total} number of short lived binary pulsar systems in the Milky Way by correcting for the pulsar beaming. This increases the total number by an order of magnitude to $\sim10^4$.  To estimate the \textit{coalescence rate} in the Milky Way, we need an estimate of the mean system lifetime to coalescence.  The above estimates determine the mean rate of neutron star coalescence events in the Milky Way. For example, a mean lifetime to coalescence of $10^8$ years leads to a coalescence rate of one per $10^4$ years if the total population is $10^4$. Finally, we ask how many Milky Way equivalent galaxies we require to observe neutron star coalescence events at a reasonable rate, say one event per week. This then defines the horizon sensitivity which was chosen for Advanced LIGO to be about 200 Mpc.

The uncertainty in the above estimate is larger than a normal  $N^{-1/2}$  sampling error because the nearest, lowest radio luminosity binary pulsar system tends to dominate the event rate. In our case the nearest system is also the shortest lifetime system observed to date. It is not possible to know whether this nearest binary pulsar is typical or anomalous. The dominance of a single system gives rise to a very large standard error. The best estimates are given by Abadie et al. in ref. \cite{ref12} which gives a predicted neutron star coalescence range between $10^{-6}$ and $10^{-4}$ per year in the Milky way galaxy. This range leads to a detectable neutron star coalescence rate of $\sim0.4$--400 per year at the Advanced LIGO predicted sensitivity.

 The neutron star coalescence event rate can be calibrated against other estimators. One is stellar evolution modelling and the second is the abundance of gold. Gold and other heavy elements have recently been shown to be created from rapid nucleosynthesis of  neutron rich ejecta  when neutron stars coalesce. Results from such estimates are broadly consistent  \cite{Piran,ref13} with the observed population estimates. However the uncertainties in all such estimates are also large, especially because the total mass of ejecta can only be predicted by numerical general relativistic  hydrodynamical codes, and depend on an unknown equation of state \cite{Rezzolla}.

The Advanced LIGO event rate estimated in ref.~\cite{ref12} predicts the most likely event rate to be 40 events per year at full Advanced LIGO sensitivity. Others predict that the rate of black hole binary coalescence limit is even larger than this.  Lee et al. \cite{LeeHM} predicted that the black hole binary coalescence events will be the first signals to be observed by ground based detectors.  Even using the worst case estimates, at full sensitivity Advanced LIGO should be able to detect one event per year. Since LIGO has already been shown to be able to detect single very rare events through blind injection testing, there is a consensus that GW detection has now reached a point at which detection can be expected within 1 or several years.

\subsection{ Advanced LIGO and advanced techniques for future enhancement}\label{sec:3}

\begin{figure}[H]
\centering
\includegraphics[scale=0.5]{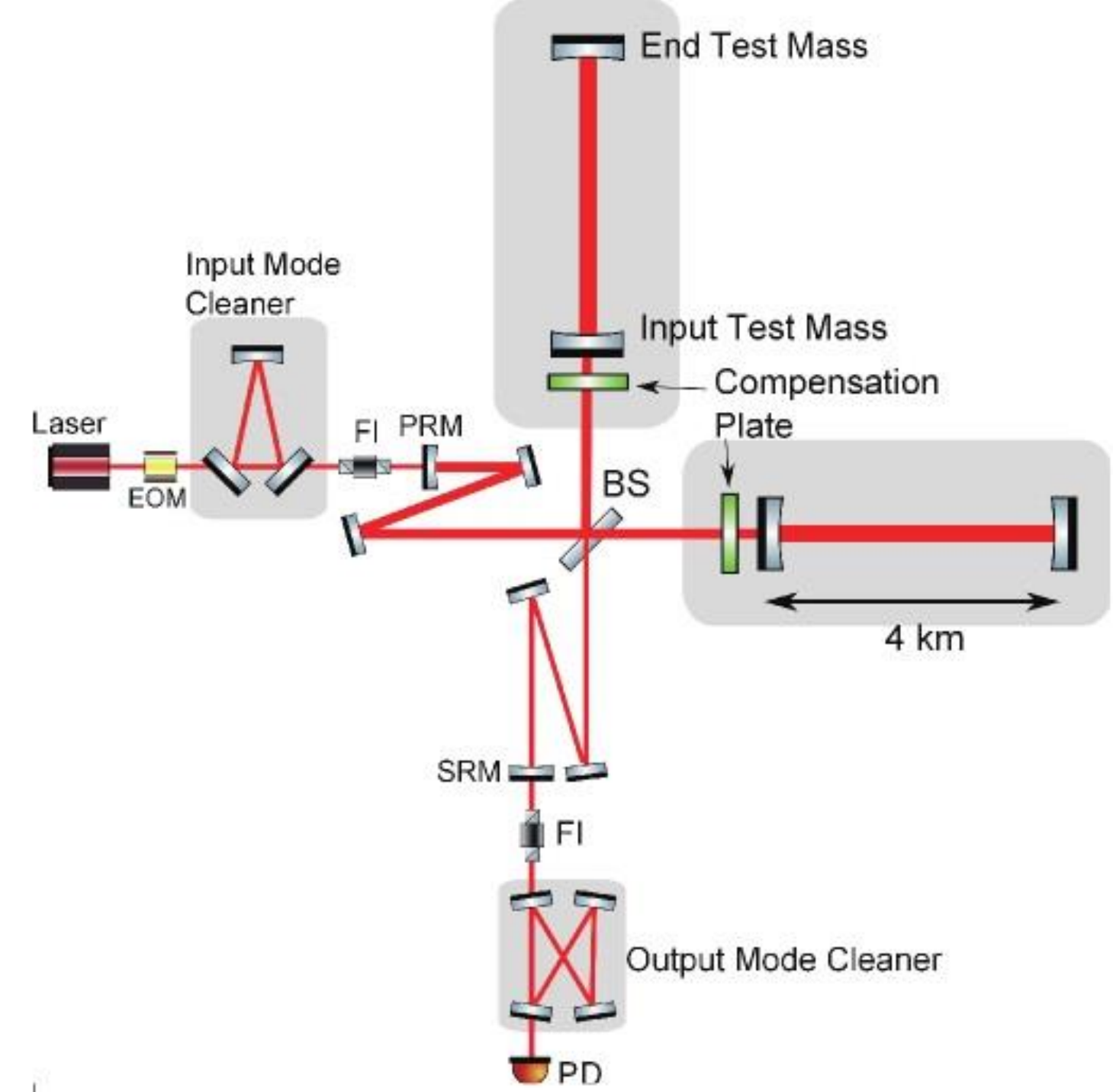}
\caption{(Color online) A simplified schematic diagram of  an Advanced LIGO-type interferometer showing the main interferometer, the power recycling mirror (PRM), the signal recycling mirror (SRM) and the compensation plates used for correcting thermal aberrations. Mode cleaners are used to suppress unwanted transverse optical modes. Faraday Isolators (FI) suppress reflected beams while electro-optic modulators add radio frequency sidebands to the laser light for the locking of laser beams to optical cavities.} 
\label{fig:figure3}
\end{figure}

 A schematic diagram of an advanced laser interferometer system such as Advanced LIGO is shown in Figure~\ref{fig:figure3}. The improvements from initial LIGO to Advanced LIGO is discussed in sect. 2 of this paper.  The design is a major upgrade of the initial LIGO detectors, combines improved vibration isolation to eliminate most of the seismic noise at low frequency, larger test masses with monolithic suspensions and higher acoustic quality factors to reduce thermal noise, and higher optical power to decrease the shot noise and bring the detector close to the standard quantum limit at ~100 Hz. The high optical power design necessitates the use of transparent reaction masses within the power recycling cavity.  In addition, numerous technical improvements are beyond the scope of this review.

Early commissioning of Advanced LIGO has already demonstrated the validity of the design, as reported in sect. 2 of this paper. Current operation is at reduced power levels, but will be increased once techniques for suppression of parametric instability are implemented. As expected, sensitivity is substantially increased even at ~10\% of maximum power to a spectral strain sensitivity of a few parts in $10^{24}$.

\begin{figure}[H]
\centering
\includegraphics[scale=1]{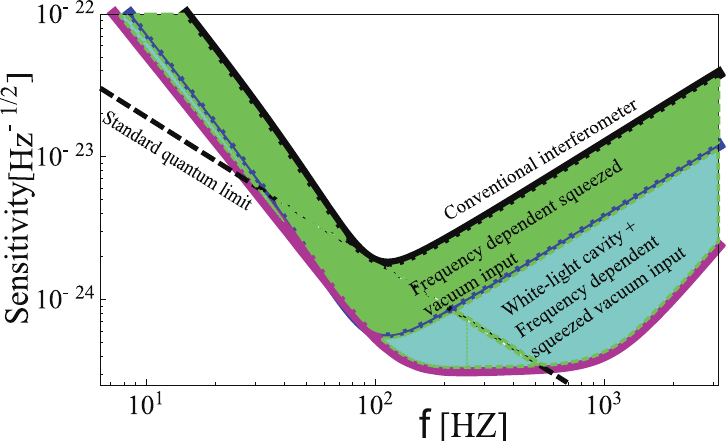}
\caption{(Color online) Sensitivity enhancement beyond the free mass standard quantum limit that can be obtained by using frequency dependant squeezing (second curve), and by combining this with white light signal recycling (lower curve). This result assumes an Advanced LIGO type interferometer with reduced mirror coating thermal noise.}
\label{newsens}
\end{figure}

There has been much effort to develop techniques to improve sensitivity beyond the standard quantum limit. These techniques include the use of squeezed light techniques and the use of optomechanics to create non-classical devices. One approach is to modify the quadrature uncertainty of the vacuum fluctuations that enter the interferometer dark port. These fluctuations which normally represent a coherent state (equal uncertainty in both signal quadratures), can be modified by a parametric amplifier, such that the quadrature uncertainty is squeezed in such a way that they contribute minimum noise to the GW strain signal sidebands.  To date, SNR enhancement by 4.3 dB at GEO and 2 dB at LIGO has been demonstrated at high frequencies. Further improvements can be attained by using lower optical losses and by using frequency dependent squeezing which changes the squeezing angle in phase space as a function of frequency so as to suppress photon shot noise at high frequency and photon radiation pressure noise at low frequency (See Figure~\ref{newsens}).

Another very promising approach to improvement is by using white light signal recycling. While squeezing changes the noise entering the interferometer, signal recycling can change the dynamics of the interferometer such that the detector absorbs more signal from the GW \cite{yiqiu}. Normally signal recycling gives rise to narrow band sensitivity improvement. White light signal recycling allows resonant power build up across a broad band of signal sideband frequencies. Realisation of this technique calls for non-classical devices, in which the nexus between bandwidth and resonant gain is overcome. Such devices can be created using optomechanics as described in~ sect. 6 in ref. \cite{nextGW}

Optomechanics creates non-classical devices by coupling very narrow linewidth mechanical resonators to laser light via radiation pressure forces. In principle, appropriately engineered devices could be used in the output optics to access approximately one order of magnitude sensitivity enhancement. Figure~\ref{newsens} shows the areas of improvement that can be obtained by modifying the output optics in the form of frequency dependent squeezing and white light signal recycling.

Work is underway to develop devices to enable the above sensitivity enhancements to be accessed, as discussed in an accompanying paper in this issue (see sect. 6 in ref. \cite{nextGW}). Optomechanics provides a path for existing detectors to achieve future improvements. Realisation of frequency dependent squeezing can be expected in the short term, but white light cavity techniques will require the development of practical thermal noise free optomechanics, a process which is likely to take some years of development.

\subsection{Need for a worldwide array of detectors for gravitational wave astronomy}\label{sec:4}
GW detection relies on widely separated detectors to minimise seismic or electromagnetic correlations between detectors. By separating detectors widely on the Earth's surface it is also possible to use time of flight triangulation to determine the location of sources on the sky. This can be achieved by incoherent comparison of signal arrival times, or optimally, by the coherent combination of all signals.

Chu et al.  \cite{chuqi12} have analysed  various networks of detectors. Assuming detectors to have equal sensitivity, one can determine the distribution of angular resolution achievable for different arrays. Monte-Carlo simulations were used to identify 200 random sky locations for binary neutron star (BNS) coalescence events. Standard masses of 1.4 solar masses for the individual neutron stars of the BNS system are used. A second order post-Newtonian approximation was used to define the BNS waveforms. The distances of those events were adjusted so that the combined SNR for the three detector LHV network is always at 10. Therefore the effects of adding one or more detectors to the  network can be fairly compared. The Wen-Chen~\cite{wen} formulae were used to calculate the localisation accuracies for different detector networks.

\begin{figure}[H]
\centering
\includegraphics[scale=0.98]{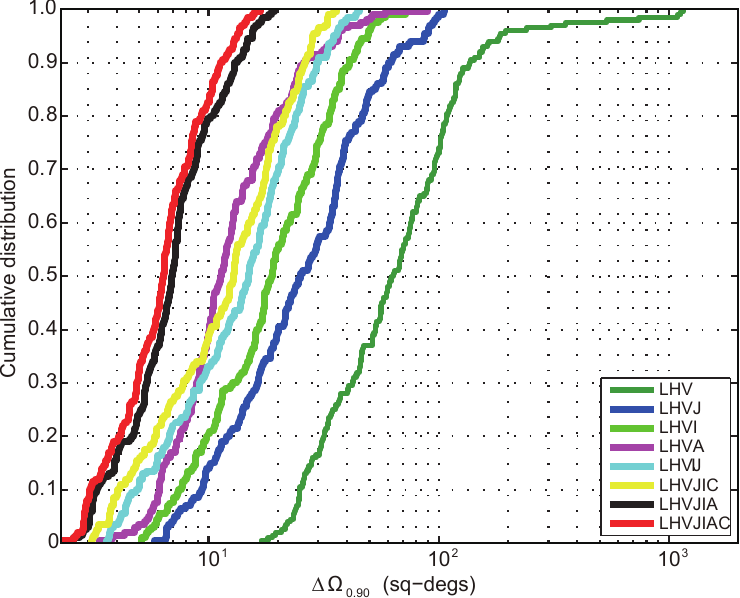}
\vspace*{-3mm}
\caption{(Color online) Cumulative distribution of error ellipses at 90\% confidence level of the 8 detector networks. L$-$Livingston (US, LIGO); H$-$Hanford (US, LIGO); V$-$Virgo (Italy); J$-$KAGRA (Japan); I$-$LIGO-India; A$-$AIGO (Australia); C$-$China. Assuming that GW sources are randomly located with respect to the detector array, the figure shows the fraction of sources with given angular resolution for different array configurarions. Angular resolution varies across the sky by an order of magnitude. A good measure of the array performance is the median value. It is clear that the addition of a southern hemisphere detector has the greatest benefit as discussed in the text.} 
\label{fig5_8in1}
\end{figure}

Figure~\ref{fig5_8in1} shows a comparison between the existing three detector arrays (LIGO and Virgo) and planned detectors in Japan (KAGRA) and LIGO-India. Finally, two new detectors are considered: a) a detector in China and b) a pair of detectors in China and Australia.  The detectors and their locations are listed in Table~\ref{tab2_7detectors}.

We can see from Figure~\ref{fig5_8in1} that adding a new detector in Australia (AIGO) alone has similar angular resolution performance to the combined effect of adding both KAGRA and LIGO-India. Chu et al.~\cite{chuqi12} studied the localisation accuracies for burst events~ with~ different~ detector~ networks ~ and

\begin{table}[H]
\caption{The detector properties considered in this paper. This table is similar to Table 1 of ref. ~\cite{chuqi12} except that one more detector at China is added. The location of the Chinese detector is selected at the position of Zhejiang province}\label{tab2_7detectors}
\begin{center}
\footnotesize
\begin{tabular}
{p{2.5cm}p{0.5cm}p{1cm}p{1cm}p{1.3cm}}
\toprule
{Detector} &{Label}&{Longitude}&{Latitude}&{Sensitivity}\\
\hline
LIGO Livingston, USA &~~L&$-90.77^\circ$&$30.56^\circ$&advL\cite{advL}\\
LIGO Hanford, USA &~~H&$-119.41^\circ$&$46.45^\circ$&advL\\
Virgo, Italy &~~V&$10.5^\circ$&$43.63^\circ$&advV\cite{advV}\\
KAGRA, Japan &~~J&$137.18^\circ$&$36.25^\circ$&advL\\
LIGO-India, India &~~I&$74.05^\circ$&$19.09^\circ$&advL\\
AIGO, Australia &~~A&$115.87^\circ$&$-31.95^\circ$&advL\\
Chinese detector, Zhejiang, China  &~~C& $120.5^\circ$&$29.2^\circ$&advL\\
\bottomrule
\end{tabular}
\end{center}
\end{table}
\vspace{-5mm}

\noindent showed a similar trend. A Chinese detector alone added to the existing/planned LHVJI network improves the network resolution only marginally from 14.92 deg$^2$ to 12.86 deg$^2$. An added Australian detector alone improves the network angular resolution by almost a factor of two, to  6.92 deg$^2$. A China-Australia pair would improve the network from 14.92 deg$^2$ to 6.36 deg$^2$. Such an improvement in localisation ability significantly reduces source confusion in multi-messenger astronomy, which becomes a greater and greater issue as the horizon distance is increased.

If a China-Australia pair of detectors consisted of next generation detectors of 8 km arm length and 4-fold greater strain sensitivity,  as discussed in ref.~\cite{nextGW}, the  improved SNR would allow the network angular resolution to be improved by a substantially larger factor.

\begin{figure}[H]
\centering
\includegraphics[scale=0.35]{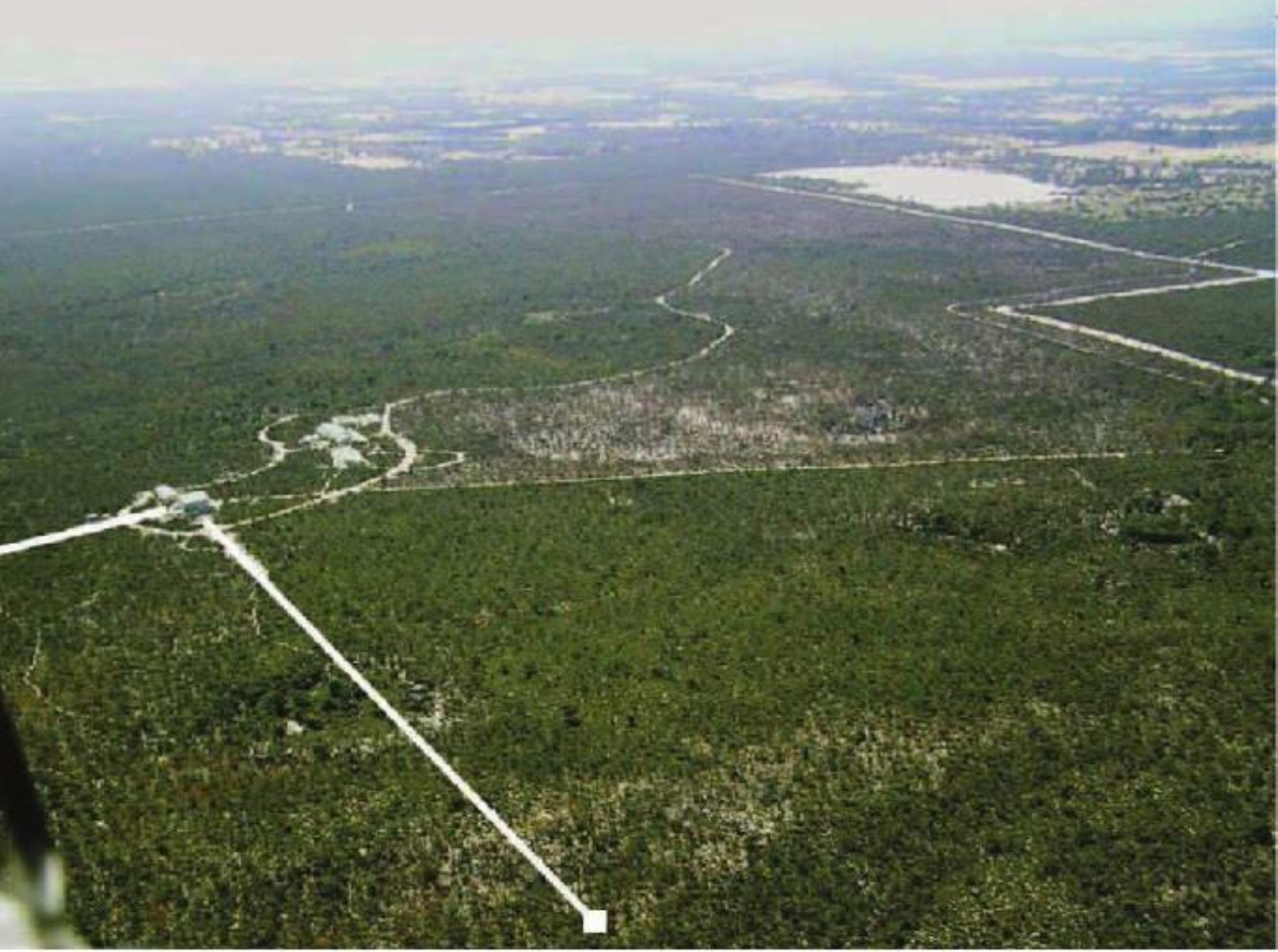}
\caption{(Color online) Aerial photos of the flat sand plain proposed as a site for a future southern hemisphere GW detector. The proposed long arms are added in this photo.} 
\label{fig6Gingin}
\end{figure}

A site for the proposed AIGO detector in Australia \cite{AIGO} has been provided by the West Australian Government,  at Gingin about 75 km north of Perth. The site is shown in Figure~\ref{fig6Gingin}.

On this site the Australian International Gravitational Observatory (AIGO)  research facility has been developed \cite{Gingin1}. It includes an 80 m research interferometer with high performance vibration isolators.  Very low loss suspension systems have been developed\cite{dumas,barriga1}.  Test masses made of sapphire (south arm) and fused silica (east arm) have been used for testing high optical power interferometry techniques. In particular it has undertaken significant studies of thermal lensing, three mode opto-acoustic interactions and methods for suppressing parametric instability. The facility includes accommodation for visiting scientists. A large public education centre called the Gravity Discovery Centre is situated beside the AIGO research facility \cite{GDC}.

\subsection{Conclusion}\label{sec:6}

We have summarised the spectacular development of technology for GW astronomy that has occurred over the past 40 years. We have shown that the first signals are likely to be detected in the near future, and we have summarised a conservative design for a future detector that could access a volume of the universe ~60 times larger than expected for the present Advanced LIGO detectors currently under construction.
There is a very strong scientific case for building a pair of  8 km interferometer detectors, one in China and one in Australia. The design and rationale for such a detector is discussed in ref. \cite{nextGW}. Such a pair of detectors would improve the angular resolution of the world-wide GW detector array by more than a factor of 2. This would greatly improve the capability of multi-messenger astronomy by reducing source confusion,  and allow a detailed census of neutron star coalescence events. The increased sensitivity would allow detection of black hole coalescence events at distances of several Gpc.  The substantial improvement in SNR for stronger events would allow exploration of black hole normal modes, \linebreak
\vspace*{-3mm}

\noindent the equation of state of neutron stars and the testing
of fundamental theorems of black hole physics.

\section{Introduction to ground based gravitational wave detection and a review of the Advanced LIGO detectors\ \ \ \ \ \ \ \ \ \ \ }

\emph{This is a segment of a larger review paper on the ``Detectors for Gravitational Waves". It constitutes half of the section on ground based detectors. In this note, we review the basics of ground based detectors of the interferometer genre, concentrating in particular on the Advanced LIGO detectors.}

\subsection{Gravitational wave basics}\label{sec:GWBasics}
GW are a natural consequence of Einstein's general theory of relativity, and their existence was predicted almost exactly 100 years ago \cite{Einstein1916,EinsteinEngle1917,Einstein1918}. Starting from the Einstein equation
\begin{equation}\label{eq:Ein}
G_{\mu \nu} = \frac{8\pi G}{c^4} T_{\mu\nu}\,.
\end{equation}
If we assume a weak gravitational field and decompose the spacetime metric into a flat Minkowski background piece $\eta_{\mu\nu}$ and a perturbative piece $h_{\mu\nu}$, then eq.~\eqref{eq:Ein} simply implies $\Box h_{\mu \nu}=0$ \cite{MTW}, with $\Box$ being the flat spacetime wave operator, therefore metric perturbations propagate as a wave. In the transverse-traceless gauge, one can further write down the explicit form of $h_{\mu \nu}$ for a wave propagating in the $z$ direction, which is
\begin{equation}\label{eq:hWave}
{h}_{\mu \nu} =\begin{bmatrix} 0 & 0&0&0 \\ 0 & h_{+} & h_{\times} & 0 \\
0 & h_{\times} & -h_+ & 0 \\ 0 & 0&0&0 \end{bmatrix} {\rm e}^{{\rm i} (\omega t - k z)},
\end{equation}
where $h_+$ and $h_{\times}$ are two real numbers indicating the amplitudes of the two polarizations of the GW, and the basis under which the matrix is written down is $\{\d t, \d x, \d y, \d z\}$.

\begin{figure}[H]
\centering
\includegraphics{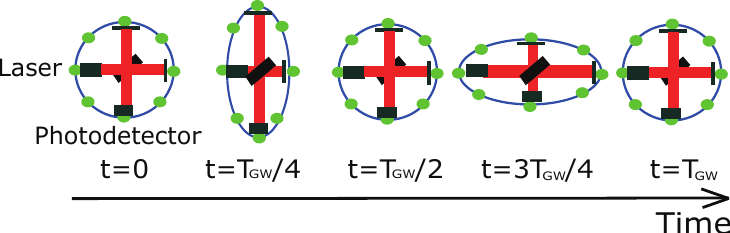}
\caption{(Color online) A cartoon depiction of the effect of GW on the two arms of a laser interferometer. The length changes in the $x$ and $y$ arms are grossly exaggerated.}
\label{fig:StrainEvo}
\end{figure}

To see how this metric perturbation wave manifests physically, let's imagine for the moment a laser interferometer consisting of two arms of (when in the absence of GW) equal lengths $L$, oriented in the $x$ and $y$ directions (see Figure~\ref{fig:StrainEvo}). Now let a GW containing only the $+$ polarization (for simplicity) passes by. The metric perturbation
would give rise to a dynamic perturbation of the arm length. But the speed of light is unchanged, so the round-trip times/accumulated phases of the laser beams would be altered by the GW. The matrix part of eq.~\eqref{eq:hWave} shows that the arm length shifts in the $x$ and $y$ directions are of opposite signs (when one is compressed, the other is stretched), so the phase difference in the two arms thus produced would add coherently. A simple integration reveals the relationship $\Delta L /L = h_+/2$, and so the GW is said to produce a strain \cite{Adhikari:2013kya}. The time dependent factor in eq.~\eqref{eq:hWave} further implies that the effect of the GW on, say, the $x$ direction alternatives between stretching and compressing in a sinusoidal fashion over time, as seen when we scan across the panels of Figure~\ref{fig:StrainEvo}.

Linear algebra further tells us that the directions that receive the largest amount of stretching and compressing ($x$ and $y$ in our example above) are the eigen-directions of the matrix
\begin{equation}
\begin{bmatrix}  h_{+} & 0 \\
0 & -h_+ \end{bmatrix}\,,
\end{equation}
or in other words the spatial transverse (to the propagation direction $z$) part of the matrix in eq.~\eqref{eq:hWave}. Had we kept the $\times$ polarization instead, the two eigen-directions would then still be orthogonal to each other, but rotated against those of the $+$ polarization by $\pi/4$. This explains the notation $h_{\times}$, and contrasts with the $\pi/2$ rotation between polarizations for the electromagnetic waves (see e.g. ref.~\cite{Nichols:2011pu} for more analogies and comparisons).  A GW propagating at an arbitrary angle to the plane of the interferometer will still produce a signal, however with a smaller amplitude as we discuss below.


\subsection{A network of ground based detectors \label{sec:Network}}
Let's now turn to the main topic of this review paper, namely the detection of GW. In this review, we will concentrate on ground based detectors.

In sect. 2.1, we have used a laser interferometer to construct the example of how a GW interacts with matter. This choice is not arbitrary, it is made because interferometers of various design parameters are the most prevalent form of GW detectors \cite{Abramovici:1992ah,Abbott:2007kv,Acernese:2008zzf,Luck:2006ug,2009CQGra..26t4020A}. We saw from the aforementioned example that GW detection is in essence a length measurement, and it has to be very accurate because the expected strain produced by detectable astronomical sources is of the order $10^{-23}$, or $0.01~ \overset{\circ}{\rm A}$ of $\Delta L$ for an $L$ of the distance between the earth and the sun. For $L=10$ km, this translates into $\Delta L=10^{-19}$ m. If we try to use laser ranging (i.e. measuring travel time of laser pulses) to observe such a miniscule distance shift, the timing resolution needs to be of the order of $3\times 10^{-28}$ s, but the current state-of-art can only offer $25\times 10^{-21}\ \text{s Hz}^{-1/2}$ \cite{Quinlan2013}.

A technologically more accessible alternative then is to use the arms of interferometers as length comparators \cite{Weiss:2003}, so instead of measuring absolute lengths, we tackle the more tractable problem of extracting relative length differentials. Presently, there is a network of second generation ground-based laser interferometer GW detectors in operation or under construction \cite{Dooley:2014iga} (See Figure 8). In the United States, a pair of Advanced LIGO (https://www.advancedligo.mit.edu/) \cite{Harry:2010zz} (which stands for ``Laser Interferometer Gravitational Wave Observatory") detectors in Hanford, Washington and Livingston, Louisiana have recently (in May 2015) been dedicated and began observing runs in September 2015. We will provide an overview of these detectors in sect.~\ref{sec:aLIGO}. In Europe, the Virgo detector is currently undergoing an upgrade into Advanced Virgo (https://wwwcascina.virgo.infn.it/advirgo/) \cite{aVIRGO:2012}. As of June 2015, its mode-cleaner has been locked, and the end of installation is expected in fall 2015. Also in Europe, the smaller GEO600 detector \cite{Dooley:2014nga} has been running while its larger siblings are down for upgrades, keeping watch on the GW universe. In parallel, it is also shouldering the task of testing many critical new technologies while being upgraded to GEO600-HF \cite{Luck:2010rt,Grote:2010zz,Affeldt:2014rza}. Looking slightly further ahead, the KAGRA (Kamioka Gravitational Wave Detector) detector (http://gwcenter.icrr.u-tokyo.ac.jp/en/) \cite{Somiya:2011np} is under construction in Japan. When completed, it will utilize cryogenic mirrors chilled to $20$ K in an underground site at the Kamioka mine. The tunnels have been completed as of 2015. Finally, in the advanced planning stage, there is LIGO-India, where one Advanced LIGO interferometer from the Advanced LIGO Construction Project will be provided by the LIGO Lab (with its UK, German and Australian partners) and installed in india.

\subsection{Detecting gravitational waves using interferometry}\label{sec:Interferometry}
\subsubsection{The basics and detector response to gravitational waves}
All of the large GW interferometers are of an essentially Michelson type \cite{Adhikari:2013kya}. Figure~\ref{fig:InterferometerCartoon} is a simplified cartoon depiction of the LIGO detector jointly operated by Caltech and MIT, as well as an enumeration of its major components (the bottom panel) categorized into the three fundamentals: mechanics, optics and electronics, as well as their combinations (e.g. opto-electronics).
We shall use it, together with its upgraded version, the Advanced LIGO, as an example, and walk through its major design features. While our review is aimed at a broader audience and thus cursory, we note that much \linebreak
\vspace*{-3mm}

\begin{figure}[H]
\centering
\includegraphics[scale=0.38]{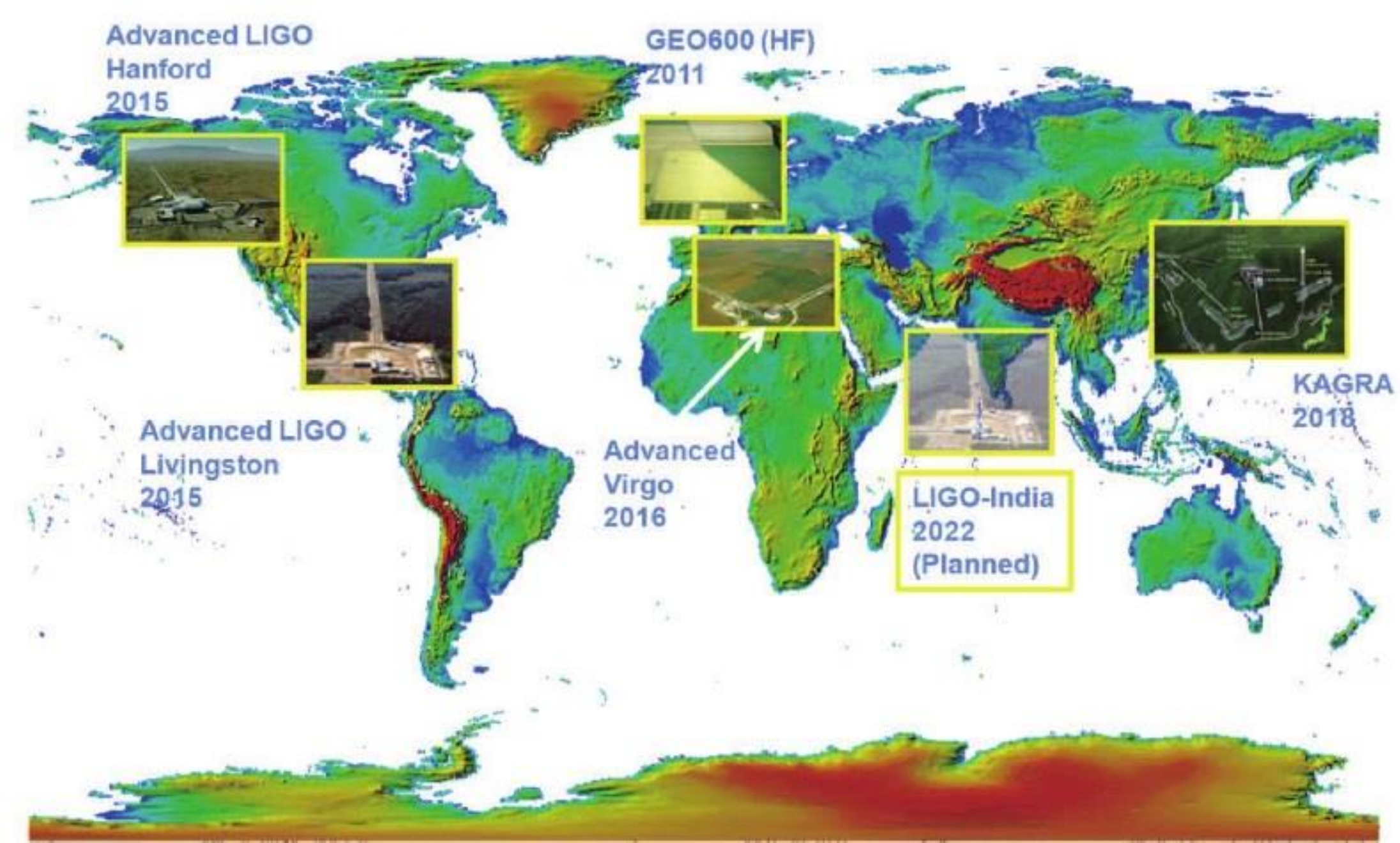}
\caption{(Color online) A network of advanced detectors presently in operation or under construction. Their (expected) construction/upgrade completion dates are shown beneath the detector names.}
\label{fig:DetectorNetwork}
\end{figure}

\noindent more comprehensive reviews on the interferometers can be found in refs.~\cite{Adhikari:2013kya,Pitkin:2011yk,Cella:2011zz,Freise:2009sf,Braginsky2008,Aufmuth:2005sv,Barish:1999vh,SaulsonBook,Weiss:1999fs,Giazotto:1989iq}.

An obvious starting point when designing a Michelson interferometer is to decide upon the arm length. One factor is the usual ``quarter wavelength" consideration, where the effect of a passing GW is most salient when its wavelength is four times that of the arm length $L_{\rm arm}$. For a GW of frequency $1$ kHz, this translates into $L_{\rm arm} \approx 75$ km. Another consideration is that since the GW produces a strain, it is desirable to have larger $L$ in order to produce a large $\Delta L$. Therefore, even though $75$ km arms are not practical from both land availability and budgetary points of view, an arm length as large as possible is still desirable. For Initial and Advanced LIGO detectors, the arm lengths are $4$ km. Beyond this, the detectors also use Fabry-Perot (FP) cavities to increase the effective arm length (see Figure~\ref{fig:InterferometerCartoon}), essentially bouncing light within the cavity many times to increase the amount of time it interacts with GW. The inclusion of the FP cavities is a significant deviation of the GW detectors from the straight-forward Michelson interferometer design. The Advanced LIGO can be classified as a Dual-Recycled Fabry-Perot Michelson Interferometer, employing both power- and signal-recycling cavities.

More precisely, a simple calculation shows that the output \linebreak
\vspace*{-3mm}

\begin{figure}[H]
\centering
\includegraphics[scale=0.65]{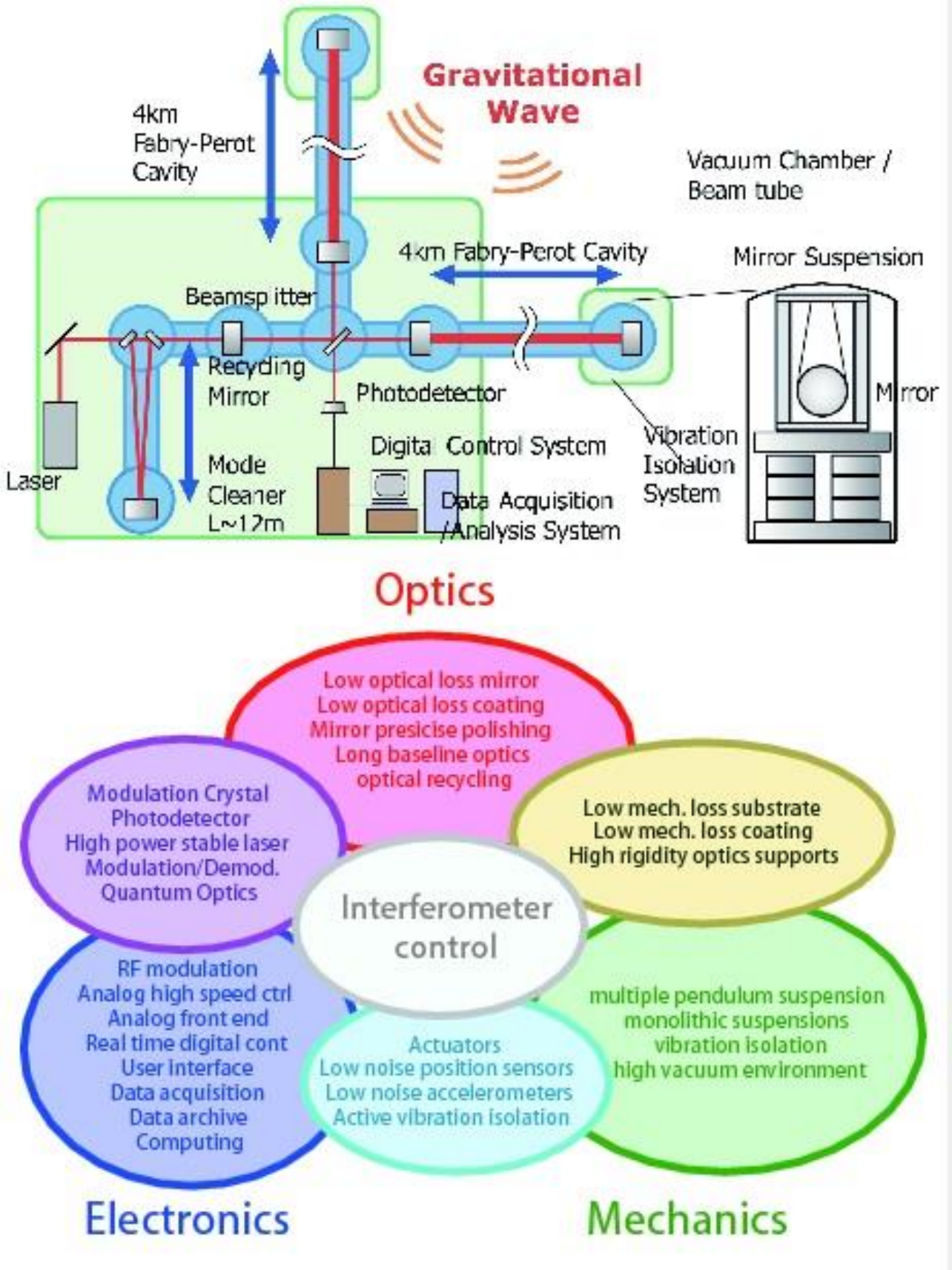}
\caption{(Color online) Top: A simplified cartoon of the LIGO interferometer. Bottom: The various components in an interferometer and how they interact.}
\label{fig:InterferometerCartoon}
\end{figure}
\begin{figure}[H]
\centering
\includegraphics[scale=0.6]{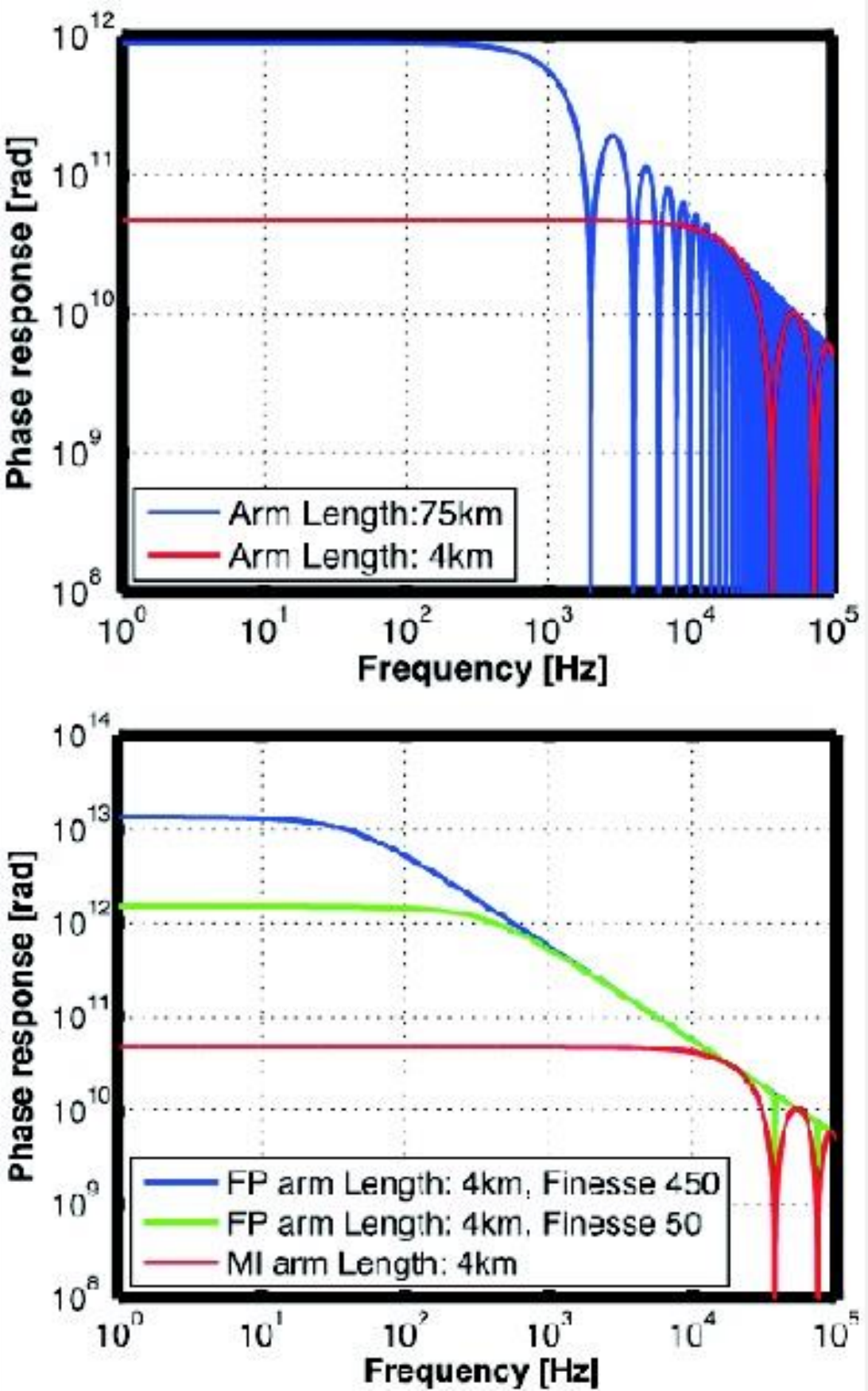}\vspace{-2mm}
\caption{(Color online) Top: the phase response $\mathcal{P}$ of Michelson interferometers with arm lengths $4$ km and $75$ km and without FP cavities. Note $\mathcal{P}$ is to be multiplied with $h_0$, so total phase difference $\Delta \phi \ll \pi$, and we do not need to be concerned with taking $\Delta \phi$ modulo $2\pi$. Bottom: when FP cavities of high finesse are introduced, the phase response receives a significant boost. }
\label{fig:PFCavity}
\end{figure}


\noindent power $P_{\rm out}$ at the output port of a Michelson interferometer is given by
\begin{equation} \label{eq:Power}
P_{\rm out} = [1-\cos(\phi_A-\phi_B)]P_{\rm in}/2\,,
\end{equation}
and so sensitive to the phase difference $\phi_A-\phi_B$ between the lasers in the two arms labeled $A$ and $B$. We would therefore desire configurations where GW generates a large phase difference $\Delta \phi$, which can be estimated to be \cite{Vinet:1988ch}
\begin{equation} \label{eq:Response}
\Delta \phi = \mathcal{P}  h_0 \e^{{\rm i}\omega t}\,, \quad
\mathcal{P} = \frac{2L\Omega}{c}\e^{-{\rm i} L \omega/c}\frac{\sin(L\omega/c)}{L\omega/c} \,,
\end{equation}
where $\omega$ is the angular frequency of the GW, $\Omega$ is the optical angular frequency, and $h_0$ is the amplitude of the GW, which for now can be seen as $h_+$ for a GW incident on the plane of the detector orthogonally (traveling in the $z$ direction, with the detector located at $z=0$ and its arms in the $x$ and $y$ directions). We plot $\mathcal{P}$ as a function of $\omega$ in the upper panel of Figure~\ref{fig:PFCavity}, and observe that $L=75$ km is indeed optimized for frequencies below $\omega = 1$ kHz, while the actual arm length of $4$ km falls somewhat short (i.e, has a lower phase response to the bandwidth of interest). Luckily, the FP cavity comes to the rescue. The amount of ``bouncing'' (rate at which light leaks out of the cavity) occurring in a FP cavity is controlled by its finesse $\mathcal{F}=\pi\sqrt{r_1 r_2}/(1-r_1r_2)$, with $r_1$ and $r_2$ being the reflectivity of the mirrors at the two ends of the cavity. When the finesse is sufficiently high, $\mathcal{P}$ can receive a significant boost, giving us a phase response comparable to a $75$ km arm length interferometer (see the bottom panel of Figure~\ref{fig:PFCavity}) even when the actual arm length is only $4$ km (albeit at the expense of frequency bandwidth; the higher the finesse, the lower frequency the interferometer response function roll-off).

\begin{figure}[H]
\centering
\includegraphics[scale=0.5]{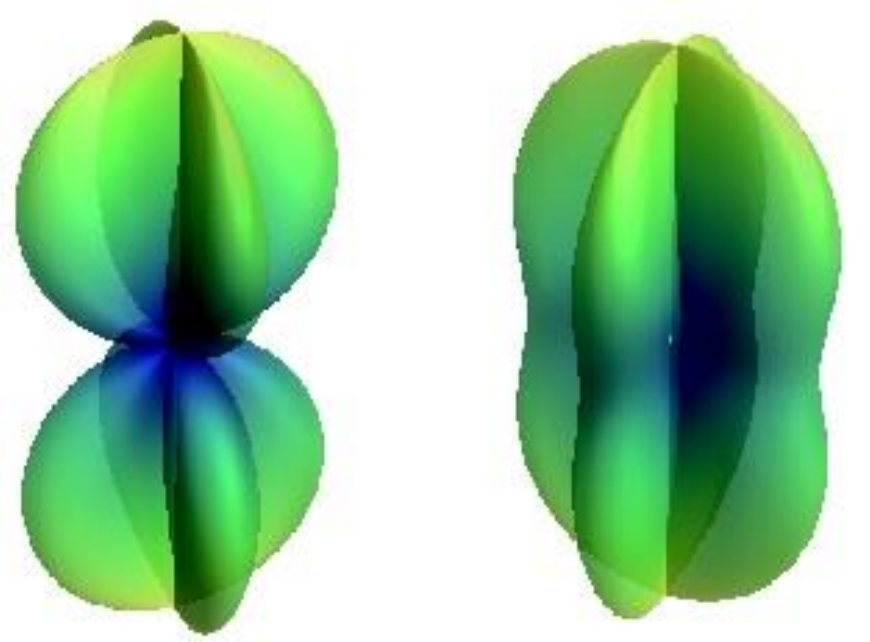}
\caption{(Color online) The antenna response of the detector to GW of $\times$ (left) and $+$ (right) polarizations, greater distance to the origin and lighter color signifies larger $|h_{yy}-h_{xx}|$ values. The interferometer arms (not shown) lie in the plane at the center of the response diagrams, oriented along the lobes of the $\times$ response, c.f.~Figure~2 in ref. \cite{Adhikari:2013kya}.}
\label{fig:Antenna}
\end{figure}

We initially simplified our treatment by assuming that the GW is of $h_+$ polarization propagating in the $z$ direction. In reality however, the GW can come in from an arbitrary direction $(\theta,\phi)$ (spherical coordinate system centred on the detector), containing both $+$ and $\times$ polarizations. In that case, we can replace the $h_0$ in eq.~\eqref{eq:Response} by $|h_{yy}-h_{xx}|$ with \cite{Christensen:1992wi}
\begin{align}
h_{xx}&= -\cos\theta \sin(2\phi) h_{\times} + (\cos^2\theta\cos^2\phi-\sin^2\phi)h_+\,, \\
h_{yy}&= \cos\theta \sin(2\phi) h_{\times} + (\cos^2\theta\sin^2\phi-\cos^2\phi)h_+ \,.
\end{align}
We plot $|h_{yy}-h_{xx}|$ as a function of direction in Figure~\ref{fig:Antenna}, which depicts the Antenna Response of the detector for GWs of the two different polarizations.  In general, an astrophysical gravitational could contain both polarization components; the interferometer response is a weighted sum of $+$ and $\times$ responses.

\subsubsection{Noises}
Having seen the basic strategy on optimizing the detectors' response to the GW, we now turn to the suppression of the noises. This is a vital component of detector design, as the observable distance is inverse proportional to the noise level. A reduction of noise by a factor of $10$ increases the observable distance by the same factor and the potential candidate events by $10^3$ (because the rate scales as the volume of space covered). Given that the estimates \cite{ref12,Belczynski:2001uc,Belczynski:2006zi,Phinney:1991ei,HawkingIsraelBook,Cutler2002,Sathyaprakash:2009xs} for the source event rates are wildly uncertain ($\approx 10^{-8}$--$10^{-5}\ \text{Mpc}^{-3}\ \text{yr}^{-1}$ for binary neutron star mergers), this could make the difference between many events per year and no event during the typical life span of a scientist.

\begin{figure}[H]
\centering
\includegraphics[scale=0.9]{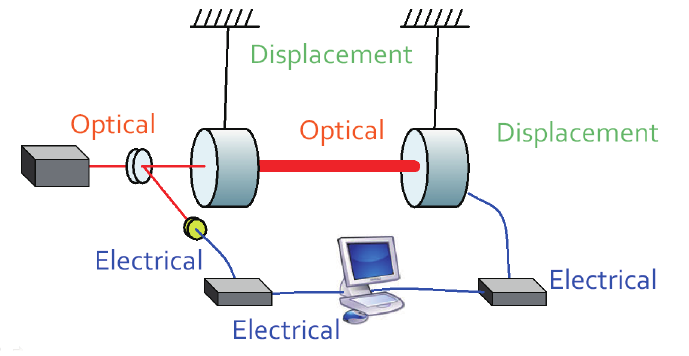}\vspace{-2mm}
\caption{(Color online) The three types of noises and the places where they enter into the measurements.}
\label{fig:NoiseClasses}
\end{figure}

\begin{figure}[H]
\centering
\includegraphics[scale=0.4]{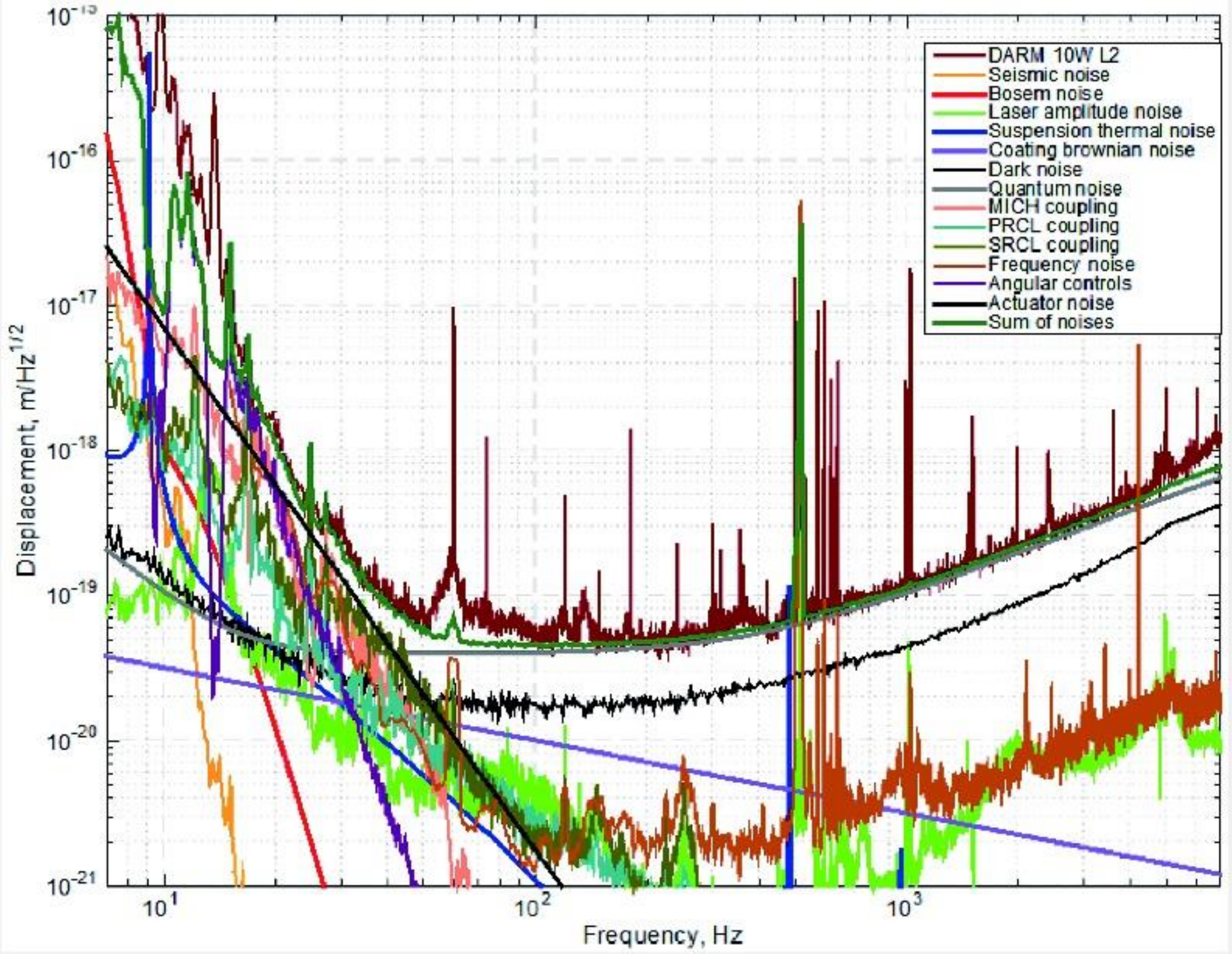}\vspace{-2mm}
\caption{(Color online) A typical Advanced LIGO noise budget, broken down into different noise components.}
\label{fig:ALigoNoise}
\end{figure}

When viewed from the basis of the three classes of interferometer components shown in Figure~\ref{fig:InterferometerCartoon}, the major types of noises are displacement noises that enter into the mechanical components, the optical noises that arise in the optics, while the electrical noises emerge in the electronics. Figure~\ref{fig:NoiseClasses} is a cartoon depiction of these three types of noises and the places where they creep into our measurements, and Figure~\ref{fig:ALigoNoise} is the Advanced LIGO Livingston noise budget, broken down into contributions from the various different types of noise sources. We will discuss the displacement and optical types in some detail below.

1. \textbf{Displacement noises}\quad
This is the uncertainty in the distance between the two mirrors in Figure~\ref{fig:NoiseClasses} due to mechanical excitations: (1) seismic motions that move the mirrors, (2) thermal noise, and (3) Newtonian gravity noise.

(1) To reduce seismic noise, one can place the mirrors on active platforms controlled by appropriate feedback systems \cite{Hensley1999,Newell1997}. These are quite good at reducing the lower frequency seismic noises. One can then place the mirror within a damped harmonic oscillator as an additional passive isolation \cite{Matichard:2014cqa,Matichard:2014laa}, which provides vibration isolation above its resonant frequency. Let $\mathsf{d}$ and $\mathcal{D}$ denote the displacement of the mirror and the ground respectively, then we can lower the peak ratio for $\mathsf{d}/\mathcal{D}$ by tuning the damping coefficient\linebreak
\vspace*{-3mm}

\noindent (the downside is that isolation at non-peak frequencies gets worse), make the $\mathsf{d}/\mathcal{D}$ curve drop with a steeper inclination into higher seismic motion frequencies by making the oscillator multi-staged \cite{Aston:2012ona} (the price to pay is the appearance of multiple resonant peaks), and produce better isolation at higher frequencies by lowering the resonant frequency (this is more complex to realize). In practice, the pros and cons of these measures are weighed, and a combination of them is deployed \cite{Giaime1996,Ponslet1998,Accadia2011,Acernese2012,
Ballardin2001,Marka:2002ct,GroteThesis,Plissi1998,
Somiya:2011np,2002CQGra..19.1591A}.

(2) A host of thermal noises affect the mirror, consisting of a bulk substrate and layers of optical coating material, as well as the suspension wires/fibers (the mirrors are suspended as pendulums, forming part of a multi-stage harmonic oscillator and can move freely in response to GW \cite{Adhikari:2013kya}). There are several types of thermal noises: Brownian motion presents itself as thermally-excited body modes in the mirror and pendulum modes in the suspension fibers; thermo-elastic noise is the mirror motion and displacement caused by temperature fluctuations \cite{2003PhLA..312..244B}; temperature fluctuations also causes fluctuations of the mirror refractive index, termed the thermo-refractive noise (which can in fact be made to largely cancel with the thermo-elastic noise \cite{Evans:2008ei,HarryBook}). The choice of mirror and suspension fiber material is important for the suppression of the remaining thermal noises. Present day substrate material such as fused silica, sapphire and silicon have very good thermal properties. However, the mirror coating material, chosen for their optical and not mechanical/thermal qualities, usually fare worse in this respect, and needs to be carefully selected and the coating layer structure closely scrutinized \cite{Bassiri:2013,Evans:2012Coating,Hong:2012jv,Flaminio:2010zzb,Kondratiev:2011kh,Harry:2006qh,Harry:2006Coating,Harry:2001iw,Rowan:2005yj,2003PhLA..312..244B,Penn:2003nh}. Indeed, the Brownian noise of the optical coating constitutes one of the limiting noise sources for Advanced LIGO.

(3) The Newtonian gravity noise is also called the gravity gradient noise. It is the motion induced via gravitational coupling by the mass density fluctuations around the detector, with the dominant type being seismic surface wave (this differs from the seismic noise discussed earlier in the type of force through which the seismic motions affect the test mass mirrors). Some strategies for mitigating this noise include \cite{Driggers:2012ac,Harms:2014soa}: going to quiet places (e.g. underground as KAGRA does), feedforward subtraction, or passive reduction by shaping local topography.

2. \textbf{Optical noises}\quad
These are optics-related noises that contaminate the readout signal. The major subclasses of optical noises are fundamental quantum noises (shot noise, radiation pressure noise), laser technical noises (frequency and intensity noises), modulation noises, and scattered light noise. In Figure~\ref{fig:ALigoNoise}, we plot the noise budget of the Advanced LIGO detector from the Livingston L1 interferometer, noting in particular that the quantum (shot) noises dominate at higher frequencies. We delve into this more deeply below.

Shot and radiation pressure noise originates directly from the quantum nature of light. The number of photons arriving at a photodetector over any period of time is statistically uncertain, and can be approximately modeled using a Poisson distribution. This results in a statistical fluctuation in output power $P_{\rm out}$ and increases the noise relative to the GW signal proportional to $\sqrt{P_{\rm out}}$.  Let $i_{\rm DC}$ be the DC photocurrent (proportional to the mean number of photons per second), and $i_{\rm shot}$ be the shot noise current (essentially the standard deviation of fluctuations in the number of photons). Following the same logic, $i_{\rm shot}$ is then proportional to $\sqrt{i_{\rm DC}}$. One notices that $i_{\rm DC}$ increases faster than $i_{\rm shot} \sim \sqrt{i_{\rm DC}}$ as we turn up the laser power, and so the percentage uncertainty in $P_{\rm out}$ drops. It is therefore beneficial to use more powerful lasers, as far as the shot noise is concerned.

Simply turning up the laser power has unfortunate side effects, however. The laser exerts pressure on the mirrors (a back action), so higher-power induced greater photon number fluctuation in the arm cavities would create a fluctuation of the back action force, causing unwanted motion of the mirror (i.e., displacement noise), so-called radiation pressure noise. A caveat is that the radiation pressure noise of the input laser itself is actually the same in both arms of the interferometer and is cancelled in the signal, it is in fact the coupling between the stable laser light and the vacuum fluctuation injected from the dark port that results in a differential power fluctuation and subsequently noise in the detector
readout \cite{Caves:1981hw,Caves:1985zz,Schumaker:1985zz,Loudon:1981qa}. Nevertheless, increased laser power produces more pronounced fluctuations of this type. An optimal trade-off between the shot and radiation pressure noises can be achieved by tuning the laser power, and the lower limit to the strain noise that can be thus obtained is named the ``Standard Quantum Limit" \cite{BraginskyBook}.  In addition, the mirrors of the interferometer absorb some of the laser light, typically 0.1\%--0.5\% of the total input power, resulting in a change in the optical characteristics (effective focal lengths and radii of curvature) of the mirrors\cite{Hello:1990,Winkler:1991} that negatively impact interferometer performance.  This is typically ably managed through adaptive optical techniques \cite{Lawrence:2004}.



\subsection{Advanced LIGO}\label{sec:aLIGO}
\begin{figure}[H]
\centering
\includegraphics[scale=0.3]{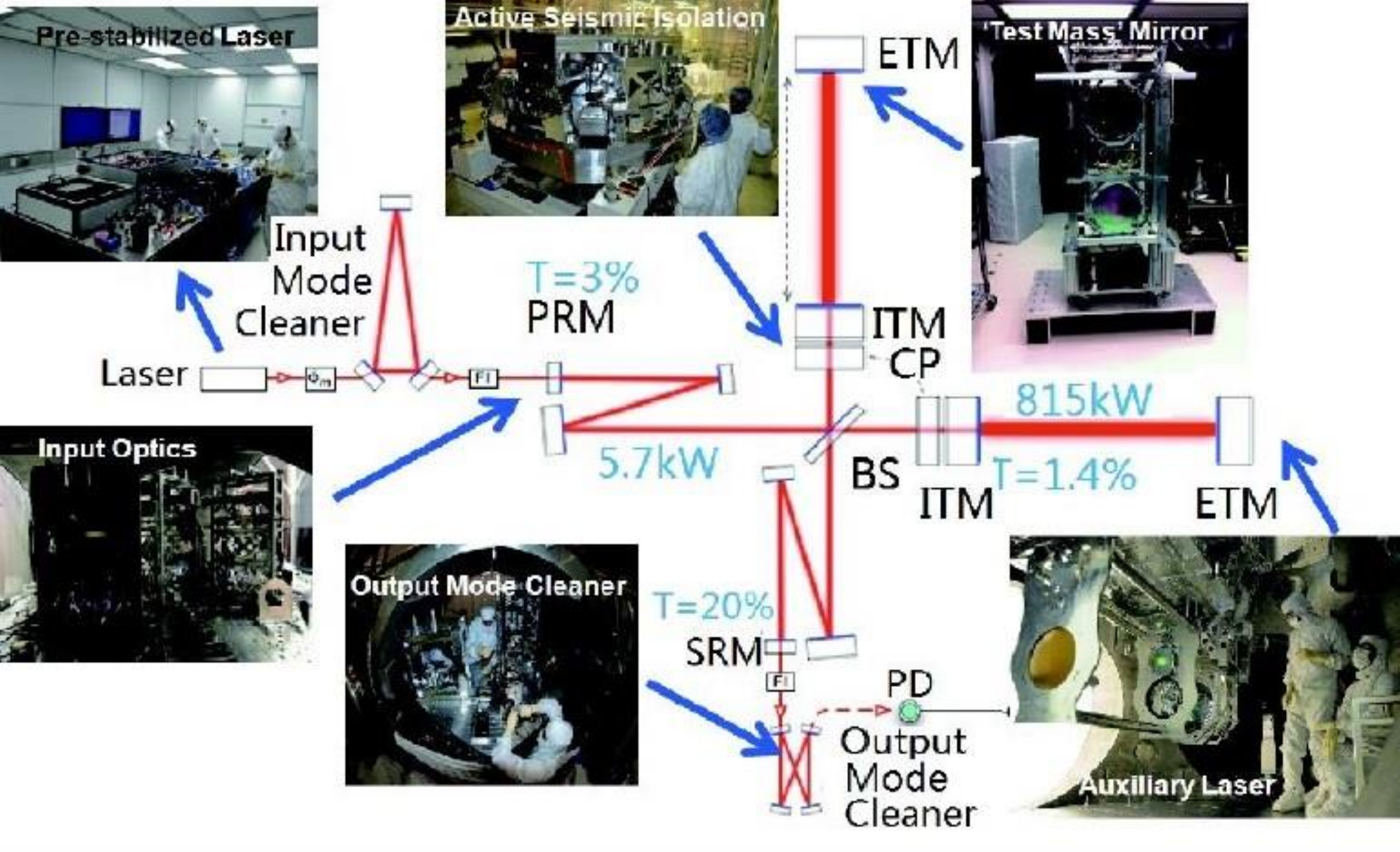}\vspace{-3mm}
\caption{(Color online) Advanced LIGO in pictures, overladen on top of a schematic of the interferometer design. }
\label{fig:ALigoPics}
\end{figure}

Lastly, we offer an overview of the Advanced LIGO project and an update on its current status. The Advanced LIGO is a complete redesign and rebuild of the LIGO interferometers. Some of the new features added include a signal-recycling \cite{Meers:1988wp,Mizuno95} mirror operated in a resonant sideband extraction mode \cite{Mizuno:1993cj,Strain:2003}. This allows to tune the interferometers' frequency response. For example, interferometers are normally operated in broadband mode, but can be tuned to a narrow-band detector with increased sensitivities at specific narrow higher frequency bands (http://lhocds.ligo-wa.caltech.edu:8000/advligo/GWINC).  Also, the laser power has been increased from initial LIGO's $10$ W to $200$ W to reduce shot noise, and the test mass mirrors have become larger ($34$ cm for Advanced LIGO as compared to $25$ cm for initial LIGO) and heavier (increased to $40$ kg from $11$ kg) to suppress thermal and radiation pressure noises respectively. The steel wires suspending the mirrors have been replaced with fused silica (the same material as the mirrors) fibres to reduce suspension thermal noise. Figure~\ref{fig:ALigoPics} is a schematic of the Advanced LIGO design,
together with photos of some of its major components. The Advanced LIGO is 10 times more sensitive than initial LIGO, thus 1000 times more of the universe will be probed, covering on the order of 100000 galaxies, extending out to $200$ Mpc (the average distance to which a 1.4 M$_{\odot}$-$1.4$ M$_{\odot}$ binary neutron star merger can be detected with a SNR of $8$). At its design sensitivity, we expect tens of detections per year for binary neutron star mergers, although there is a large uncertainty in the rates as noted above.

As a project, the Advanced LIGO was funded by the National Science Foundation in April 2008 in the amount of $205$M USD, and an additional $17$M USD was contributed by partners in Germany (Max Planck Albert Einstein Institute), UK (Science and Technology Facilities Council) and Australia (Australian Research Council). The Advanced LIGO Construction Project finished on March 31, 2015, with a formal dedication of the occurring on May 19th, 2015 at the Hanford facility. One interferometer at Hanford and one at Livingston are operational and Advanced LIGO's first observational run began in September 2015. Components for a third interferometer have been assembled and put in storage for future installation in India. The funding for the LIGO-India project has been referred to the Cabinet of the Prime Minister of India, and is currently awaiting approval.

\begin{figure}[H]
\centering
\includegraphics[scale=0.45]{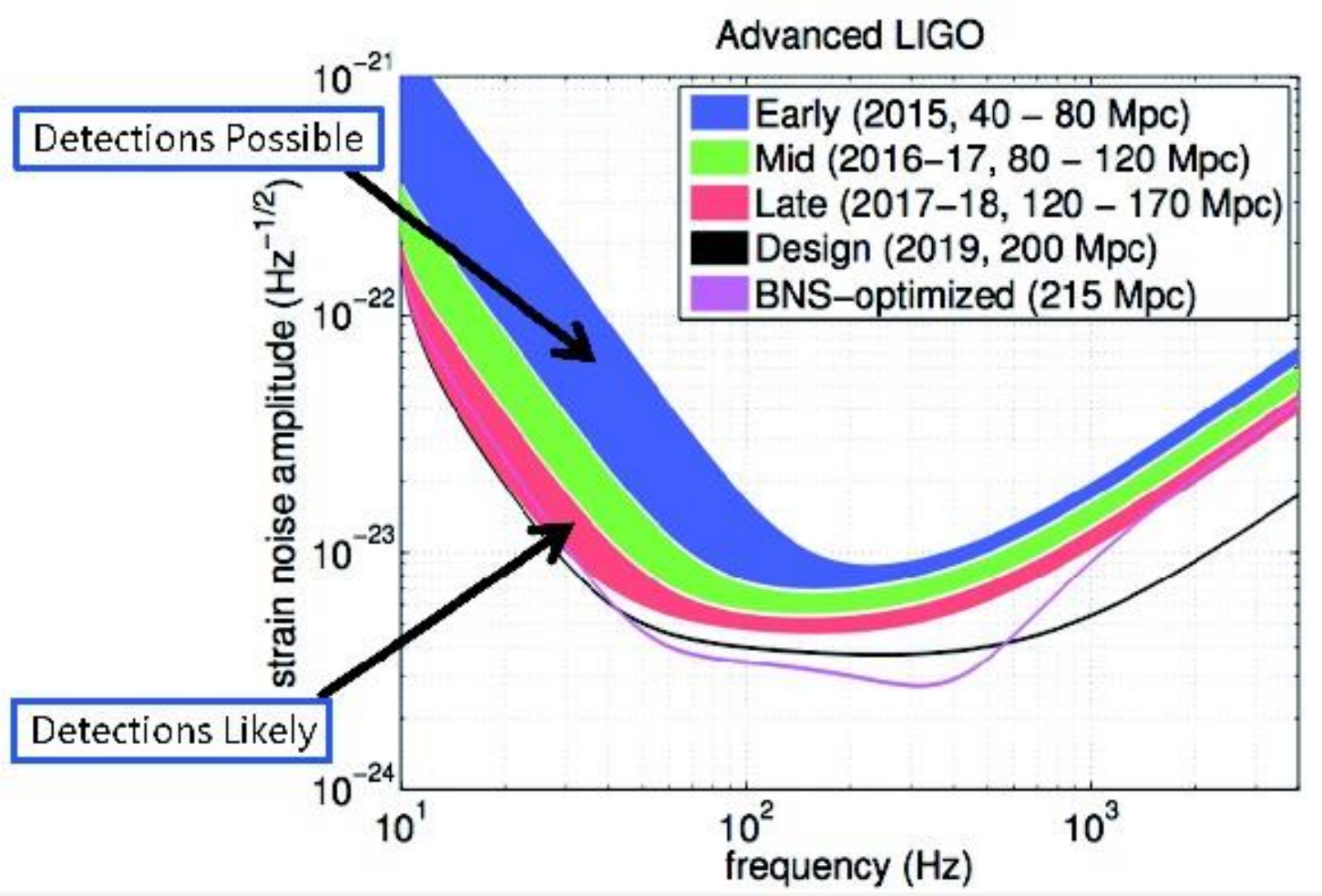}\vspace{-3mm}
\caption{(Color online) The anticipated sensitivity progression for Advanced LIGO over the next five years. Taken from ref. \cite{aasi:2014}.}
\label{fig:ALigoSchedule}
\end{figure}

In the future, improvements will be applied gradually, leading up to the design sensitivity being reached in around 2019. The anticipated time for reaching design sensitivity is shown in Figure~\ref{fig:ALigoSchedule}.
We are keeping our fingers crossed for a timely first detection, and an ensuing blossoming of GW astronomy.

\section{Cryogenic interferometers for enhanced gravitational wave detector sensitivity}
\subsection{Introduction}\label{introduction}

Mechanical thermal noise is one of the main limitation to
\linebreak
\vspace*{-3mm}

\noindent the sensitivity of laser interferometer GW detectors such as
LIGO \cite{Abbott:2007kv}, Virgo \cite{2Accadia}, GEO \cite{Luck:2006ug} and KAGRA \cite{Somiya:2011np}. This displacement noise is a direct consequence of the fluctuation-dissipation theorem according to which any mechanical system affected by some form of energy dissipation will be driven by a stochastic force whose amplitude depends on the system temperature and the dissipation amplitude \cite{5 Callen}. While the total random motion of the system should agree with the well-known expression $E=1/2kT$, the spectral distribution of this displacement will depend on the details of the dissipation mechanism and of the mechanical response of the system to the stochastic force. As a general rule the motion of a mechanical system with lower losses (i.e. larger mechanical quality factor) will be more concentrated at the resonant frequencies of the system while the motion amplitude out of the resonances will be smaller.

\begin{figure}[H]
\centering
\includegraphics[scale=0.4]{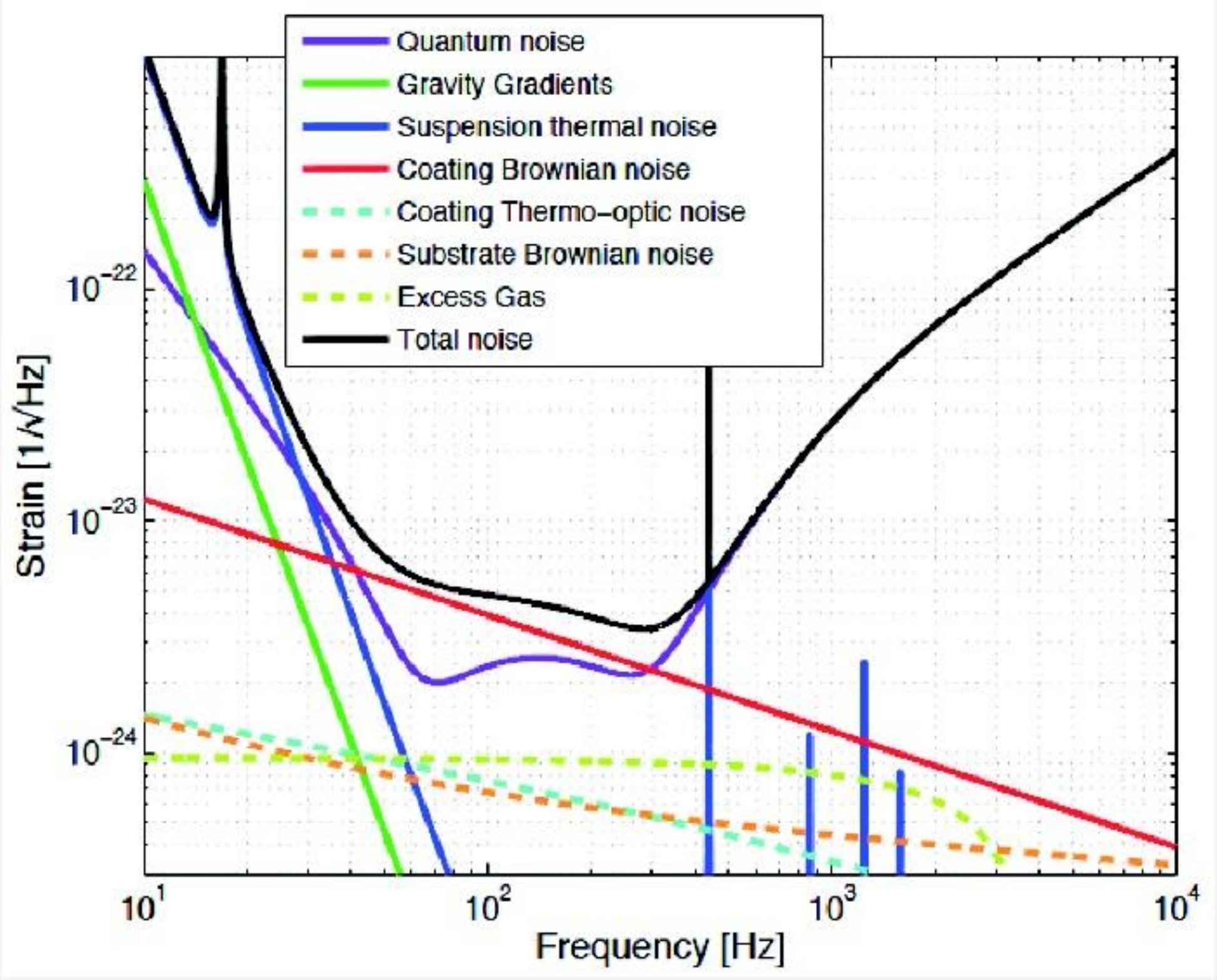}
\caption{(Color online) Advanced Virgo design sensitivity. Coating thermal noise is the limiting factor over a large part of the frequency band. Substrate thermal noise and coating thermos-optic noise are also shown.}
\label{fig1:Virgosens}
\end{figure}

The mirrors of laser interferometer GW detectors as well as their suspensions are mechanical system affected by energy losses and as such are subject to thermal noise \cite{6Saulson}. In the case of the mirror suspension the dominant effect are the mechanical losses in the fibers used to suspend the mirror as well as any loss in the contact points between the fibers and the mirror itself. The case of the mirrors is more complex and several mechanism have been considered. The mechanical losses inside the coatings providing the required mirror reflectivity are the dominant effect \cite{7Gretarsson,Harry:2006Coating} but in some cases also the losses inside the substrate materials can be important \cite{9Levin,10Penn}. Thermo-elastic damping is one of the fundamental loss mechanism affecting any type of material. It is related to the dissipation of energy taking place when a material vibrates and heat is transferred from one place to another in the system \cite{6Saulson}. This noise, known as thermo-elastic noise, increase with the temperature and increases with the thermal expansion coefficient \cite{11Braginsky}. It can be a major limitation to the sensitivity of laser interferometers depending on the material used and the operating temperature. Since the interferometer is sensing the optical properties of the mirror, opto-mechanical effects such as the variation in the index of refraction of the coating or of the substrate due to the temperature fluctuations is also a potential noise source whose amplitude depends on the mirror temperature \cite{12Braginsky2}. Figure \ref{fig1:Virgosens} shows the Advanced Virgo sensitivity \cite{2Accadia} and the impact of the various type of thermal noises.

In order to reduce the thermal noise in laser interferometer GW detectors several approaches have been pursued. In general one can reduce the effect of thermal noise by building laser interferometer with longer arms, heavier mirrors and larger laser beams, but this will certainly impact the cost. As a consequence research has been conducted to reduce the friction on mirror suspension adopting the so called monolithic design, where the wire suspending the mirror are fibers made of the same material as the mirror substrates (typically fused silica) and soldered to the latter. On the other side a lot of research has been done to study and reduce the losses in the materials composing the mirror substrate and the mirror coating. So far the reference solution has been to use substrates made of high quality fused silica \cite{10Penn}. As for the coating the best solution found so far is to use a multilayer stack made of fused silica and titania dope tantala layers since this minimize both mechanical and optical losses \cite{Harry:2006qh,Flaminio:2010zzb}. An alternative approach is to reduce thermal noise by operating the interferometer mirror at cryogenic temperature i.e. decreasing the temperature from 300 K to about 20 K. The latter is the solution being pursued by the KAGRA project \cite{Somiya:2011np} and also envisaged by the Einstein Telescope (ET) \cite{15Abernathy}. In the rest of this paper we briefly introduce the main challenges that have to be faced to operate an interferometer at cryogenic temperature. We then describe the present design for the cryogenic mirrors and suspensions of KAGRA. Finally we briefly describe the approach envisaged by the ET.

\subsection{The cryogenic challenges}\label{sec_cryo}

The general argument that by reducing the mirror temperature from 300 K to 20 K thermal noise will be reduced by about a factor about four does not applies since the \linebreak
\vspace*{-3mm}

\noindent amplitude of mechanical losses inside materials are usually temperature dependent. For instance it is well know that mechanical losses inside bulk fused silica increase at low temperature \cite{16Vacher}. A somewhat similar increase has been seen in optical coatings made of SiO$_2$, Ta$_2$O$_5$ and titania doped Ta$_2$O$_5$ even if in the case of coatings different behaviors have been observed \cite{17Martin,18Hirose}.

On the other hand mechanical losses in crystalline materials tend to decrease at low temperature \cite{19Nawrodt}. For this reason materials such as sapphire and silicon look promising as they can be produced in large size. So far the size is somewhat limited by the optical properties. In fact both in the case of silicon and of sapphire it has not been possible to achieve the level of optical absorption obtained in fused silica (nowadays well below 1 ppm/cm). In the case of Czochralski grow silicon the absorption is in the range of several 100's ppm/cm. For floating zone grown silicon, absorption as low as 5 ppm/cm has been measured but size is limited to about 10 cm diameter \cite{20Degallaix}. In the case of sapphire the best absorption is around 30 ppm/cm with sizes in the range of 20 cm diameter \cite{21Hirose2}. The bottom line is that, at present, an interferometer operated at low temperature will have to use smaller mirror compared to the 35 cm diameter fused silica mirrors that are used at room temperature. As a consequence the laser beam size will have to be smaller and so the thermal noise, which increase as $1/\sqrt{beam size}$ will tend to increase \cite{Harry:2006Coating}. As an example Advanced LIGO and Advanced Virgo will have the beam size on the mirrors in the range of 5 to 6 cm while KAGRA will use a beam size of 3.5 cm.

One of the main challenge when attempting to build a cryogenic interferometer is the delicate compromise to be found between the requirement of having low mechanical losses in the mirror suspension and the need to efficiently extract the heat from the mirror in order to cool it down. In this regard the optical absorption in the mirror plays a crucial role since it determines the heat to be extracted. The mechanical losses in the suspension is the other crucial parameter. The goal is to find a monolithic suspensions with good mechanical losses at low temperature and which allow to extract the heat from the mirror.
The vibrations from the cooling system is one of the other challenges. The two cooling techniques considered so far are the use of He circulation or the use of pulse tube cryo-coolers. Both are potential source of vibrations even if in general pulse tube are noisier unless special design is used. The use of He circulation raises the question of safety especially if the detector is located underground as it is the case for KAGRA and for the ET.
Finally the cooling time is also a challenge for a cryogenic interferometer as it determines the down time for the detector maintenance or repair. Apart from the interventions during the detector commissioning phase which might require to heat up the mirror, according to the estimate of the water adsorption by mirror surface, it will be required to heat up the mirrors at least once a year during the operation phase. For this reason the minimization of the cooling time is very important.

\subsection{The KAGRA solutions}

The KAGRA detector currently being built in the Kamioka mine in Japan \cite{Somiya:2011np,22Uchiyama}, will be the first km scale laser interferometer GW detector, whose mirrors will be operated at cryogenic temperature. The goal is to maintain their temperature at 20 K. The mirror substrates will be made of sapphire while amorphous silica and tantala pentoxide layers will be used for the high reflectivity coating. The sapphire crystals substrates will be 22 cm diameter and 15 cm thick. The polishing of a test sample has shown that is possible to achieve flatness rms below 1 nm as it is the case for fused silica substrates \cite{21Hirose2}. The most critical parameter is the absorption in the sapphire which is required to be 30 ppm/cm (the large crystals received so are about a factor of two above this value).

\begin{figure}[H]
\centering
\includegraphics[scale=0.6]{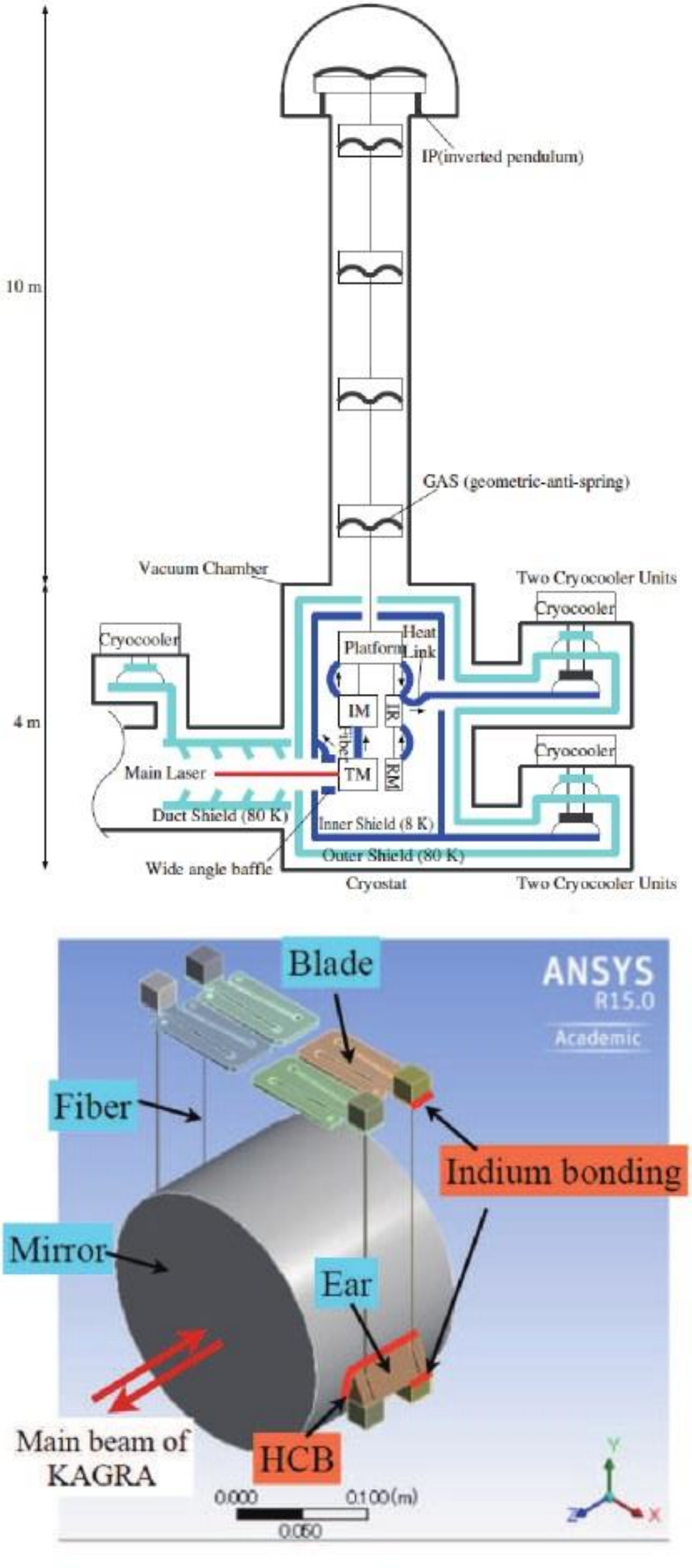}\vspace{-2mm}
\caption{(Color online) The configuration of the cryostat for KAGRA (top) and the design of the sapphire monolithic suspension to suspend the sapphire mirror inside the cryostat.}
\label{fig2:cryo}
\end{figure}

The KAGRA laser power will deliver 180 W at a wavelength of 1064 nm. Assuming 80 W entering the recycling cavity and a recycling gain of 10, the power transmitted through the Fabry-Perot input mirrors will be 400 W \cite{23Aso}. If 30 ppm/cm are absorbed, the heat deposited in the mirrors will be 0.4 W. This is the power to be extracted in the steady state to keep the mirror at 20 K. In order to do so the cryostat design includes two inner shields: an outer one at 80 K and an inner one at 8 K (see Figure~\ref{fig2:cryo} \cite{24Sakakibara}). Four cryo-coolers will be used. The first stages of the cryo-coolers are devoted to the cooling of the outer shield. Two of the cryo-cooler second stages cool the inner shield and the other two cool the mirror payload itself. This design should allow to extract the amount of power deposited by the laser in the mirrors and cool down the mirror to 20 K provided a sufficiently heat conduction path is made between the mirror and the cryo-coolers. In addition two 5 m long cryogenic ducts allow the laser beam to enter the cryostat and while reducing the 300 K radiation input to the mirror by a factor of a thousand compared to the input one would have from the tube in the absence of the ducts \cite{25Sakakibara2}.

The mirrors will be suspended by means of a monolithic suspension made of sapphire (see Figure~\ref{fig2:cryo} \cite{26Chen}). As explained in the previous section, its design is primarily driven by the need to cool down the mirror while keeping its thermal noise as low as possible. The compromise has been made to choose sapphire fibers 35 cm long and 1.6 mm in diameter with a nail head at both ends \cite{27Khalaidovski}. The lower fiber heads are attached to sapphire prisms (called ears) their selves attached to the side of the mirrors by hydroxide catalysis bonding. The prism shape is such that the fiber bending point is slightly above the mirror center of mass to avoid instabilities. On their upper end the fiber heads are bonded to sapphire blade springs attached to the suspension upper stage (so called intermediate mass). The sapphire blades are there to provide some compliance in the vertical direction and compensate for any difference in the fiber lengths. This design allows lowering the vertical resonance of the suspension to about 15 Hz. Both fiber ends are bonded by means of a thin Indium intermediate layer. The indium bonding has been chosen since it allows debonding in case the fiber breaks. Extensive studies were performed to verify the strength, thermal conductivity and mechanical quality factor of the bondings. The fibers mechanical quality factor was measured and found to be around $10^7$ at 20 K a value which allows reaching the thermal noise requirements. The fibers thermal conductivity was measured to be larger than 5500 ${\rm W}/({\rm m}\cdot {\rm K})$ \cite{27Khalaidovski} and so sufficient to extract the heat provided the diameter of 1.6 mm is used. A complete prototype of the monolithic sapphire payload is currently being built \cite{28Kumar}.

The cooling time is also an important parameter since the mirrors will have to be heated up and cooled down once a year to remove the water deposited on the optical surface due \linebreak
\vspace*{-3mm}

\noindent to adsorption. Radiation cooling will allow to reach 150 K. Below 150 K the efficiency of radiation cooling becomes insufficient and another methods should be used to extract the heat. Several solutions have been considered. One is to use a heat switch, i.e. something similar to a finger moved remotely and establishing a thermal contact between the cryo-cooler and the mirror. This reliability of the switch is the main concern of this solution since the thermal contact is uncertain as well as the risk of having such a switch stacked under vacuum at cryogenic temperature. Another option is to use He as thermal exchange gas. While efficient, this technique, which was used in resonant bar detectors, requires to confine the He inside the cryostat in order to be able to pump it out after cooling is completed. This requires to have remotely controlled doors closing the two beam accesses to the cryostat. Again the risk connected with having these relatively large pieces of hardware moving under vacuum at cryogenic temperature make this solution difficult to implement. Finally the option of optical cooling was considered; while being promising its efficiency is still too far from the requirement. In conclusion KAGRA will use permanent heat links connecting the cryo-coolers with the payload recoil masses (see Figure~\ref{fig2:cryo}). It should be noted that these permanent heat links will be required anyway to extract the heat in the steady state. Moreover, in order to accelerate the cooling time, it is planned to coat all the payload (with the exception of the mirror, of course) with diamond like carbon (DLC). According to the simulations this solution allows increasing the radiation cooling efficiency and thus reducing the total cooling time from nearly two months to 39 d. Experimental tests done with a half size prototype of the payload coated with DLC confirmed the validity of the simulation \cite{25Sakakibara2}.

The heat links are a potential path for the transmission of the cryostat vibrations. This is one of the main challenge whenever cryostats are used in conjunction with an apparatus requiring a very low level of vibrations. To study this effect the cryostat vibration have been measured at the factory and combined with the expected seismic noise at the site (which is considerably lower than the one at the factory). The estimate of the cryostat vibrations is then used as input in a mechanical model of the heat links combined with a model of the suspended payload to estimate the mirror vibrations. The simulations show that even if the heat links are connected to the recoil mass of the intermediate mass, and so not directly connected to the mirror, several of the cryostat vibration peaks can limit the interferometer displacement noise. The model shows that this effect can be reduced below the noise requirement if the heat links go through another mass seismically isolated independently from the main mirror. The effect of this mass is to short circuit the heat links vibrations \cite{29Chen2}. In this area more research and development will be needed.

\begin{figure}[H]
\centering
\includegraphics[scale=1]{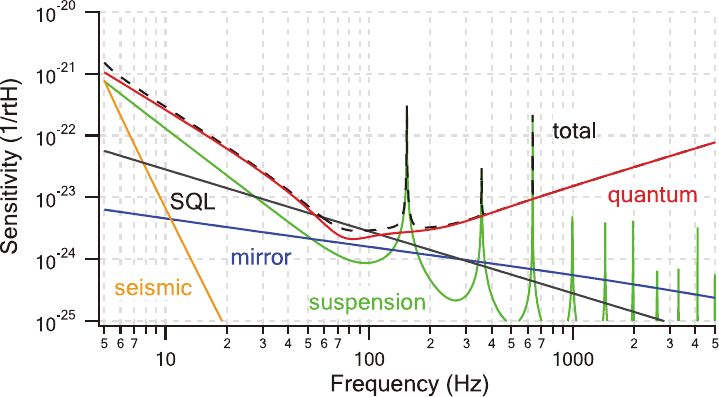}\vspace{-2mm}
\caption{(Color online) The planned sensitivity for KAGRA. Thanks to the reduction of thermal noise, the quantum noise is the main limitation over the frequency band.}
\label{fig3:KAGRAsens}
\end{figure}

By using these solutions, the KAGRA project is planned to reach the sensitivity shown in Figure~\ref{fig3:KAGRAsens}. This sensitivity should allow detecting the coalescence of neutron stars up to distances of about 150 Mpc. ~The main~ difference~ with~ the

\noindent sensitivity of laser interferometer GW detectors operated at room temperature is that quantum noise is the limiting factor over most of the detector bandwidth.

\subsection{The Einstein Telescope approach}

The ET is a third generation GW detector for which the construction of a new research infrastructure is being proposed in Europe \cite{15Abernathy}. The goal is to improve by one order of magnitude in sensitivity over the detectors currently being prepared (Advanced LIGO, Advanced Virgo and KAGRA). To this purpose a new underground infrastructure able to host three nested laser interferometers with arms 10 km in length is proposed. Other ingredients of this new detector will be larger mirrors, larger laser beams and the use of frequency dependent squeezed states of light.
The use of cryogenics to operate the interferometer at low temperature is also part of the ET conceptual design but the approach is different. In order to cope with the contrast between the need of higher laser power to improve the sensitivity at high frequencies and the use of cryogenics to reduce the thermal noise, more relevant in the low frequencies, the ET design foresees the combination of two interferometers covering two different frequency bands. The principle is known as the xylophone approach \cite{30Hild}. The main characteristics of the two interferometers are summarized in Table \ref{1.3tab1_sources} (extracted from ref. \cite{15Abernathy}).

The high frequency detector uses high power laser to reduce the shot noise, large mirror masses to cope with the radiation pressure noise and frequency dependent squeezing to reduce further the quantum noise. Given the frequency band of interest and the high intra-cavity laser power, this interferometer is operated at room temperature.

The low frequency detector will be cryogenic and operated at 10 K. In order to reduce the heat load in the mirror only 3 W are injected into the interferometer. This will result into 18 kW inside the Fabry-Perot arm cavities (to be compared with to about 0.5 MW in advanced detectors) thus helping the reduction of radiation pressure noise effect. The reduction of the intra-cavity power being insufficient to attain the desired sensitivity it is necessary to use mirrors with very large mass (200 kg).

\begin{table}[H]
\caption{Interferometer parameters of the Einstein Telescope conceptual design \cite{15Abernathy}}\label{1.3tab1_sources}
\begin{center}\vspace{-2mm}
\footnotesize
\begin{tabular}{lcc}
\hline
\hline
Parameter &ET-D-HF & Et-D-LF \\
\hline
Arm length & 10 km & 10 km\\
Input power (after IMC)& 5000 W& 3 W\\
Arm power &3 MW  &18 kW\\
Temperature& 290 K  & 10 K\\
Mirror material& fused silica  & silicon\\
Mirror diameter/thickness& 62 cm /30 cm  & min 45 cm/T\\
Mirror mass& 200 kg  & 211 kg\\
Laser wavelength& 1064 nm  & 1550 nm\\
SR-phase& tuend (0.0)  & detued (0.6)\\
SR transmittance& $10\%$  & $20\%$\\
Quantum noise suppression& freq. dep. squeez.  & freq. dep. squeez.\\
Filter cavities& $1\times10$ km  &$2\times10$ km \\
Squeezing level& 10 dB (effective)  & 10 dB (effective)\\
Beam shape& LG$_33$  & TME$_00$\\
Beam radius& 7.25 cm  & 9 cm\\
Scatter loss per surface& 37.5 ppm  & 37.5 ppm\\
Seimic isolation& SA, 8 m tall  & mod SA, 17 m tall\\
Sesmic (for $f>1$ Hz)& $5\times10^{-10}$/m/$f^2$  & $5\times10^{-10}$/m/$f^2$ \\
Gravity gradient substraction &none &none\\
\hline
\hline
\end{tabular}
\end{center}
\end{table}

Using the Czochralski crystal growth method it is now possible to obtain silicon single crystals of this mass in the shape of cylinder with 45 cm diameter. Unfortunately these crystals do not have the best purity and as a consequence their optical absorption is in the range of several 100 s of ppm/cm. Lower absorption has been measured in silicon crystals grown with the floating zone method but unfortunately their size is limited to about 10 cm diameter. In both cases the optical quality is not well known as these crystals are mainly developed for electronic applications so that an intense R$\&$D program is required to reduce their optical absorption and to investigate optical properties such as homogeneity, birefringence, internal defects etc.

\begin{figure}[H]
\centering
\includegraphics[scale=1]{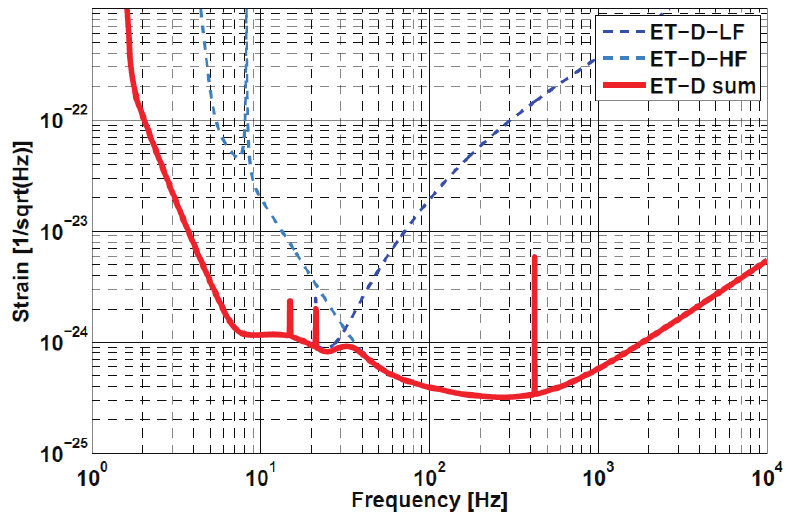}
\caption{(Color online) Sensitivity of the ET in the `xylophone' configuration.}
\label{fig4:ETsens}
\end{figure}

On Figure~\ref{fig4:ETsens} the sensitivity of each ET detector is shown
together with the overall sensitivity of the combined detectors. The design allows to decrease the best sensitivity in the
range of a few parts in $10^{-24} /\sqrt{\rm Hz}$ and to extend the band down to a few Hz. The latter result could be obtained thanks to the cryogenic operation.

\subsection{Perspectives in the use of cryogenics for gravitational wave detectors}

Despite the research performed and the progresses made, thermal noise continue to be one of the fundamental limitation to the sensitivity of laser interferometer GW detectors. The use of cryogenics is the most direct way to reduce it but several challenges have to be faced.

The KAGRA project will be the first ~laser ~interferometer


\noindent with km-scale arms to be operated at cryogenic temperature. As a consequence several issues were studied and a design based on a monolithic sapphire suspension is available. The forthcoming operation of KAGRA will not only increase the capability of the worldwide GW detectors network but will also pave the road toward the realization of the ET.

Both studies show that the mirrors substrate material and in particular its optical absorption plays a crucial role as it sets the refrigeration needs. Both sapphire and silicon requires an intense R$\&$D program to improve their optical quality. The ET design sensitivity somehow assumes that this issue will be solved. It is interesting to note that even in this case, the coating thermal noise will limit the high frequency detector. This noise will also play a role in the low frequency detector unless the coating mechanical losses at low temperature will become comparable to the one reached today at room temperature. In this regards the recent development of crystalline coatings with lower mechanical losses both at room temperature and at cryogenic temperature seems very promising to reduce further the thermal noise limitation \cite{31Cole}.

\section{Detecting nanohertz gravitational waves with pulsar timing arrays}
\emph{Complementary to ground-based laser interferometers, pulsar timing array experiments are being carried out to search for nanohertz GWs. Using the world's most powerful radio telescopes, three major international collaborations have collected $\sim$10-year high precision timing data for tens of millisecond pulsars. In this section we give an overview on pulsar timing experiments, GW detection in the nanohertz regime, and recent results obtained by various timing array projects.
}

\subsection{Introduction}\label{sec:intro}

One hundred years ago, Albert Einstein completed his general theory of relativity, radically revolutionizing our understanding of gravity. In this theory, gravity is no longer a force, but instead an effect of spacetime curvature. The mass and energy content of spacetime creates curvature that in turn dictates the behavior of objects in spacetime. Generally speaking, when a massive object is accelerating, it produces curvature perturbations that propagate at the speed of light. Such \emph{ripples} in the fabric of spacetime are called GWs.

Forty years ago, Hulse and Taylor \cite{ref7} discovered the binary-pulsar system PSR B1913+16. Subsequent observations in the following years showed that its orbital period was gradually decreasing, at a rate that was entirely consistent with that predicted by general relativity as a result of gravitational radiation \cite{Taylor82,Weisberg10}. This provided the most convincing evidence of the existence of GWs and earned Hulse and Taylor the Nobel prize in Physics in 1993.


It was realized in late 1970s that precision timing observations of pulsars can be used to detect very low frequency GWs ($\sim 10^{-9}$--$10^{-7}$ Hz; \cite{Sazhin1978,Detweiler1979}). Based on a calculation by Estabrook $\&$ Wahlquist \cite{Estabrook75} for the GW detection using Doppler spacecraft tracking, for which the principle is similar to pulsar timing, Detweiler \cite{Detweiler1979} explicitly showed that ``\textit{a GW incident upon either a pulsar or the Earth changes the measured frequency and appears then as a anomalous residual in the pulse arrival time}". In 1983, Hellings $\&$ Downs \cite{Hellings_Downs83} used timing data from four pulsars to constrain the energy density of any stochastic background to be $\lesssim10^{-4}$ times the critical cosmological density at $\lesssim10^{-8}$ Hz (see, refs. \cite{RomaniTaylor83,Bertotti83} for similar results). Almost simultaneously to these works, Backer et al.  \cite{Backer82MSP} discovered the first millisecond pulsar PSR B1937+21. Because of its far better rotational stability than any previously known pulsars and relatively narrower pulses, timing observations in the following years improved the limit on stochastic backgrounds very quickly and by orders of magnitude \cite{Davis85,Rawley87,Stineb90,Kaspi94}.

Measurements with a single pulsar can not make definite detections of GWs whose effects may be indistinguishable from other noise processes such as irregular spinning of the star itself. By continued timing of an array of millisecond pulsars, i.e., by constructing a pulsar timing array (PTA), GWs can be searched for as correlated signals in the timing data. Hellings $\&$ Downs \cite{Hellings_Downs83} first did such a correlation analysis to put limits on stochastic backgrounds. Romani \cite{Romani89} and Foster $\&$ Backer \cite{Foster_Backer90} explored the greater scientific potential of PTA experiments: (1) searching for GWs; (2) providing a time standard for long time scales; and (3) detecting errors in the solar system ephemerides.

\begin{figure}[H]
\centering
\includegraphics{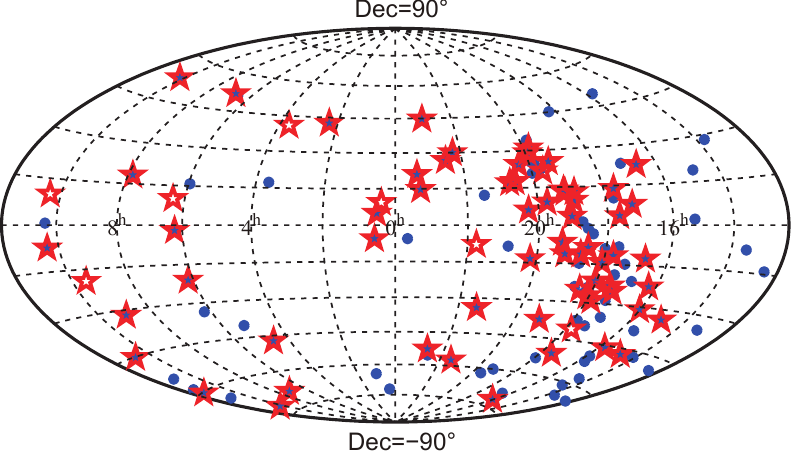}\vspace{-2mm}
\caption{(Color online) Distribution of millisecond pulsars on the sky in equatorial coordinates, containing 70 IPTA pulsars (red stars) and all presently known pulsars that have pulse periods $P<15$ ms and $\dot{P}<10^{-19}\,{\rm{s}}{\rm{s}}^{-1}$ (blue dots; as found in the ATNF Pulsar Catalogue version 1.53 \cite{ATNF05Pulsar}).}
\label{fig:SkyMSP}
\end{figure}

There are three major PTAs currently operating: the Parkes PTA (PPTA (http://www.atnf.csiro.au/research/pulsar/ppta/); \cite{PPTA2013,PPTA13CQG}) was set up in 2004 using the Parkes radio telescope in Australia; the European PTA (EPTA (http://www.epta.eu.org/); \cite{EPTA}) was initiated in 2004/2005 and makes use of telescopes in France, Germany, Italy, the Netherlands and the UK; and the North American Nanohertz Observatory for Gravitational Waves (NANOGrav (http://nanograv.org/)); \cite{NANOGrav}) was formed in 2007 and carries out observations with the Arecibo and Green Bank telescopes. We summarize in Table \ref{tb:PTAinfo} information about the telescopes used and the number of pulsars monitored by these PTAs. Recently the three PTA collaborations were combined to form the International PTA (IPTA (http://www.ipta4gw.org/); \cite{IPTA,IPTAdick13,Maura_IPTA14}). Looking into the future, GW detection with pulsar timing observations is one of major science goals for some powerful future radio telescopes, such as the planned Xinjiang Qitai 110 m radio telescope (QTT; which is a fully-steerable single-dish telescope \cite{QTT}) and the Five-hundred-meter Aperture Spherical Telescope (FAST; which is expected to come online in 2016 \cite{FAST11,GeorgeFAST14}) in China, and the planned Square Kilometer Array (SKA) and its pathfinders (see ref. \cite{Lazio13CQG} and references therein). Figure \ref{fig:SkyMSP} shows the distribution of IPTA pulsars on the sky along with those presently known millisecond pulsars that may be useful for PTA research with future telescopes.

In this section, we first introduce the basics of pulsars and \linebreak
\vspace{-3mm}

\begin{table}[H]
\begin{center}
\caption{Information about pulsar timing array projects\hspace{22mm}}\label{tb:PTAinfo}
\vspace{-1mm}
\footnotesize
\begin{tabular*}{0.41\paperwidth}
{lcccc}
\hline
     &  {Telescope} & {Diameter} (m)  & {Country}   & { Pulsars}$^{\rm{a)}}$ \\
\hline
  PPTA &  Parkes & 64 & Australia   & 20 \\
\hline
 &  Effelsberg & 100 & Germany   & \\
   &  Lovell & 76.2 & UK  &   \\
{EPTA}   &  Nancay & $94^{\rm{b)}}$ & France  &    {22}\\
   &  Sardinia & 64 & Italy  &    \\
   &  Westerbork & $96^{\rm{b)}}$ & Netherlands  &    \\
\hline
\raisebox{-1.50ex}{NANOGrav} & Arecibo & 305 & \raisebox{-1.50ex}{USA}  &\raisebox{-1.50ex} {22} \\
   & GBT & 100 &   &  \\
\hline
\end{tabular*}
\end{center}
\vspace{-2mm}
\footnotesize
 a) We present the number of pulsars that have been timed for more than 5 years for each project here (see Table 3 in ref. \cite{IPTAdick13} for more information). It is worth pointing out that a number of pulsars were recently added to the timing arrays and some pulsars were dropped from these arrays. The total number of pulsars that are currently being observed for IPTA is 70, of which 13 are timed by two timing arrays and 10 by all three arrays \\
b) Values of circular-dish equivalent diameter \cite{EPTA10}
\end{table}

\noindent pulsar timing techniques in sects. \ref{sec:psr} and \ref{sec:pt} respectively. In sect. \ref{sec:noise} we highlight some major noise sources that affect PTA's sensitivities to GWs. In sect. \ref{sec:GWpta} we describe how a PTA responds to GWs. Sect. \ref{srcPTA} contains a review on GW sources and related results derived from the latest analysis of PTA data. Finally we discuss future prospects in sect. \ref{sec:conclu}.

\subsection{Pulsars}
\label{sec:psr}
Neutron stars are born as compact remnants of core collapse supernovae during the death of massive main sequence stars with masses of around $8$--$25\, {\rm M}_{\odot}$. A pulsar is a highly magnetized, spinning neutron star. It can be detected as it emits beams of electromagnetic radiation along its magnetic axis that is misaligned with its rotational axis. Such beams of radiation sweep over the Earth in the same way lighthouse beams sweep across an observer, leading to pulses of radiation received at the observatory. Because of their exceptional rotational stability, pulsars are powerful probes for a wide range of astrophysical phenomena. As mentioned above, long-term timing observations of PSR B1913+16 provided the first observational evidence of the existence of GWs. Timing observations of PSR B1257+12 led to the first confirmed discovery of planets outside our solar system \cite{ExoplantPT92}. The double pulsar system PSR J0737$-$3039A/B, with both neutron stars having been detected as radio pulsars \cite{Burgay03Nat,Lyne04Sci}, enabled very stringent tests of general relativity and alternative theories of gravity in the strong-field regime \cite{Kramer06Sci}.

\begin{figure}[H]
 \centering
 \includegraphics{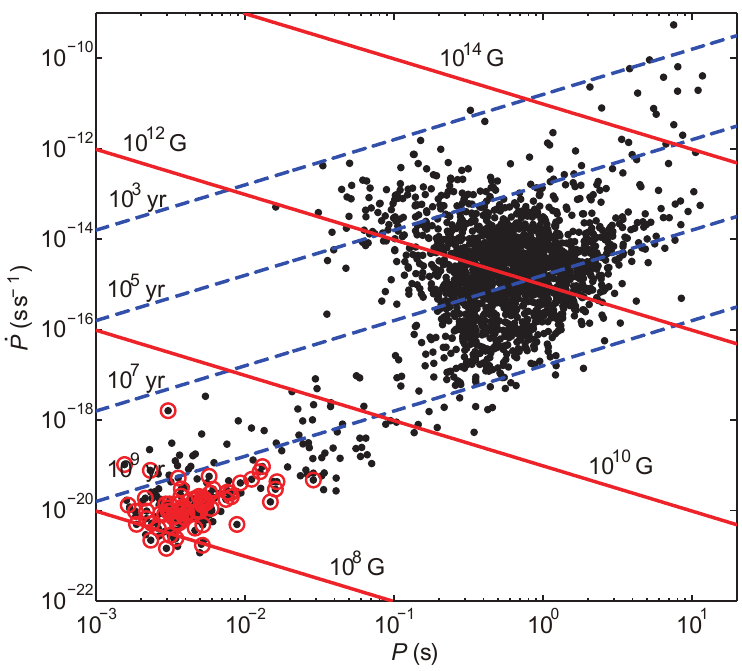}
  \caption{(Color online) The period ($P$) vs. period derivative ($\dot{P}$) diagram for all known pulsars (black dots; data taken from the ATNF Pulsar Catalogue version 1.53). Red (grey) circles represent the millisecond pulsars currently being timed by the IPTA project. Also shown are lines of constant characteristic age $\tau=P/2\dot{P}$ (dash) assuming that pulsars are spinning down solely because of magnetic dipole radiation \cite{LorimerKramer05}, and of constant inferred surface magnetic field $B_{0}=3.2\times 10^{19}\,\sqrt{P\dot{P}}\,{\rm{Gauss}}$ (solid; \cite{DickTaylor77}). Two distinct populations are apparent in this diagram: (a) \emph{normal pulsars}, with $P\sim$ 0.1--4 s and $B_{0}\,\sim10^{11}$--$10^{13}$ Gauss; (b) \emph{millisecond pulsars}, with $P\sim$ 3 milliseconds and $\dot{P}\sim 10^{-20}\,{\rm{s}}{\rm{s}}^{-1}$.}
  \label{fig:PSRppdot}
\end{figure}

The first pulsar was discovered by Hewish and Bell et al. \cite{Pulsar68} in 1967. Since then over 2500 pulsars have been discovered$^{1)}$\footnote{1) See the ATNF Pulsar Catalogue website {http://www.atnf.csiro.au/research/pulsar/psrcat/} for up-to-date information}, with spin periods ranging from about 1 millisecond to 10 s. It is generally believed that pulsars are born with periods of order tens of milliseconds but quickly ($\sim10$ Myr) spin down (because of the loss of rotational energy) to periods of order seconds. This energy loss could be due to a variety of mechanisms, such as the magnetic dipole radiation, emission of relativistic particle winds and even GWs. As pulsars spin down, they eventually reach a point where there is insufficient energy to power electromagnetic radiation. However, for pulsars in binary systems, it is very likely that pulsars are spun up as mass and angular momentum are accreted from their stellar companions. Such an accretion process is observed in X-ray binaries. These pulsars, usually named as \emph{millisecond pulsars}, have spin periods of about several milliseconds and much lower spin-down rates. Figure \ref{fig:PSRppdot} shows the distribution of all known pulsars in the period-period derivative diagram.

\subsection{Pulsar timing techniques}
\label{sec:pt}
Much of the science based on pulsar observations makes use of the ``pulsar timing" technique \cite{LorimerKramer05,TEMPO2Edwards,LomDemorest13}, which involves measurement and prediction of pulses' times of arrival (TOAs). Individual pulses are generally not useful in this \linebreak
\vspace{-5mm}

\noindent regard as they are unstable and mostly too weak to observe. The average pulse profile over a large number of pulses is stable for a particular pulsar at a given observing wavelength, and therefore very suitable for timing experiments.

The first step in pulsar timing is to measure the \emph{topocentric} pulse arrival times with clocks local to the radio observatories. Data collected with the telescope are \emph{de-dispersed} to correct for frequency-dependent dispersion delays due to the ionized interstellar medium. These data are then \emph{folded} with the period derived from previous observations to form the mean pulse profile. This profile is then correlated with a standard template, either an analytic function or simply a very high signal-to-noise ratio observation, to record the pulse arrival time at the observatory.

The measured TOAs are further transformed to the pulse emission time via a \emph{timing model}, from which the pulse phase of emission is computed. The rotational phase $\phi(t)$ of the pulsar as a function of time $t$ (measured in an inertial reference frame of the pulsar) can be represented as a Taylor series:
\begin{equation}
\phi(t) = \phi(t_{0})+f(t-t_{0})+\frac{1}{2}\dot{f}(t-t_{0})^{2}+\cdots,
\label{PSRphase}
\end{equation}
where $t_0$ is an arbitrary reference time, $f={\rm{d}}\phi/{\rm{d}}t$ is the spin frequency, and $\dot{f}$ is the frequency derivative. A number of corrections are applied when converting the topocentric TOAs to the pulsar frame. Such corrections include:

1. Clock corrections, which account for differences in the observatory time and a realization of Terrestrial Time (e.g., the International Atomic Time).

2. pulse delay induced by Earth's troposphere.

3. the Einstein delay, i.e., the time dilation due to changes in the gravitational potential of the Earth, the Earth's motion, and the secular motion of the pulsar or that of its binary system.

4. the Roemer delay, i.e., the vacuum light travel time between the observatory and the solar system barycenter, and for pulsars in binaries between the pulsar and the binary system's barycenter.

5. Shapiro delays, i.e., gravitational time delays due to the solar system objects and if applicable the pulsar's companion.

6. Dispersion delays caused by the interstellar medium, the interplanetary medium and the Earth's ionosphere.

The timing model, which describes the above corrections and the pulsar's intrinsic rotational behavior, predicts the rotational phase of the pulsar at any given time as observed from the solar system barycenter. Basic parameters of a pulsar timing model include the spin period, spin-down rate, right ascension and declination of the pulsar, the dispersion measure (discussed later in sect. \ref{sec:noise}), and Keplerian orbital parameters if the pulsar is in a binary system. Measured TOAs are compared with predictions based on the timing model, and the differences are called \emph{timing residuals}. The (pre-fit) timing residual for the $i$-th observation is calculated as \cite{TEMPO2}:
\begin{equation}
R_{i} = \frac{\phi(t_{i})-N_{i}}{f},
\label{TRsdefi}
\end{equation}
where $N_{i}$ is the nearest integer$^{2)}$\footnote{2) Here note that $\phi(t)$ is measured in \emph{turns} equal to $2\pi$ radians} to each $\phi(t_{i})$. One can see the key point in pulsar timing is that every single rotation of the pulsar is unambiguously accounted for over long periods (years to decades) of time.

A linear least-squares fitting procedure is carried out to obtain estimates of timing parameters, their uncertainties and the post-fit timing residuals. In practice, this is done iteratively: one starts from a small set of data and only includes the most basic parameters (with values derived from previous observations) so that it is easier to coherently track the rotational phase. Parameter estimates are then improved by minimizing the timing residuals and additional parameters can be included for a longer data set.

The fitting to a timing model and analysis of the timing residuals can be performed with the pulsar timing software package \textsc{TEMPO2} \cite{TEMPO2,TEMPO2Edwards,TEMPO2III}, which is freely available on the internet for download (www.sf.net/projects/tempo2/). More recently, an alternative method based on Bayesian inference was also developed \cite{LentatiBayes14,BayesPT14}.

\subsection{Noise sources in pulsar timing data}
\label{sec:noise}

Timing residuals generally come from two groups of contribution: (1) un-modelled deterministic processes, e.g., an unknown binary companion or a single-source GW; and (2) stochastic processes, e.g., the intrinsic pulsar spin noise and a GW background. Before the discussion of GW detection with PTAs, we briefly discuss some major noise processes in pulsar timing data here.

\subsubsection{Radiometer noise}
Radiometer noise arises from the observing system and the radio sky background (including the atmosphere, the cosmic microwave background and synchrotron emission in the Galactic plane). It can be quantified as \cite{LorimerKramer05}:
\begin{equation}
\sigma_{\rm{rad.}} \approx \frac{W}{S/N}\approx \frac{WS_{\rm{sys}}}{S_{\rm{mean}}\sqrt{2\Delta \nu t_{\rm{int}}}}\sqrt{\frac{W}{P-W}}\, ,
\label{Sigma-radio}
\end{equation}
where $W$ and $P$ are the pulse width and period respectively, $S/N$ is the profile signal-to-noise ratio, $S_{\rm{sys}}$ is the system equivalent flux density which depends on the system temperature and the telescope's effective collecting area, $S_{\rm{mean}}$ the pulsar's flux density averaged over its pulse period, $\Delta \nu$ and $t_{\rm{int}}$ are the observation bandwidth and integration time respectively. Radiometer noise can be reduced by using low-noise receivers, observing with larger telescopes, and increasing observing time and bandwidth. One reason to fold individual pulses to obtain an average profile is to reduce the radiometer noise; the reduction is equal to the square root of the number of pulses folded. Radiometer noise is an additive Gaussian white noise and is formally responsible for TOA uncertainties. From eq. (\ref{Sigma-radio}), one can see that bright, fast spinning pulsars with narrow pulse profiles allow the highest timing precision.

\subsubsection{Pulse jitter noise}
Pulse jitter manifests as the variability in the shape and arrival phase of individual pulses. The mean pulse profile is an average over a large number of single pulses. Although it is stable for most practical purposes, there always exists some degree of stochasticity in the phase and amplitude of the average pulse profile. Pulse jitter noise is intrinsic to the pulsar itself, and thus can only be reduced by increasing observing time, i.e., averaging over more single pulses. Pulse jitter is also a source of white noise, which has been found to be a limiting factor of the timing precision for a few very bright pulsars \cite{Shannon14jitter,1713global}. For future telescopes such as FAST and SKA, jitter noise may dominate over the radiometer noise for many millisecond pulsars \cite{GeorgeFAST14}. Improvement in the timing precision for the brightest pulsar PSR J0437$-$4715 was recently demonstrated with the use of some mitigation methods for pulse jitter noise \cite{Oslow11,Oslow13}.

\subsubsection{``Timing noise"}
It has long been realized that timing residuals of many pulsars show structures that are inconsistent with TOA uncertainties \cite{Blandford84,Hob06noise}. Such structures are collectively referred to as \emph{timing noise}. Timing noise is very commonly seen in normal pulsars and has also become more prominent in a number of millisecond pulsars as timing precision increases and the data span grows. The exact astrophysical origins of timing noise are not well understood. It is mostly suggested to be related to rotational irregularities of the pulsar and therefore it is also usually called as spin noise \cite{Hob10noise,RyanS10noise}. For millisecond pulsars, power spectra of timing residuals can typically be modelled as the sum of white noise and red noise. For red noise, a power law spectrum with a low-frequency turnover appears to be a good approximation (see, e.g., Figure 11 in ref. \cite{PPTA2013} for analyses of 20 PPTA pulsars).

The presence of red noise results in problems to the pulsar timing analysis as the standard least-squares fitting of a timing model assumes time-independent TOA errors. Blandford et al. \cite{Blandford84} analytically showed the effects of timing noise on the estimates of timing model parameters and suggested the use of the noise covariance matrix to pre-whiten the data for improved parameter estimation. More recently, Coles et al. \cite{Coles2011} developed a whitening method that uses the Cholesky decomposition of the covariance matrix of timing residuals to whiten both the residuals and the timing model. By doing so, noise in the whitened residuals is statistically white and the ordinary least-squares solution of a timing model can be obtained. van Haasteren $\&$ Levin \cite{vHasLevin13} developed a Bayesian framework that is capable of simultaneously estimating timing model parameters and timing noise spectra.

\subsubsection{Dispersion measure variations}
Because of dispersion due to the interstellar plasma, pulses at low frequencies arrive later than at high frequencies. Specifically, this dispersion delay is given by
\begin{equation}
\Delta_{\rm{DM}}\approx (4.15\, {\rm{ms}})\, {\rm{DM}}\, \nu_{\rm{GHz}}^{-2},
\label{Deff}
\end{equation}
where $\nu_{\rm{GHz}}$ is the radio frequency measured in GHz, and dispersion measure (DM, measured in pc cm$^{-3}$) is the integrated column density of free electrons between an observer and a pulsar. Because of the motion of the Earth and the pulsar relative to the interstellar medium, the DM of a pulsar is not a constant in time. Such DM variations introduce time-correlated noise in pulsar timing data.

Pulsars in a timing array are usually observed quasi-simultaneously at two or more different frequencies. For example, the PPTA team observes pulsars at three frequency bands -- 50 cm ($\sim$700 MHz), 20 cm ($\sim$1400 MHz), and 10 cm ($\sim$3100 MHz) -- during each observing session (typically 2--3 d). This makes it possible to account for DM variations and thus reduce the associated noise with various methods \cite{NANOGrav2012,MikeDM2013,KJLee14dm}. The method currently being used by the PPTA is described in Keith et al. \cite{MikeDM2013}, which was built on a previous work by You et al. \cite{YouXP07}. In this method timing residuals are modelled as the combination of a (radio-)wavelength-independent (i.e., common-mode) delay and the dispersion delay. Both components are represented as piece-wise linear functions and can be estimated through a standard least-squares fit. A key feature of this method is that GW signals are preserved in the common-mode component. Ultra wide-band receivers are in development by various collaborations in order to better correct for noise induced by DM variations (along with other benefits such as increasing timing precision, studies of interstellar medium, and etc.).

\subsubsection{Interstellar scintillation}
Interstellar scintillation refers to strong scattering of radio waves due to the spatial inhomogeneities in the ionized interstellar medium \cite{Narayan92}, analogous to twinkling of stars due to scattering in the Earth's atmosphere. There are multiple effects associated with interstellar scintillation that cause time-varying delays in measured TOAs, with the dominant one being pulse broadening from multipath scattering \cite{Stinebring13,CordSha10}. Various mitigation techniques have been developed for this type of noise \cite{CordSha10,Demor11Cyclic,LiuK14}. Generally speaking, noise induced by interstellar scintillation is a Gaussian white noise, and can be reduced by increasing observing time and bandwidth. For millisecond pulsars that are observed at current radio frequencies by PTAs, the effects of scintillation are predicted to be small. However, when pulsars are observed at lower frequencies, or more distant (and more scintillated) pulsars are observed, these effects can become more important \cite{CordSha10,Cordes15DM}.

\subsubsection{Correlated noise among different pulsars}
The above noise processes are generally thought to be uncorrelated among different pulsars. In a PTA data set that we hope to detect GWs, some correlated noise may be present. For example, (1) instabilities in Terrestrial Time standards affect TOA measurements of all pulsars in exactly the same way, i.e., clock errors result in a \emph{monopole} signature in a PTA data set; (2) the solar system ephemerides, which provide accurate predictions of the masses and positions of all the major solar system objects as a function of time, are used to convert pulse arrival times at the observatory to TOAs referenced at the solar system barycenter. Imperfections in the solar system ephemerides induce a \emph{dipole} correlation in a PTA data set. Indeed, it has been demonstrated that PTAs can be used (1) to search for irregularities in the time standard and thus to establish a pulsar-based timescale \cite{George_Clock12}, and (2) to measure the mass of solar system planets \cite{Champ10}.

\subsection{Gravitational waves and pulsar timing arrays}
\label{sec:GWpta}
The effects of GWs in a single-pulsar data may be indistinguishable from those due to noise processes as discussed in the previous subsection. Indeed, even without any such noise, GWs that have the same features as those due to uncertainties in the timing model parameters would still be very difficult to detect with only one pulsar. Therefore analysis of single-pulsar data can be best used to constrain the strength of potential GWs \cite{JenetWen04,YiShuXu14MN}.

A PTA is a Galactic-scale GW detector. If one wishes to have an analogy to a laser interferometer, pulsars in the timing array are ``test masses"; pulses of radio waves act as the laser; and the pulsar-Earth baseline is a single ``arm". Millisecond pulsars in our Galaxy, typically $\sim$kpc (or thousands of light years) away, emit radio waves that are received at the telescope with extraordinary stability. A GW passing across the pulsar-Earth baseline perturbs the local spacetime along the path of radio wave propagation, leading to an apparent redshift in the pulse frequency that is proportional to the GW strain amplitude. Let us first consider the special case where a linearly polarized GW propagates in a direction perpendicular to the pulsar-Earth baseline, the resulting timing residual is given by
\begin{equation}
r(t)=\int_{0}^{L/c}h\left(t-\frac{L}{c}+\tau\right){\rm{d}}\tau,
\label{TRlin0}
\end{equation}
where $L$ is the pulsar distance and we adopt the plane wave approximation$^{3)}$\footnote{3) For sources that are close enough ($\lesssim$ 100 Mpc), it may be necessary to consider the curvature of the GW front. This, in principle, would allow luminosity distances to GW sources to be measured via a parallax effect \cite{DengXH2011}}. With the definition of ${\rm{d}}A(t)/{\rm{d}}t=h(t)$, the timing residual takes the following form:
\begin{equation}
r(t)=\Delta A(t)=A(t)-A\left(t-\frac{L}{c}\right).
\label{TRptet}
\end{equation}
Here we can see that $A(t)$ results from the GW induced spacetime perturbation incident on the Earth (i.e., the \emph{Earth term}), and $A(t-\frac{L}{c})$ depends on the GW strain at the time of the radio wave emission (i.e., the \emph{pulsar term}). Typical PTA observations have a sampling interval of weeks and span over $\sim$10 yr, implying a sensitive frequency range of $\sim$1--100 nHz. Therefore PTAs are sensitive to GWs with wavelengths of several light years, much smaller than the pulsar-Earth distance.

In the general case where a GW originates from a direction $\hat{\Omega}$, the induced timing residuals can be written as:
\begin{equation}
\label{TR1}
r(t,\hat{\Omega}) = F_{+}(\hat{\Omega})\Delta A_{+}(t) + F_{\times}(\hat{\Omega}) \Delta A_{\times}(t),
\end{equation}
where $F_{+}(\hat{\Omega})$ and $F_{\times}(\hat{\Omega})$ are antenna pattern functions as given by \cite{Wahlq87}
\begin{align}
F_{+}(\hat{\Omega}) =& \frac{1}{4(1-\cos\theta)}\{(1+\sin^2 \delta)\cos^2 \delta_p \nonumber\\
&\times\cos[2(\alpha-\alpha_p)]- \sin2\delta \sin2\delta_p\cos(\alpha-\alpha_p) \nonumber\\
&+ \cos^2 \delta (2-3\cos^2 \delta_p)\},
\label{Fp}
\end{align}
\begin{align}
F_{\times}(\hat{\Omega}) =& \frac{1}{2(1-\cos\theta)}\{\cos \delta \sin 2\delta_p \sin(\alpha-\alpha_p)\nonumber\\
& - \sin \delta \cos^2 \delta_p \sin[2(\alpha-\alpha_p)]\} ,
\label{Fc}
\end{align}
where $\cos\theta = \cos\delta \cos\delta_p \cos(\alpha-\alpha_p)+\sin\delta \sin\delta_p$, $\theta$ is the opening angle between the GW source and pulsar with respect to the observer, and $\alpha$ ($\alpha_p$) and $\delta$ ($\delta_p$) are the right ascension and declination of the GW source (pulsar) respectively. The source-dependent functions $\Delta A_{+,\times}(t)$ in eq. (\ref{TR1}) are given by
\begin{equation}
\label{TR2}
\Delta A_{+,\times}(t) = A_{+,\times}(t)-A_{+,\times}(t_p)
\end{equation}
\begin{equation}
\label{TpTe}
t_p = t-(1-\cos\theta)\frac{L}{c}.
\end{equation}
The forms of $A_{+}(t)$ and $A_{\times}(t)$ depend on the type of source that we are looking for.

\begin{figure}[H]
  \centering
  \includegraphics{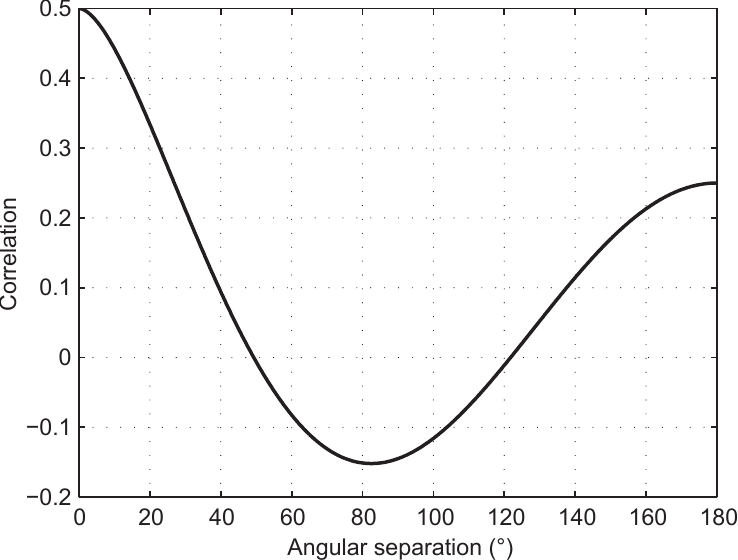}
  \caption{(Color online) The Hellings-Downs curve, which depicts the expected correlation in timing residuals due to an isotropic stochastic background, as a function of the angular separation between pairs of pulsars.}
  \label{fig:HDcurve}
\end{figure}

\vspace*{3mm}
{\textbf{The Hellings-Downs curve}}\\
An isotropic stochastic background will produce a correlated signal in PTA data sets. Such a correlation uniquely depends on the angular separation between pairs of pulsars, as given by \cite{Hellings_Downs83}
\begin{align}
\zeta(\theta_{ij}) =& \frac{3}{2}\frac{(1-\cos\theta_{ij})}{2}\ln\left[\frac{(1-\cos\theta_{ij})}{2}\right]\nonumber\\
& -\frac{1}{4}\frac{(1-\cos\theta_{ij})}{2}+\frac{(1+\delta_{ij})}{2} ,
\label{HDcorr}
\end{align}
where $\theta_{ij}$ is the angle between pulsars $i$ and $j$, and $\delta_{ij}$ is 1 for $i=j$ and 0 otherwise. Figure \ref{fig:HDcurve} shows the famous Hellings-Downs curve as given by eq. (\ref{HDcorr})---it is a factor of $3/2$ larger than the original result of ref. \cite{Hellings_Downs83}. This is because $\zeta(\theta_{ij})$ is normalized to $1$ for the autocorrelation of the stochastic background induced timing residuals for a single pulsar. In Figure \ref{fig:HDcurve} the correlation function takes a value of $0.5$ at zero angular separation as the autocorrelation due to pulsar terms is neglected.

\subsection{Gravitational wave sources and recent observational results}
\label{srcPTA}
Potential signals that could be detectable for PTAs include: (1) stochastic backgrounds. The primary target is that formed\linebreak
\vspace*{-3mm}

\noindent by the combined emission from numerous binary supermassive black holes distributed throughout the Universe. Background signals from cosmic strings (e.g., ref. \cite{Sanidas12}) and inflation (e.g., refs. \cite{ZhaoWen13,TongML14}) have also been studied in the context of PTAs; (2) continuous waves, which can be produced by individual nearby binaries; (3) bursts with memory associated with binary black hole mergers; (4) and all other GW bursts. Below we give a brief overview of these sources and summarize some recent astrophysical results.

\subsubsection{Stochastic backgrounds}
The stochastic background from the cosmic population of supermassive binary black holes has been the most popular target for PTA efforts. Generally speaking the signal amplitude depends on how frequently these binaries merge in cosmic history and how massive they are. Both of these quantities are poorly constrained observationally. Assuming that all binaries are in circular orbits and evolve through gravitational radiation only, the characteristic amplitude spectrum of this background is given by \cite{RAstronJago95,Jaffe_Backer03,Wyithe_Loeb03,WenZL09,Sesana13GWB,Ravi2012,McWilliams14}
\begin{equation}
h_{\rm c}(f)=A_{\rm{yr}}\left(\frac{f}{f_{\rm{yr}}}\right)^{-2/3},
\label{hcGWBsmbbh}
\end{equation}
where $A_{\rm{yr}}$ is the dimensionless amplitude at a reference frequency $f_{\rm{yr}}=1\, {\rm{yr}}^{-1}$. The fraction of the cosmological critical energy density (per logarithmic frequency interval) contained in the GW background is related to the amplitude spectrum through $\Omega_{\rm{GW}}(f)=(2\pi^{2}/3H_{0}^{2})A_{\rm{yr}}^{2}f_{\rm{yr}}^{2}(f/f_{\rm{yr}})^{2/3}$ where $H_0$ is the Hubble constant. Various models predict a similar range of $A_{\rm{yr}}$, most likely to be $\sim 10^{-15}$ (see, e.g., refs. \cite{WenZL09,Sesana13GWB,Ravi2012}). An exception is the recent model in ref. \cite{McWilliams14} whose prediction is two to five times higher. Recent studies that include the effects of environmental coupling and orbital eccentricities indicate a reduced signal at below $\sim 10$ nHz \cite{Enoki07,Sesana13CQG,Ravi14GWB}.

\begin{figure}[H]
  \centering
  \includegraphics{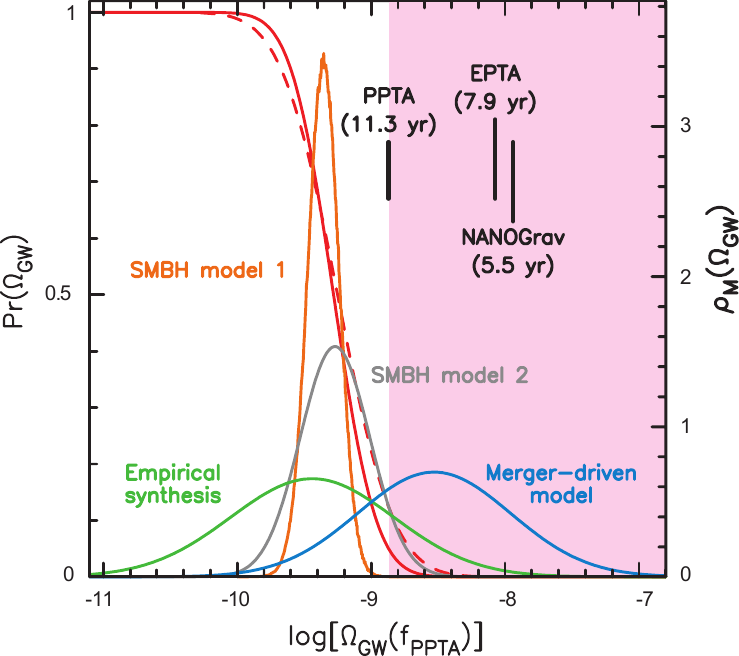}
  \caption{(Color online) Upper limits on the fractional energy density of the GW background $\Omega_{\rm{GW}}(f)$, at a frequency of 2.8 nHz, as compared against various models of the supermassive binary black hole background. The red (grey) solid and dashed lines show the probabilities ${\rm{Pr}}(\Omega_{\rm{GW}})$ that a GW background with energy density $\Omega_{\rm{GW}}(f)$ exists given the PPTA data, assuming Gaussian and non-Gaussian statistics respectively. Three vertical lines mark the $95\%$ confidence upper limits published by NANOGrav \cite{NANOGrav2012}, EPTA \cite{EPTAlimit} and PPTA \cite{PPTA2013Sci}, where the times next to these lines correspond approximately to the observing spans of the data sets. The shaded region is ruled out with $95\%$ confidence by the PPTA data. Gaussian curves represent the probability density functions $\rho_{M}(\Omega_{\rm{GW}})$ given different models: a merger-driven model for growth of massive black holes in galaxies \cite{McWilliams14}, a synthesis of empirical models \cite{Sesana13GWB}, semi-analytic models (SMBH model 1; \cite{PPTA2013Sci}) based on the Millenium dark matter simulations \cite{Millennium,MillenniumII} and a distinct model for black hole growth (SMBH model 2; \cite{Kulier15}). Figure taken from ref. \cite{PPTA2013Sci}.}
  \label{fig:PPTAlimitGWB}
\end{figure}

Jenet et al. \cite{Jenet05} suggested that it is possible to make a detection of the binary black hole background if 20 pulsars are timed with a precision of $\sim$ 100 ns over $\gtrsim$ 5 years. Three PTAs have searched for such a background signal assuming it is isotropic, leading to increasingly more stringent upper limits on the background strength \cite{Jenet2006,YardleySGWB,EPTAlimit,NANOGrav2012,EPTA15GWB,PPTA2013Sci}. The most constraining limit published to date ($A_{\rm{yr}} <2.4 \times 10^{-15}$) comes from the PPTA collaboration by Shannon et al. \cite{PPTA2013Sci}. As shown in Figure \ref{fig:PPTAlimitGWB}, this limit ruled out the most optimistic model of ref. \cite{McWilliams14} with $90\%$ confidence and is in tension with other models at $\sim50\%$ confidence.

Recently methods have also been proposed to search for a more general anisotropic background signal \cite{Cornish13,Minga13Aniso,Taylor13Aniso,Gair14}. Using the 2015 EPTA data, Taylor et al. \cite{Taylor15Aniso} placed constraints on the angular power spectrum of the background from circular, GW-driven supermassive black hole binaries and found that the data could not update the prior knowledge on the angular distribution of a GW background.

\subsubsection{Continuous waves}
Individual supermassive binary black holes, especially the most nearby and/or massive ones, could provide good opportunities for detection of continuous waves. Unlike compact binaries in the audio band, supermassive binary black holes detectable for PTAs are mostly in the early stage of inspiral and therefore emit quasi-monochromatic waves. For an inspiralling circular binary of component masses $m_1$ and $m_2$, the GW strain amplitude is given by \cite{Thorne87}
\begin{equation}
h_{0}=2\frac{(G M_{\rm c})^{5/3}}{c^{4}}\frac{(\pi f)^{2/3}}{d_{\rm L}}
\label{h0},
\end{equation}
where $d_{\rm L}$ is the luminosity distance of the source, and $M_{\rm c}$ is the chirp mass defined as $M_{\rm c} = M \eta ^{5/3}$, with $M= m_1+m_2$ the total mass and $\eta = m_1 m_2 /M^{2}$ the symmetric mass ratio. After averaging over the antenna pattern functions given by eqs. (\ref{Fp}) and (\ref{Fc}) and the binary orbital inclination angle, the Earth-term timing residuals induced by a circular binary is $\sim h_{0}/2\pi f$.

Before the establishment of major PTAs, Jenet et al. \cite{JenetWen04} developed a framework in which pulsar timing observations can be used to constrain properties of supermassive binary black holes and applied the method to effectively rule out the claimed binary black hole system in 3C 66B \cite{3C66B03}. In recent years growing efforts have gone into investigating the detection prospects \cite{Sesana2009,KJLee2011,Chiara12PRL,Ravi14Single,Rosado15} of, and designing data analysis methods \cite{Babak2012,Ellis2012,EllisBayesian,Taylor14,YanWang14,PPTAcw14,Zhu15Single} for, continuous waves.

\begin{figure}[H]
  \centering
  \includegraphics[scale=0.4]{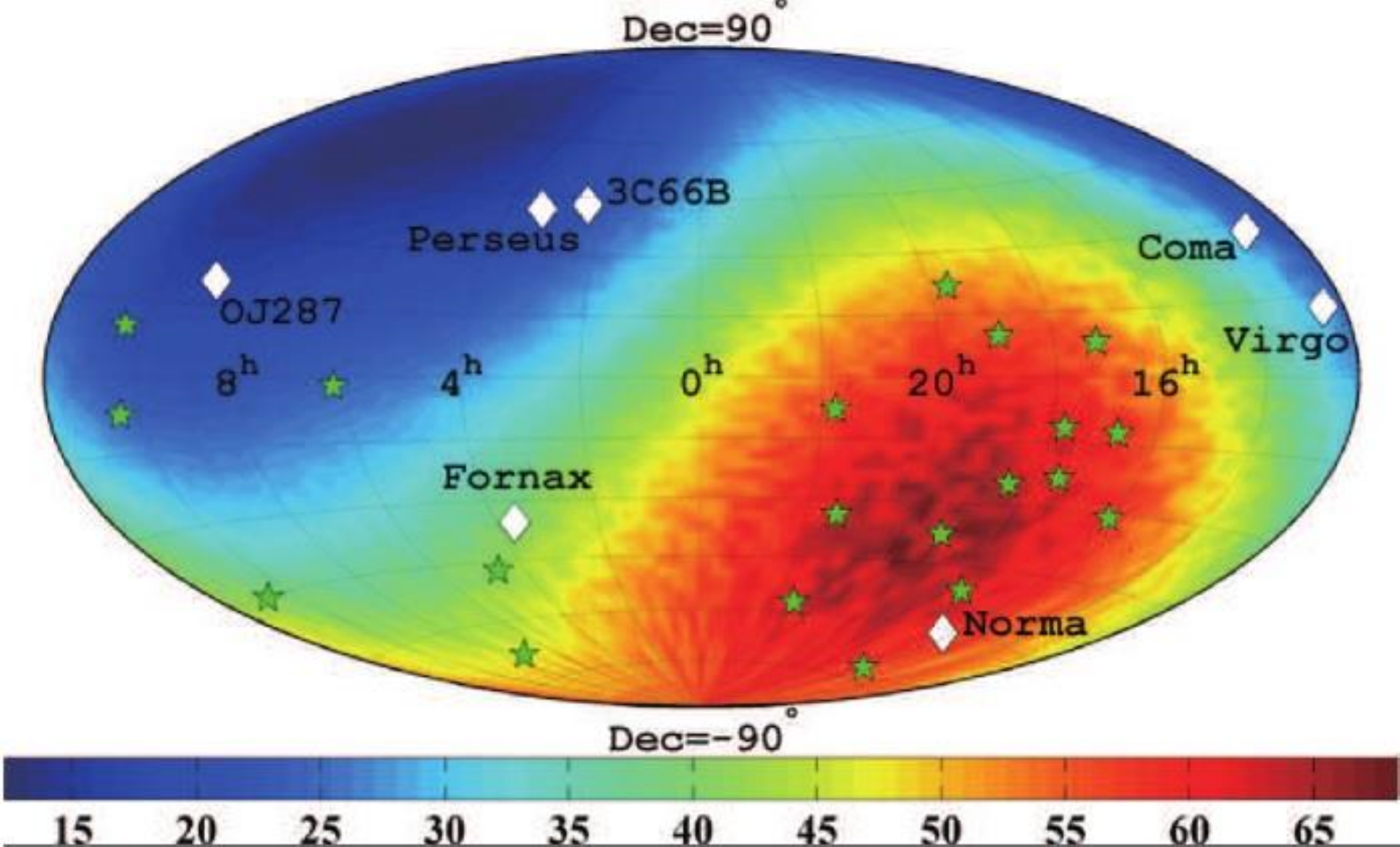}
  \caption{(Color online) Sky distribution of luminosity distance (in Mpc) out to which a circular binary black hole of chirp mass $10^{9}~ {\rm M}_{\odot}$ and orbital period of 5 nHz could be detected with the PPTA data set. Sky locations of the 20 PPTA pulsars are indicated by ``$\star$". White diamonds mark the location of possible supermassive black hole binary candidates or nearby clusters: Virgo (16.5 Mpc), Fornax (19 Mpc), Norma (67.8 Mpc), Perseus (73.6 Mpc), 3C 66B (92 Mpc), Coma (102 Mpc) and OJ287 (1.07 Gpc). Figure originally published in ref. \cite{PPTAcw14}.}
  \label{fig:SkySenDR1}
\end{figure}

Yardley et al. \cite{Yardley2010} calculated the first sensitivity curve of a PTA to this type of sources using an earlier PPTA data set presented in ref. \cite{Verbiest09}. Recently both PPTA \cite{PPTAcw14} and NANOGrav \cite{NANOcw14} conducted searches for continuous waves in their corresponding real data sets. Because of its excellent data quality, PPTA has achieved by far the best sensitivity for continuous waves. Figure \ref{fig:SkySenDR1} shows a sky map of PPTA's sensitivities to circular binary black holes of chirp mass $10^{9}~{\rm M}_{\odot}$ and orbital period of $\sim$6 years \cite{PPTAcw14}. Unfortunately most nearby galaxy clusters or binary black hole candidates are within the less sensitive sky region. Zhu et al. \cite{PPTAcw14} also presented the most stringent upper limit (published to date) on the GW strain amplitude ($h_0<1.7\times 10^{-14}$ at 10 nHz) and on the local merger rate density of supermassive binary black holes (fewer than $4 \times 10^{-3}\ {\rm{Mpc}}^{-3}\ {\rm{Gyr}}^{-1}$ for $M_{\rm c}\geqslant10^{10}~{\rm M}_{\odot}$).

\subsubsection{Gravitational wave memory}
A GW memory is a permanent distortion in the spacetime metric \cite{Memory87,Favata09}. Such effects can be produced during mergers of supermassive binary black holes and cause instantaneous jumps of pulse frequency. For a single pulsar, this is indistinguishable from a glitch event. With a timing array, GW memory effects can be searched for as simultaneous pulse frequency jumps in all pulsars when the burst reaches the Earth. It has been suggested that GW memory signals are in principle detectable with current PTAs for black hole mass of $10^{8}~{\rm M}_{\odot}$ within a redshift of 0.1 \cite{Seto09,GWM-RvH,Pshirkov10,Dusty14GWM}. However, the event rate is highly uncertain. Current estimates are very pessimistic, predicting only $0.03$ to $0.2$ detectable events every 10 years for future PTA observations based on the SKA \linebreak
\vspace*{-3mm}

\noindent \cite{CordesJenet12,Ravi14Single}. Actual searches in existing PTA data sets---see ref. \cite{GWM_PPTA} for PPTA and ref. \cite{NANO15Memory} for NANOGrav---have set upper limits on the memory event rate, which remain orders of magnitude above theoretical expectations.

\subsubsection{Gravitational wave bursts}
Potential burst sources of interest to PTAs include the formation or coalescence of supermassive black holes, the periapsis passage of compact objects in highly elliptic or unbound orbits around a supermassive black hole \cite{1.4.1}, cosmic (super) string cusps and kinks \cite{Vilenkin81,Damour00,Siemens06}, and triplets of supermassive black holes \cite{TripleBH10}. Finn $\&$
Lommen \cite{1.4.1} developed a Bayesian framework for detecting and characterizing burst GWs (see also ref. \cite{Deng14PRD}). Zhu et al. \cite{Zhu15Single} recently proposed a general coherent (frequentist) method that can be used to search for GW bursts with PTAs. No specific predictions have been made for detecting GW bursts with PTAs in the literature. There has been no published results on searches for bursts using real PTA data.

\subsection{Summary and future prospects}
\label{sec:conclu}
With their existence predicted by general relativity one century ago, GWs have not yet been directly detected. However, it is widely believed that we are on the threshold of opening the gravitational window into the Universe. In the audio band (from 10 to several kilo Hz), 2nd-generation laser interferometers such as Advanced LIGO are about to start scientific observations and a detection of signals from compact binary coalescences (e.g., binary neutron star inspirals) is likely within a few years (see Sect. 2). In the nanohertz frequency range, PTA experiments have achieved unprecedented sensitivities and started to put serious constraints on the cosmic population of supermassive black hole binaries.

The sensitivity of a PTA can be improved by a) increasing the data span and observing cadence, b) including more pulsars in the array, and c) reducing the noise present in the data. Regarding the first two factors, the combination of data sets from three PTAs to a single IPTA data set offers the most straightforward benefit. Other ongoing efforts include optimization of observing strategies, searches for millisecond pulsars, characterization of various noise processes and corresponding mitigation methods, and development of advanced instrumentations. In the longer term, pulsar timing observations with FAST and SKA will provide advances in all aspects of PTA science, not only leading to the detection of GWs but also allowing detailed studies of the nanohertz gravitational Universe.


\section{Searching for gravitational waves in the CMB}
\emph{We briefly review the efforts for the detection of cosmological GWs into the CMB anisotropies of the early Universe.
We describe the progresses over the last year, the current upper limit to tensor modes in cosmology coming from their direct search into CMB
polarization, and the expectations from near and long term probes.}

\subsection{Introduction}
\label{i}

The $\Lambda$CDM cosmological concordance model, now supported by all observations including the latest Planck measurements
of the CMB (see ref. \cite{planck_mission_paper_2015} and references therein), is composed by a non-relativistic (Cold) Dark Matter component
($26\%$), known particles (baryons and leptons, $4\%$), as well as a Cosmological Constant responsible for about $70\%$ of the
entire cosmic energy density. In the early phases of expansion, the $\Lambda$CDM model assumes an era of accelerated expansion,
known as inflation, in which an almost scale invariant spectrum of Gaussian density perturbations is generated, responsible for
the structures we observe today. Many open problems concerning this model are related to the exact nature of the dark cosmological
components, as well as to the details of the very early Universes, in order to discriminate between different inflationary scenarios.
The latter aspect is the subject of the present contribution to the Next Generation Gravitational Wave Antenna conference
which took place at the Kavli Institute in Beijing, in May 2015.

According to the inflationary process, if the latter is early enough, a production of cosmological GWs would take
place along with the density ones. A key parameter for describing them is the tensor to scalar ratio, i.e. the ratio between
the amplitude in tensor (GWs) power and scalar (density) ones:
\begin{equation}
r=\frac{A_{T}}{A_{S}}\ .
\label{r}
\end{equation}

Cosmological GWs do excite one mode of linear polarization in the CMB. The latter is characterized by anisotropies
in the temperature ($T$), gradient modes of the linear polarization tensor ($E$) and curl component, the $B$-modes, which is where
the cosmological GWs imprint their contribution. The duration of the the decoupling process of photons, the origin
of the CMB, determines the angular scale at which the $B$-modes from GWs should be detected, which is about $1^{\circ}$
in the sky. On the other hand, at the arcminute scale, the dark matter traced by luminous objects would gravitationally lens the
CMB photons, causing a transfer of power from the $E$ to $B$ modes. The latter effect is most relevant for investigating the
dark cosmological components, in particular in cross-correlation with LSS surveys \cite{bianchini_etal_2015}.
An intensive experimental effort is ongoing in the attempt to detect the $B$-modes on all angular scale, which is the subject of this
contribution. In  sect. 5.2 we describe the main phenomenology determining the angular power spectrum of CMB anisotropies.
In sect. \ref{bmo} we describe the status of $B$-mode observations at the time of the conference. Finally, in sect. \ref{einialt},
we describe the expectations in the near, intermediate and far term.

\subsection{CMB anisotropies}
\label{cmba}

\begin{figure}[H]
\centering
\includegraphics[scale=1]{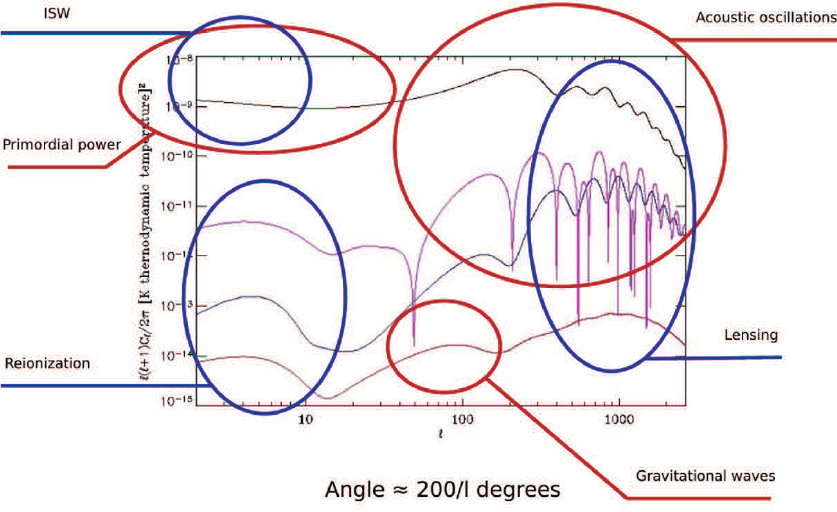}
\caption{(Color online) CMB power spectra for $\Lambda$CDM cosmologies, with a contribution from cosmological gravitatioanl waves corresponding to
$r=0.1$, along with the main effects in their relevant angular domain. The Black line represent the $T$ component. The $E$ is blue,
while the $TE$ correlation is magenta. The $B$-modes are in dark red.}
\label{cmb_spectra}
\end{figure}

We describe here the main phenomenology of the CMB $T$, $E$ and $B$-modes in terms of their angular power spectra. In Figure
\ref{cmb_spectra} we show the power spectra of a typical $\Lambda$CDM model, corresponding to the Planck best fit in terms
of the main cosmological parameters, and differing for the value of $r$, which is set equal to 0.1. Effects at decoupling are
marked in red, while blue labels mean secondary anisotropies, i.e. those imprinted on the line of sight to us. The horizon scale
at decoupling, marking the distinction between super-horizon and sub-horizon perturbations at the corresponding epoch, corresponds
to about $l=200$, which is about 1 degree in the sky. The $T$ modes feature basically unperturbed primordial power on larger scales,
while being subject to oscillations on smaller ones, characterized by competition between radiation pressure and gravitational
potentials provided by the dark matter. Since polarization is sourced by the local quadrupole at decoupling, the acoustic oscillations
are transferred to $E$ modes, as well as their correlation with $T$, named $TE$ components. The $B$-modes feature a peak at
$l\simeq 100$ due to cosmological GWs, as well as the leak of power from $E$ to $B$ due to gravitational lensing,
causing a peak, resembling the $E$ acoustic oscillations, centered at about $l=1000$. On large scales, secondary effects
coming from the Integrated Sachs-Wolfe and reionization, which we do not treat here, are also highlighted.

\subsection{$\bm B$-mode observations}
\label{bmo}

An extraordinary progress has occurred in the year 2014. PolarBear has published the first detection
of auto-spectra of $B$-modes at the arcminute scale. That was found consistent with predictions from cosmological
gravitational lensing within the $\Lambda$CDM concordance model \cite{polarbear_b_modes}, and, to date, it represents the first detection
of cosmological $B$-modes from one single experiment, at the auto-spectrum level. Immediately after,
an excess on the degree angular scales, which could be ascribed to cosmological GWs, was
announced by the BICEP2 collaboration \cite{bicep2_b_modes}. The Planck collaboration,
by analyzing the data of the polarization sensitive at 353 GHz, was able to estimate the level of
Galactic contamination from thermal dust to the BICEP2 measurements, finding that, within uncertainties,
the latter are to be attributed to the foreground emission \cite{planck_b_modes_dust}. The BICEP2 and
Planck collaborations produced a join analysis in the sky region which was observed by BICEP2, using also
the data from the Keck array, resulting in an upper limit on the value of $r$ coming for the first time from direct $B$-mode
measurements \cite{bicep2_planck_b_modes}:
\begin{equation}
r\le 0.12\ {\rm (95\%\ confidence\ level).}
\label{r_limit}
\end{equation}

\begin{figure}[H]
\centering
\includegraphics[scale=0.5]{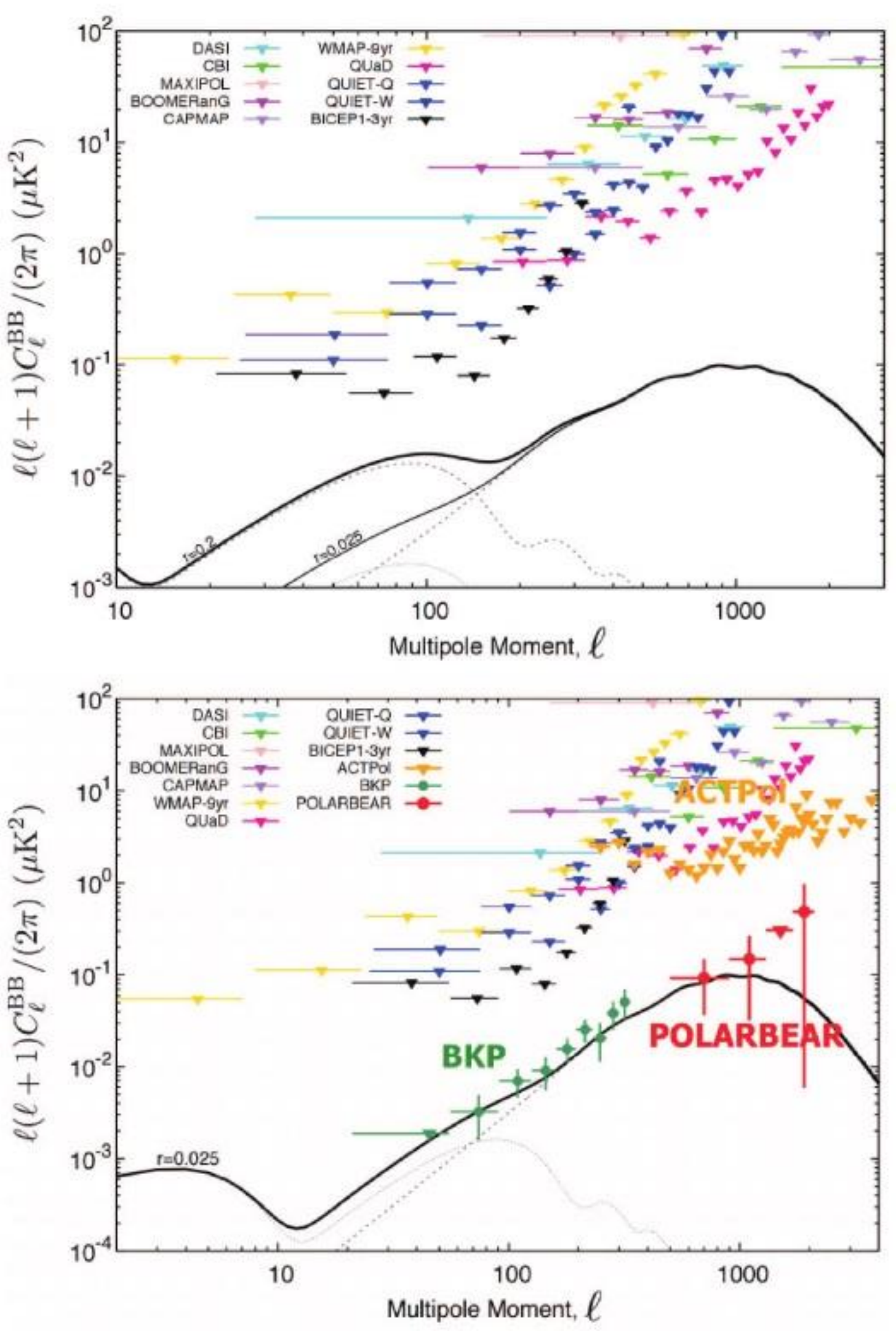}
\caption{(Color online) Experimental data at the beginning of 2014 (left) and now (right). Courtesy of Yujii Chinone of the University of California Berkeley.}
\label{b_modes_2014}
\end{figure}

Recently $B$-mode measurements by the ACTpol \cite{actpol_b_modes} and SPTpol \cite{sptpol_b_modes} observatories have been reported.
In Figure \ref{b_modes_2014} we show the measurements on $B$-modes at the beginning of 2014 and at the time of the Conference. The upper limits
turned into actual measurements, extending from the degree angular scales by the~ BICEP-Keck-Planck data~ (labeled~ by

\noindent BKP) to the  PolarBear
ones at the arcminute scale. Also plotted, the theoretical curves for $\Lambda$CDM models with various levels of $r$, as shown.

\subsection{Expectations in the near, intermediate and long term}
\label{einialt}

At this moment, operating probes are mounting new detectors to go multi-frequency, in order to be able to monitor and
subtract the
Galactic foregrounds. We quote here a fraction of the most important efforts, a complete list being available at lambda.gsfc.nasa.gov/product/expt.
The PolarBear program will turn into the PolarBear-2 program taking data at 90, 150 and 220 GHz in 2016, while two other telescopes
will be built, constituing the Simons Array (bolo.berkeley.edu/polarbear). The same frequency range is being probed by the
BICEP and Keck (bicepkeck.org) arrays, as well as the ACTpol (www.princeton.edu/act) and SPTpol (pole.uchicago.edu)
telescopes. Ground measurements, as the quoted ones, are likely to process, in the near term, from now to a few years, the frequency interval extending
from 90 to 220 GHz. The limit at low frequency comes from the relevance of the polarized synchrotron Galactic foreground, which is at this moment only
known on large angular scales thank to the Planck \cite{planck_foreground_component_separation} and WMAP \cite{bennett_etal_2013} observations.
At high frequency, the atmosphere is not negligible for these measurements. On the other hand, it is necessary to monitor lower and higher frequencies,
where the synchrotron and the dust represent very relevant diffuse Galactic foregrounds, respectively. At low frequency, and in the intermediate term,
the CLASS (sites.krieger.jhu.edu/class) will be extending observations on large portion of the sky down to 40 GHz. A similar program
will be the LSPE balloon (planck.roma1.infn.it/lspe); both probes also observe below 300 GHz on the high frequency side. At higher frequencies
baloon programmes from SPIDER (cmb.phys.cwru.edu/ruhl$_{-}$lab/spider.html) and EBEX (groups.physics.umn.edu/cosmology/ebex) have observed
up to 410 GHz, and are currently analyzing the data. Ambitious programs from satellites, with frequency ranges probing the high and low
frequency tails, exist. In the intermediate term, a satellite dedicated to the large angular scales will be LiteBird (litebird.jp/eng).
In the long term, many frequency bands will be observed by the PIXIE (asd.gsfc.nasa.gov/pixie) satellite, with the goal of determining with great
accuracy the high frequency foreground properties in terms of large scale distribution, composition and frequency scaling. Also in the long term, the idea
of a nearly ultimate satellite, CORE+ (www.core-mission.org), with the goal of mapping the $B$-modes on the whole sky down to the arcminute scale,
are being proposed. Within these efforts, the ultimate threshold of detection for $r$ is claimed of order $10^{-3}$.

\section{Roadmap for
gravitational wave detection in space---a preliminary study}
\subsection{Introduction}

In 2008, under the auspices of the National Microgravity
laboratory of the Institute of Mechanics, Chinese Academy of
Sciences (CAS), an (unofficial) gravitational physics  consortium
comprising a number of institutes and universities both within and
outside the CAS was established. The primary objective of the consortium is to
coordinate and promote research in the  detection of GW in space in China. A roadmap was soon worked out  in order to
build up  the expertise and required technologies step by step for
future prospective Chinese mission. As
a first step of the roadmap, a geodesy mission to monitor the
temporal variation of the Earth gravity field using low-low
satellite to satellite tracking by means of laser interferometry
will be developed. This will enable us to acquire the key
technologies at a lower level of precision and at the same time
assemble a core team for further development. As a geodesy mission
will only test laser interferometry in space and is less stringent
in requirement in inertial sensor related technologies, a LISA
Pathfinder (LPF) type mission is also required at some stage to test
the inertial sensor and the related dragfree technologies to the
level of precision required by GW detection. As the
two proof masses in an LPF type mission may be regarded naturally as
a one dimensional gravity gradiometer, in recent years, some work
has also been ongoing to explore possible additional scientific
benefits of an LPF type mission, apart from testing the key
technologies in space.

Supported by the Xiandao (pioneer explorer) program of the National
Space Science Center of the CAS, the general relativity group of the
Morningside Center of Mathematics and Institute of Applied Maths,
CAS have been undertaking the task of doing preliminary study on
the prospective missions outlined in the roadmap. The purpose of
the present article is to present an overview of the work we have
been doing in this area.

The outline of the present article is described as follows. In sect. 6.2,  we will outline a feasible design and its primary science driver for a mission to detect GWs in space.
This is then followed in sect. 6.3 by a sketch of a geodesy mission study we have been doing. In sect. 6.4, we will discuss the
precision measurement of the Earth's gravitomagnetic field  as possible additional science for a LISA Pathfinder type mission.
Some brief remarks will be made in sect. 6.5 to conclude this section.

\subsection{Gravitational wave detection in space}
Concurrent to the mission study in satellite gravity beginning in 2009 which will be described in sect. 6.3,
a preliminary study was also made on the scientific potential of a spaceborne gravitational detector whose most sensitivity
frequency band centers around 0.01 or 0.1 Hz. The main theme of the study was to look at possible GW sources around 0.01 Hz which was
not entirely well understood at then. With  ALIA \cite{BenderALIA, BenderBegelman} adopted as a representative mission concept around the 0.01 Hz band, further analysis revealed that there is a
rich source of intermediate mass black hole binaries at high redshift and the detection would shed light on the black hole-galaxy coevolution
during the structural formation process of the Universe \cite{8thLISA}.

During 2011--2013,  a subsequent, more indepth study was made \cite{LISA_JPCS, Xiandao}.  This time technological constraints were taken into account
and we strived to obtain a balance between merits in science and viability in technologies. With this guideline in mind, after examining a few representative mission options,
a mission concept with the
following  baseline design parameters (Table \ref{parameters}) and sensitivity (Figure 27) is favoured as a blueprint for more in depth study in the near future.

The capability of the mission designs to detect lighter seed black holes at high redshifts is illustrated in Figure \ref{range1} \cite{LISA_JPCS}.

A semi-analytic Monte Carlo simulation was also
carried out to understand the cosmic black hole merger histories starting from intermediate
mass black holes at high redshift as well as the possible scientific merits of the mission design
 in probing the light seed black holes and their coevolution with galaxies in early
Universe (Figure 29) \cite{LISA_JPCS}.

Moreover, the mission design is also capable of probing
IMRI (intermediate mass ratio inspiral)  in dense star
clusters in the local Universe. Presented in Table \ref{1.6talbe 2} is some event rate calculations  for IMRIs in local Universe \cite{LISA_JPCS}.
It should be remarked that  estimate is model dependent and subject to many uncertainties and  we should not
attach too much importance to the precise numbers. Instead, it illustrates  the advantage of shifting slightly the most sensitive region of the measurement band
to a few hundredth Hz. As far as  IMRIs are concerned, the
 event rate goes up as the cubic of the improvement in sensitivity.

The result of the study suggests that,  by choosing the armlength of the interferometer to be three
million kilometers and shifting the sensitivity
floor to around one-hundredth Hz, \linebreak\vspace*{-3mm}

\begin{table}[H]
\caption{Baseline design parameters\hspace*{60mm}}\label{parameters}
\footnotesize
\begin{tabular}{lllll}
\toprule
$L$ (m) & $D$ (M) & $P$ (W) &  $S_{\rm posi} \left(\frac{\textrm{pm}}{\sqrt{\textrm{Hz}}}\right)$  & $S_{\rm acc} \left(\frac{{\textrm m}~{\textrm s}^{-2}}{\sqrt{\textrm{Hz}}}\right) $ \\
\hline
 $3\times10^9$ & 0.45--0.6 & 2 & 5--8 & $3\times10^{-15}\,(>0.1\ \textrm{mHz})$ \\
 \bottomrule
\end{tabular}
\end{table}

\begin{figure}[H]
\centering
\includegraphics[scale=1]{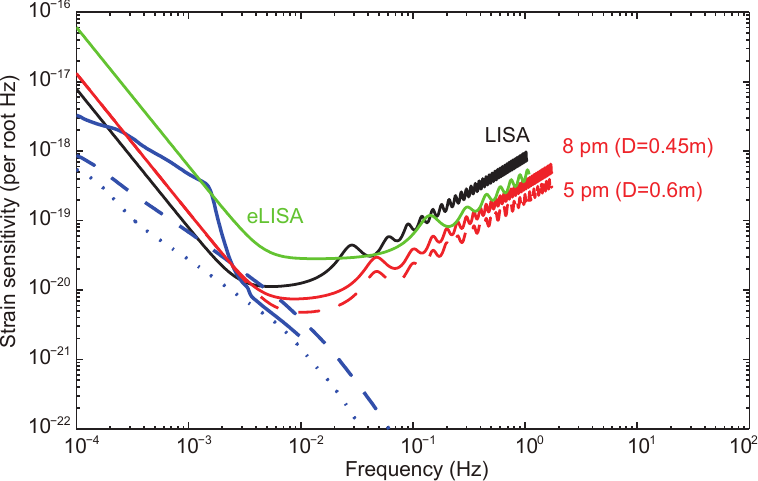}\vspace*{-2mm}
\caption{(Color online) Sensitivity curves of the mission designs, taken into consideration  the confusion
 noise generated by both galactic and extra-galactic compact binaries converted from those by Bender  and Hils \cite{HilsBender} and Farmer and Phinney \cite{FarmerPhinney}.
}
\label{sensitivity}
\end{figure}

\begin{figure}[H]
\centering
\includegraphics[scale=1]{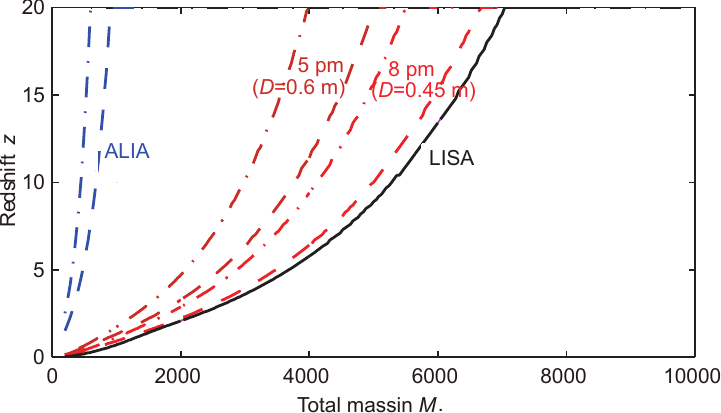}\vspace*{-2mm}
\caption{(Color online) All-angle averaged detection range under a
single Michelson threshold SNR of 7 for 1:4 mass ratio IMBH-IMBH
binaries, one year observation prior to merger. For each mission option, both upper and lower confusion noise levels (represented by the dashed curve and dotted dashed curve respectively) due to extragalactic compact binaries are considered.}\label{range1}
\end{figure}

\begin{figure}[H]
\centering
\includegraphics[scale=0.28]{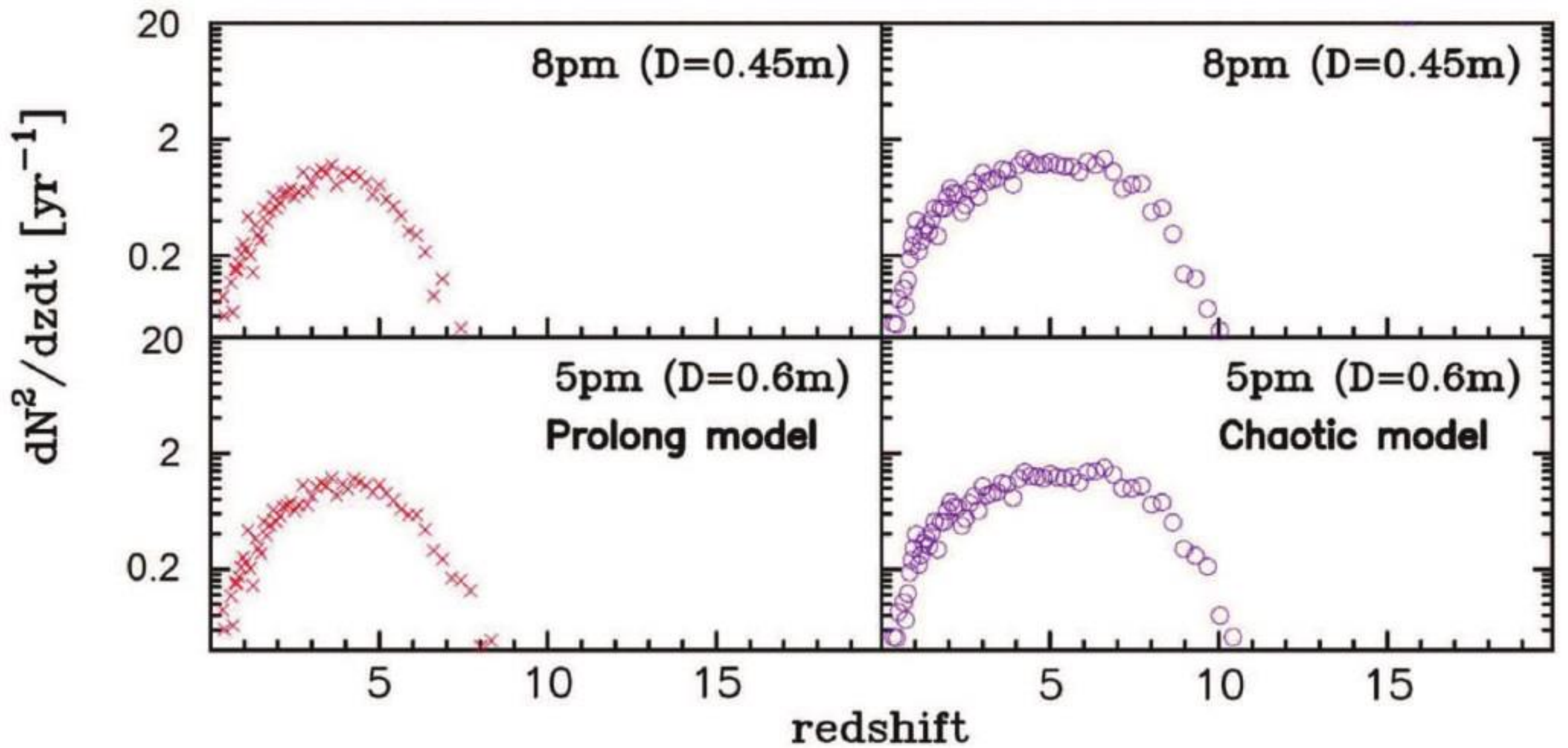}\vspace*{-2mm}
\caption{(Color online) Event rates given by the area enclosed by the curves.}
\label{detectionrate}
\end{figure}

\begin{table}[H]
\caption{Expected globular cluster harbored IMBH-BH final inspirals and mergers detectable for year-long observation, scaled in unit of
$\frac{f_{\rm tot}}{0.1}\frac{\nu_0}{10^{-10}~{\textrm{yr}}^{-1}}\frac{10~{\rm M}_{\odot}}{\mu}$}\label{1.6talbe 2}\vspace*{-5mm}
\begin{center}\doublerulesep 0.1pt \tabcolsep 4pt
\footnotesize
\begin{tabular}{ccc}
\toprule
Mission option & Upper level of  confusion &Lower level of confusion \\
 \hline
5 pm $(D=0.6$ m) &$\sim90$ &$\sim130$   \\
8 pm $(D=0.45$ m)&$\sim26$& $\sim32$   \\
\bottomrule
\end{tabular}
\end{center}
\end{table}

\noindent together with
a very moderate improvement on the position noise budget, there are certain mission options
capable of exploring light seed, intermediate mass black hole binaries at high redshift that are
not readily accessible to eLISA/LISA \cite{Arun, NGO_1}, and yet the technological requirements seem to within
reach in the next few decades for China \cite{LISA_JPCS, Xiandao}.

\subsection{Low-low satellite to satellite tracking using laser interferometry}

In 2009, commissioned by the National Space Science  Center, Chinese
Academy of Sciences,  a comprehensive study was undertaken in order
to understand  various aspects of a future low-low satellite to
satellite tracking mission with microwave ranging replaced by laser
interferometry.

During the feasibility study, different groups
undertaking different tasks within the mission study suggested
possible mission designs for future satellite gravity missions. Due
to the limitation imposed by aliasing generated by the atmosphere
and ocean currents, perhaps in a way not too surprising, subject to
minor variations, all groups came up with very similar mission
design (see for instance ref. \cite{Zheng}, compare also with that of the NGGM mission \cite{Gruber, Anselmi} whose misson design parameters are given in  Table \ref{1.6talbe 3}.

The critical difference between the two mission designs is the attitude of the orbit. For Option 1 in Table \ref{1.6talbe 3}, thruster (dragfree) technology is required
to maintain the orbit at such a low attitude.

\begin{figure}[H]
\centering
 \includegraphics[scale=0.95]{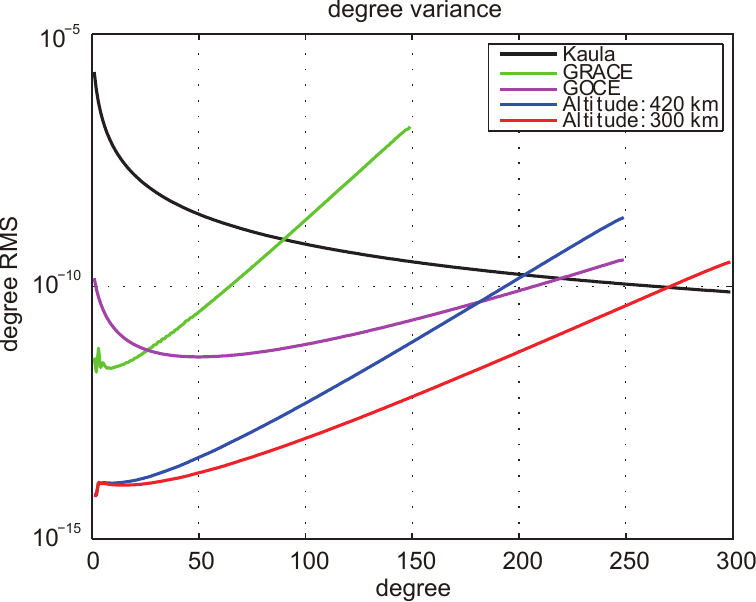}\vspace*{-2mm}
 \caption{(Color online) Static gravity field recovery in terms of spherical harmonic degree variance for different altitudes. Range: 100 km, laser metrology: $10\ \textrm{nm}/\sqrt{\textrm{Hz}}$.}
 \label{alti}
\end{figure}

\begin{figure}[H]
\centering
 \includegraphics[scale=0.95]{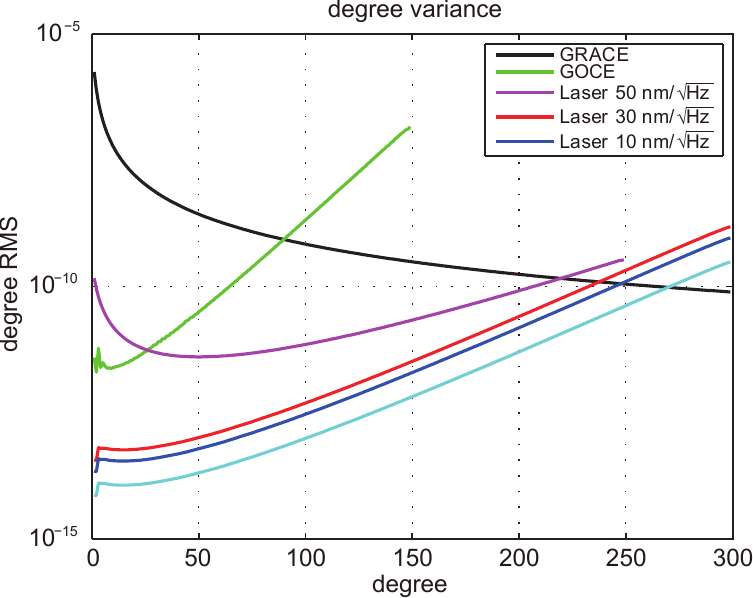}\vspace*{-2mm}
 \caption{(Color online) Static gravity field recovery in terms of spherical harmonic degree variance for different laser accuracy. Altitude: 300 km, range: 100 km.}
 \label{accu}
\end{figure}

By means of the semi-analytic method \cite{Sneeuw}, the capability of static gravity field recovery of
the mission designs is illustrated in Figures \ref{alti} and  \ref{accu}.

\begin{figure}[H]
 \includegraphics[scale=0.31]{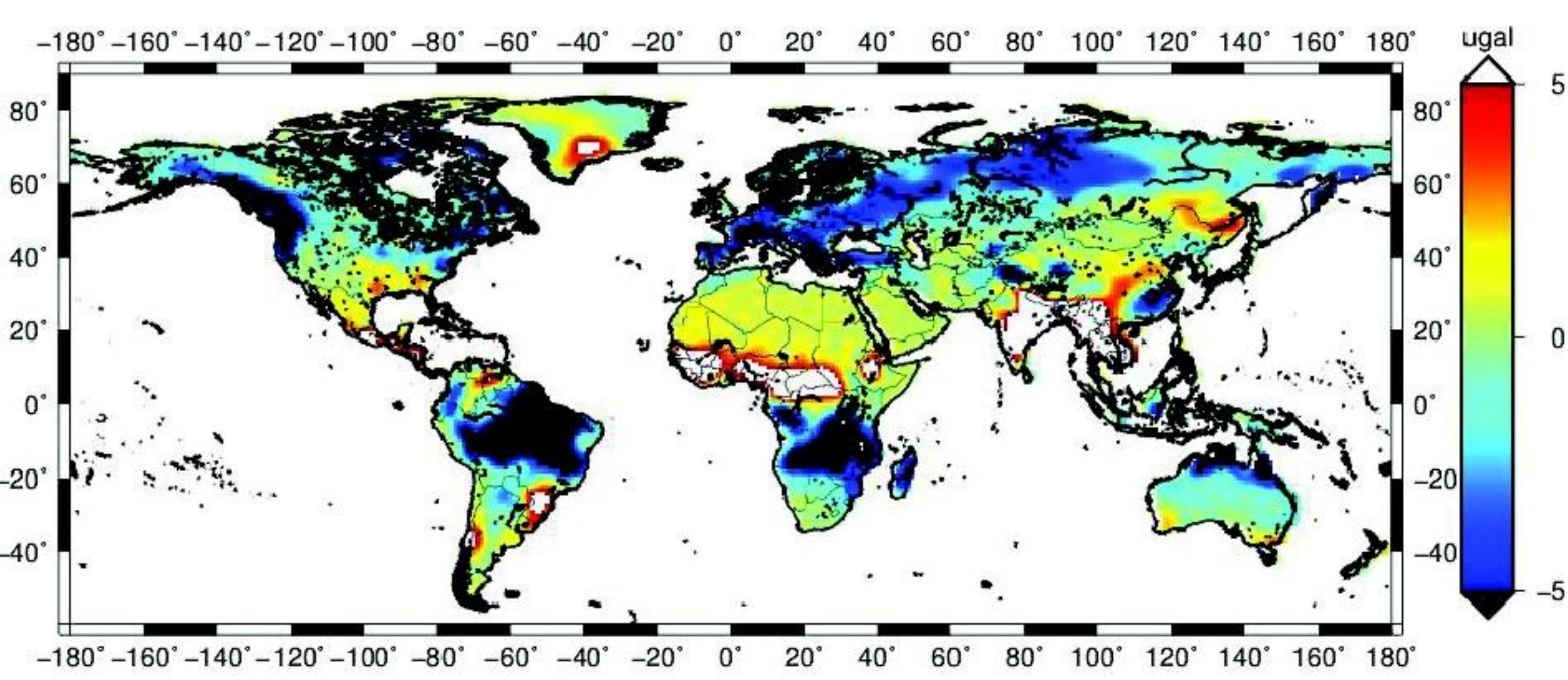}\vspace*{-2mm}
 \caption{(Color online) GLDAS model in terms of gravity anomaly (units in $\mu$gal) for October, 2009, up to harmonic degree 90.}
 \label{6.3figure32}
\end{figure}

For the capability to track temporal variation of the Earth gravity field, using the GLDAS model and taking into account of AOD aliasing,
the hydrological signal recovery is displayed in Figures \ref{6.3figure32}--\ref{fig:amss1} for the two mission design considered.

 Work is still ongoing to understand the various aspects of the mission concepts (see for instance refs. \cite{Gaowei, Xupeng}).

\subsection{Precision measurement of gravitomagnetic field in terms of gradiometry}\label{}

Apart from detection of GWs, another out-standing problem \linebreak

\begin{tablehere}
\caption{Baseline design parameters for two representative mission options}\label{1.6talbe 3}
\vspace{-3mm}\footnotesize
\begin{center} \doublerulesep 0.1pt \tabcolsep 5pt
\begin{tabular}{ccccccc}
\toprule
&Mission &  & Accelero- & Laser & Satellite & Inter \\
&duration  & Orbit & meter (mHz & (mHz & altitude & satellite\\
 & &  & to 3 Hz) & to 3 Hz) & & range\\\hline
1&10-year  & Polar & ${10}^{-10}$ & 10--30  & $300\ \text{km}$ & $50-$\\
 & &  & ${\textrm{ms}}^{-2}\ {\textrm{Hz}}^{-1/2}$ & $\textrm{nm}\ {\textrm{Hz}}^{-1/2}$  & & $200\ \textrm{km}$\\
2&10-year  & Polar & ${10}^{-9}$ & 10--30  & $420-$ & $50-$\\
 & &  & ${\textrm{ms}}^{-2}\ {\textrm{Hz}}^{-1/2}$ & $\textrm{nm}\ {\textrm{Hz}}^{-1/2}$  & $450\ \textrm{km}$ & $200\ \textrm{km}$\\
 \bottomrule
\end{tabular}
\end{center}
\end{tablehere}

\begin{figure}[H]
 \includegraphics[scale=0.31]{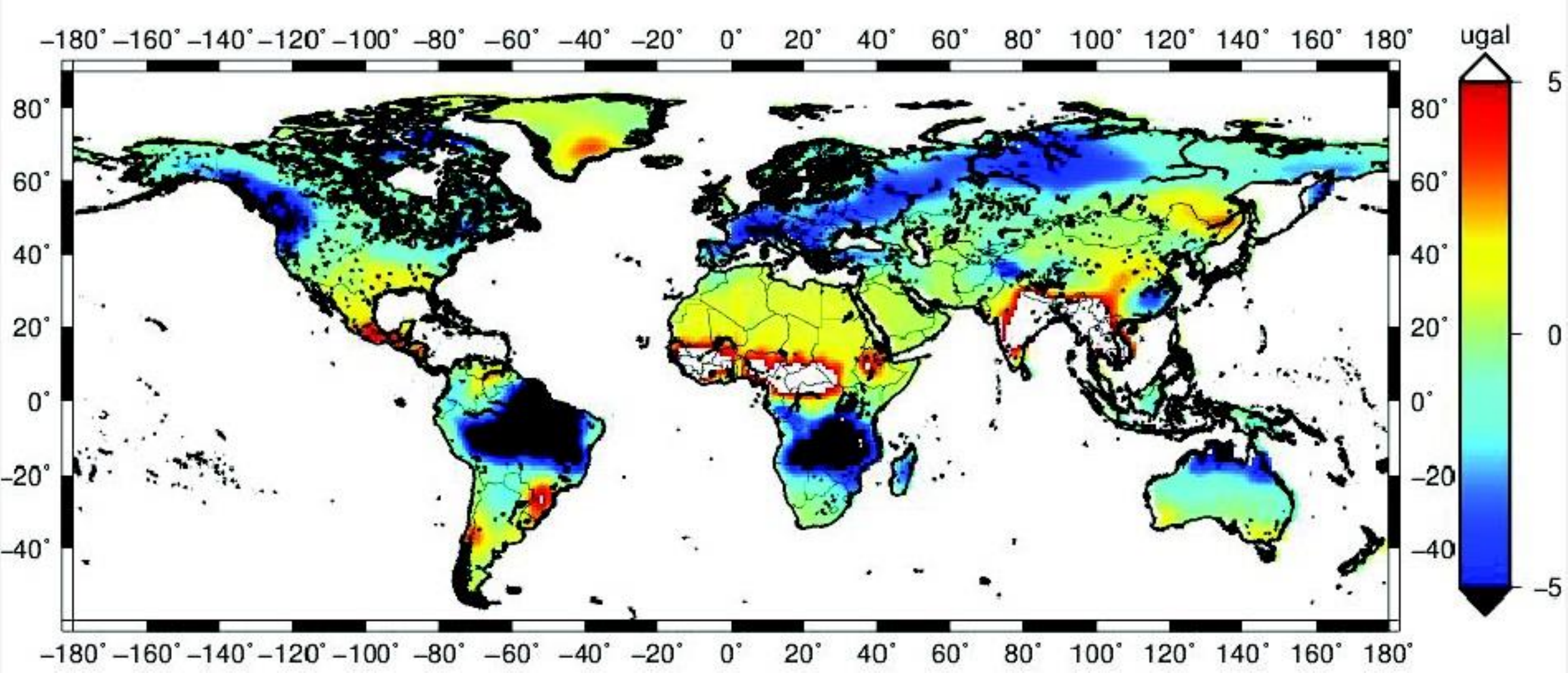}\vspace*{-2mm}
 \caption{(Color online) Recovered GLDAS anomaly for October, 2009, up to harmonic degree 90. Altitude at 300 km with 400 km Gaussian filter.}
 \label{fig:amss1}
\end{figure}

\begin{figure}[H]
  \includegraphics[scale=0.31]{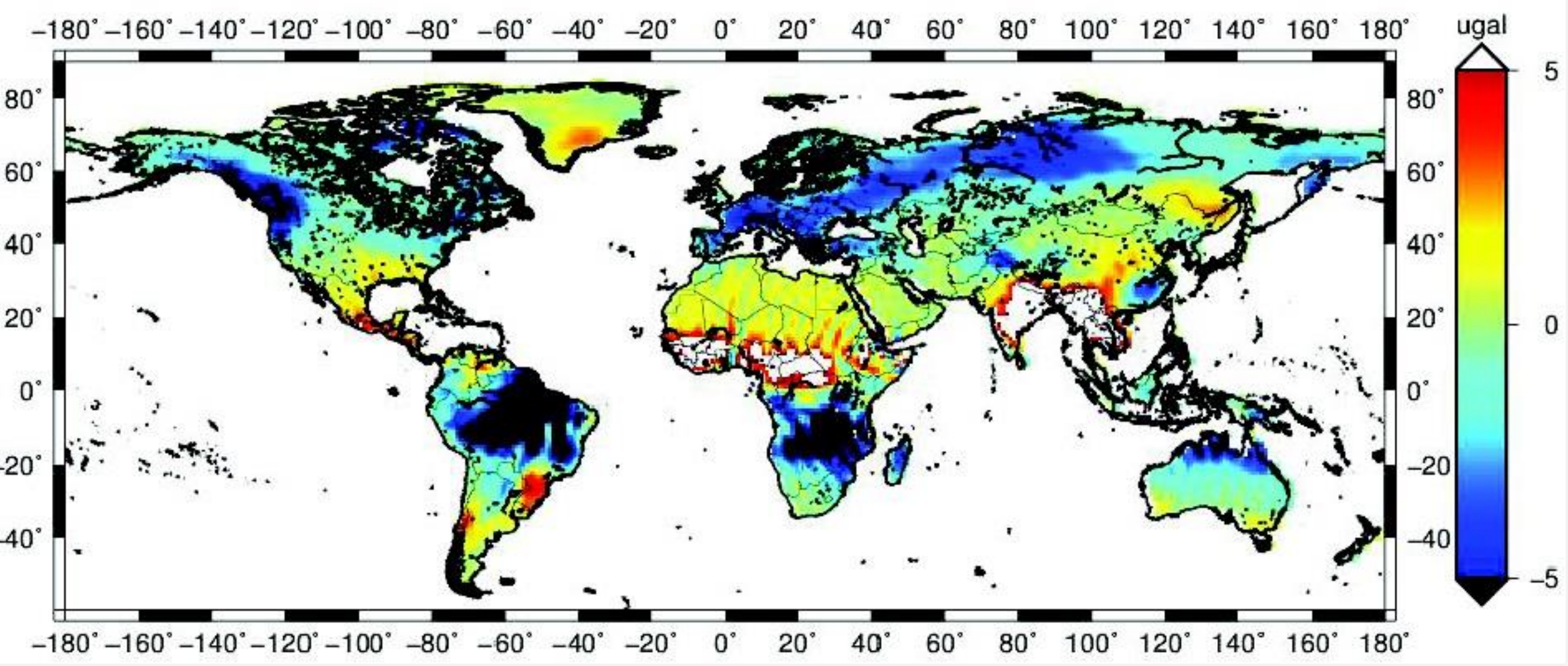}\vspace*{-2mm}
 \caption{(Color online) Recovered GLDAS anomaly for October, 2009, up to harmonic degree 90. Altitude at 450 km with 400 km Gaussian filter.}
 \label{fig:amss1}
\end{figure}

\noindent  in experimental relativity is the detection of gravitomagnetic
field \cite{Thorne1988, Ciufolini1995} of a rotating body which may be regarded as a test of general relativity
at the planetary scale.
 As the GM field is part of the spacetime curvature,
it will contribute to the tidal forces acting on a family of nearby free falling particles.
It is  natural to contemplate to detect its presence through the measurement of  the tidal force gradient
generated by the GM field \cite{Braginskii1980,  Mashhoon1989}.
As the two free falling test masses (TMs) in the LISA Pathfinder type mission naturally constitute
a one dimensional gravity gradiometer, apart from testing the technologies
for GW detection, we also try to look at whether
such mission is also capable of measuring GM field around a planet.

Consider an ideal case in which the Earth is modelled as a uniform rotating
spherical body. Let $(t, x_i), i=1,2,3$ be the Earth centered coordinates and $\bm e^{}_{(a)}$, $a=1,2,3$ be the Earth pointing orthonormal three frame attached to the
center of mass of the spacecraft. Units in which $c = G = 1$ will be adopted.

\begin{figure}[H]
\centering
\includegraphics[scale=1.2]{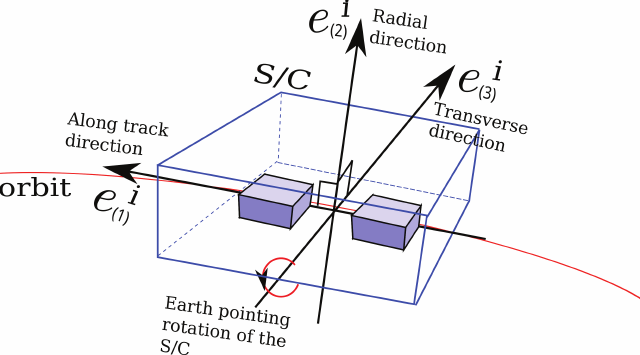}
\caption{(Color online) Two TMs are housed in the along track direction following a nearly circular orbit. The spacecraft may be viewed as a gyroscope rolling about the transverse direction $\bm e^{}_{(3)}$.}
\label{sc-frame}
\end{figure}

Given an Earth pointing, drag-free spacecraft orbiting the Earth in a nearly circular orbit (radius $a\approx 1000$ km) with constant inclination $i$. Two TMs are housed in the along track direction separated by a distance $d\approx 50$ cm (see Figure \ref{sc-frame}). Let $\delta^i$ be the relative displacement of the two TMs.
In a time scale short compared with the Lense-Thirring precession of the orbital plane and subject to $\delta^i\ll d  $, at the 1 PN level, the geodesic deviation (Jacobi) equation
describing the relative displacement $\delta^i$ may be simplified to become (see ref.
\cite{xupengLT1} for details)
\begin{eqnarray}
\ddot{\delta}^{1}(t)&=&-2\omega\text{\ensuremath{\dot{\delta}^{2}}(\ensuremath{t})}+\frac{6J\omega d\cos i}{a^{3}},\nonumber\\
\ddot{\delta}^{2}(t)&=&2\omega\dot{\delta}^{1}(t)+3\omega^{2}\text{\ensuremath{\delta^{2}}}(t)-\frac{9dJ
  t \omega^{2}\cos i}{a^{3}},\label{d3}\\
\ddot{\delta}^{3}(t)&=&-\omega^{2}\text{\ensuremath{\delta^{3}}}(t)+\frac{3dJ\omega\sin
    i\cos(\omega t)}{a^{3}}.\nonumber
\end{eqnarray}
At the Newtonian level, we recover the  well known  Clohessy Wiltshire \cite{Clohessy1960} (Hill) equation from eq. (\ref{d3}).

In the direction transverse to the orbit plane, at the Newtonian level, it is known that the relative motion of the two\linebreak
\vspace*{-2mm}

\begin{figure}[H]
  \centering
  \includegraphics[scale=1.3]{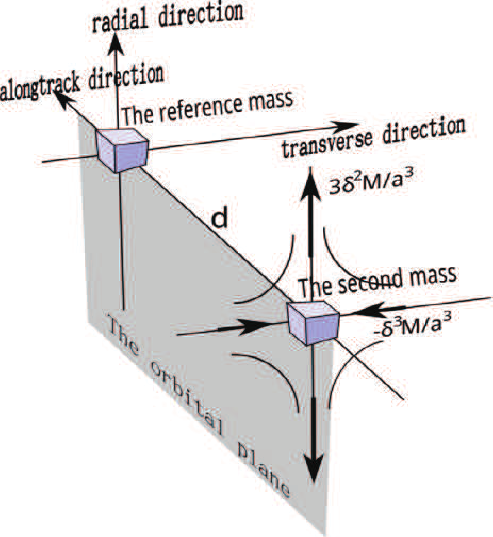}
  \caption{(Color online) For a circular orbit, take one TM as reference, the motion of the second TM in the transverse direction is equivalent to that of a forced harmonic oscillator with natural frequency which matches that of the orbital frequency.}
  \label{tidal}
\end{figure}

\begin{figure}[H]
  \centering
  \includegraphics[scale=0.3]{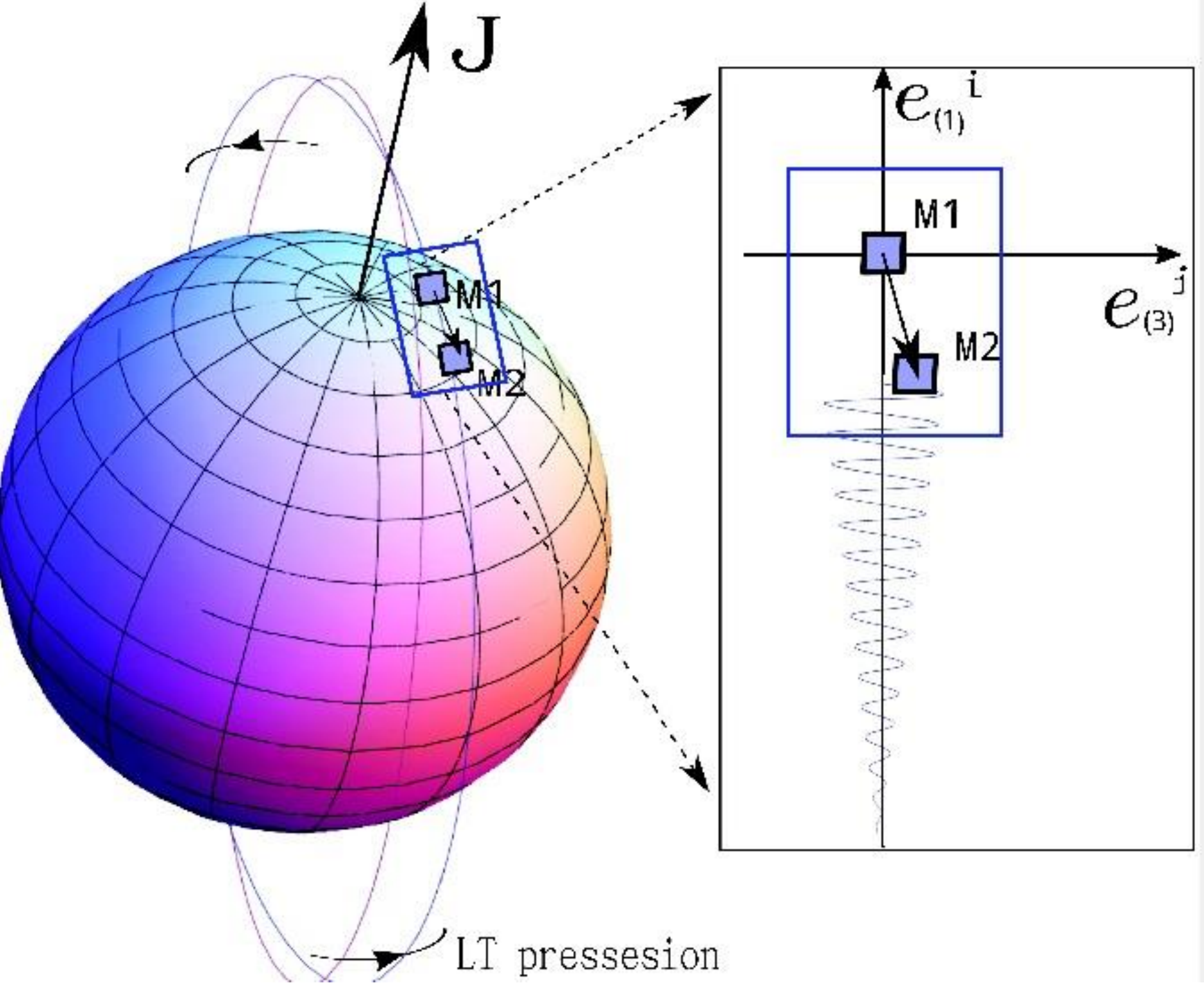}
  \caption{(Color online) The illustration of the GM signal.\hspace*{18mm}}
  \label{signal}
\end{figure}

\noindent TMs resembles that of a simple
harmonic oscillator (see Figure \ref{tidal}). The presence of a gravitomagnetic field serves to generate a Lorentz type force
acting on the TMs and results in a forced harmonic motion. Up to a constant, the frequency of the force matches the frequency $\omega=\sqrt{\frac{M}{a^3}}$ determined by the central
gravitational potential and this gives rise to  a resonant forced harmonic oscillation. Further numerical simulations indicate that the resonant
oscillation prevails for more general elliptic orbits and with Earth gravity field multiples taken into account. The amplitude of the oscillation will
not grow unbounded when nonlinearity of the Lense-Thirring precession of the orbit plane takes effect, but this will occur in a time scale much longer than the mission lifetime  of around
one year. By integrating the PN generalization of the CW equations in eq. (\ref{d3}), we further find that the growing oscillation amplitude of the TMs in the transversal direction is given by (see Figure \ref{signal})
\begin{equation}
s_{\rm GM}(t)=\frac{3GdJ\sin i\sin(\omega t)}{2c^{2}a^{3}}t. \label{eq:signal}
\end{equation}
For our orbit choice, the signal has frequency about $0.1$ mHz and its magnitude will reach a few nanometers within 1--2 d.

\begin{figure}[H]
  \centering
  \includegraphics[scale=1.3]{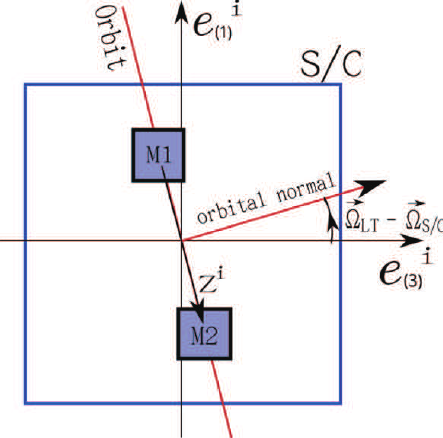}
  \caption{(Color online) Due to the differential precession between the spacecraft (or the local frame) and the orbital plane, the two TMs will oscillate relative to the spacecraft along the transverse direction.}
  \label{diff}
\end{figure}

To understand the physical picture underlying the signal readout given in eq. (\ref{eq:signal}), we note that gravitomagnetic field, apart from generating a Lense-Thirring precession of the orbital plane,
also generates a precession of the orthonormal frame $\bm e^{}_{(a)}$ \cite{Schiff1960}. In particular, the results in the precession of the plane perpendicular to $\bm e^{}_{(3)}$
that contains the TMs. As a result, the relative displacement of the two TMs is actually generated by the differential precession between the orbital plane and the plane perpendicular to $\bm e^{}_{(3)}$ (see Figure \ref{diff}). If we regard $\bm e^{}_{(a)}$ as a gyroscope resembling that in the GPB
\linebreak
\vspace*{-2mm}

\noindent experiment \cite{gpb} and the orbit itself is also a gigantic gyroscope, then we may see that a very distinctive feature of the proposed measurement scheme is that, unlike GPB or LAGEOS/LARES \cite{lageos1,lageos2,lares} which measures precession with respect to an globally defined inertial reference, it measures the differential precession of the two gyroscopes.

With the theoretical foundation of the measurement scheme worked out above, there remain many key issues to be resolved or understood in order to implement the measurement scheme in a physically realistic situation (see for instance ref.  \cite{xupengLT2}). The study is still ongoing and we hope to report further progress in the near future.

\subsection{Concluding remarks}

An overview is given on the work we have been doing in the past
few years in relation to detection of GW in space.
Needless to say there is no end to such study. It is envisaged that
more in depth study will be undertaken in the next few years to
contribute to the development of GW detection in
China.

\section{Space missions for gravitational wave detection}
\subsection{Introduction}
For detection of GWs in the low frequency (100 nHz--0.1 Hz) band and in the middle frequency band (0.1 Hz--10 Hz), space interferometric detectors are needed. GW sources in these bands are abundant. For the critical technologies required to detect GWs in the range of 0.1 mHz to 1 Hz, the European Space Agency (ESA), at the time of writing, is in the final stage of preparing LISA Pathfinder (Laser-Interferometric Space Antenna Pathfinder) for launch. The mission will demonstrate critical technologies for future GW observatories, in a space environment. These technologies include: inertial sensing, drag-free attitude control, and interferometry with free-falling mirrors. LISA Pathfinder is scheduled to launch at the end of this year. The success of LISA Pathfinder will prepare the technological basis for space laser interferometers ~both in Earth orbit and in solar orbit.

On 28 November 2013, ESA announced its new vision to study the invisible universe. The hot and energetic Universe and the search for GWs will be the focus of ESA$'$s next two large science missions L2 and L3---the ``Hot and Energetic Universe'' for L2 and ``The Gravitational Universe'' for L3$^{4)}$\footnote{4) http://www.esa.int/Our\_Activities/Space\_Science/ESA\_s\_new\_vision\_to\_study\_the\_invisible\_Universe}.

The ESA L3 mission is likely to have a launch opportunity in 2034$^{4)}$. It is expected that eLISA/NGO (evolved LISA/New Gravitational-Wave Observatory) GW mission or its further evolved version will be the major candidate. Since  it will take one year to transfer to the science orbit, a starting time for the science phase is likely to be in 2035.

Motivated by the multiplicity of astrophysical GW sources for eLISA/NGO mission and other solar-orbit GW missions, and because of the 20 year period to 2035,  there is currently a  renaissance of Earth-orbiting GW mission proposals. Sufficiently numerous sources exist to justify an Earth-orbiting GW mission indicating that these mission proposals are worth considering before eLISA begins. Such a mission would be a natural precursor to a solar-orbit mission.

\textit{An essential technology for achieving sufficient GW sensitivity of space laser interferometers is Time Delay Interferometry. Time delay interferometry (TDI) is needed to suppress laser frequency noise. In the next subsection, we review and discuss TDI.}

\subsection{Time delay interferometry}
In space-based laser-interferometric GW antennas, the arm lengths vary according to the orbital dynamics of the independent spacecraft. In order to attain the requisite sensitivity, laser frequency noise must be suppressed below the secondary noises such as the optical path noise, acceleration noise etc. For suppressing laser frequency noise in this environment, it is necessary to use time delay interferometry (TDI) in the analysis to match the optical path length of different beams closely. The better the match of the optical path lengths, the better the cancellation of the laser frequency noise and the easier it is to achieve the requisite sensitivity. In case of exact match, the laser frequency noise is fully cancelled, as in the original Michelson interferometer.

The TDI concept was first used in the study of ASTROD (Astrodynamical Space Test of Relativity using Optical Devices) mission concept\,\cite{4Ni1997,5Ni2003,6Ni1997b}. In deep-space interferometry, long distances are invariably involved. Due to long distances, laser light is attenuated to a great extent at the receiving spacecraft. To transfer the laser light back or to another spacecraft, amplification is needed. The procedure is to phase lock the local laser to the incoming weak laser light and to transmit the local laser light back to another spacecraft. Liao et al. \cite{7Liao2002,8Liao2002b} have demonstrated the phase locking of a local oscillator with 2-pW laser light in laboratory. Dick et al.\,\cite{9Dick2008} have demonstrated phase locking to 40-fW incoming weak laser light. The power requirement feasibility for both e-LISA/NGO and ASTROD-GW (ASTROD  optimized for Gravitational Wave [GW] detection) is met with these developments. In the 1990s, Ni et al. \cite{4Ni1997,5Ni2003} used the following two TDI configurations during the study of ASTROD interferometry and obtained numerically the path length differences using Newtonian dynamics.

These two TDI configurations are the unequal arm Michelson TDI configuration and the Sagnac TDI configuration for 3 spacecraft formation flight. The principle is to have two split laser beams to go to Path 1 and Path 2 and interfere at their end path. For unequal arm Michelson TDI configuration, one laser beams starts from spacecraft 1 (S/C1) directed to and received by spacecraft 2, and optical phase locking the local laser in spacecraft 2 (S/C2); the phased locked laser beam is then directed to and received by spacecraft 1, and optical phase locking another local laser in spacecraft 1; and so on following Path 1 to return to (S/C1):

Path 1: S/C1$\rightarrow$S/C2$\rightarrow$S/C1$\rightarrow$S/C3$\rightarrow$S/C1.

\noindent The second laser beam starts from S/C1 also, but follows the route:

Path 2: S/C1$\rightarrow$S/C3$\rightarrow$S/C1$\rightarrow$S/C2$\rightarrow$S/C1,

\noindent
to return to S/C1 and to interfere coherently with the first beam. If the two paths has exactly the same optical path length, the laser frequency noises cancel out; if the optical path length difference of the two paths are small, the laser frequency noises cancel to a large extent. In the Sagnac TDI configuration, the two paths are:

Path 1: S/C1$\rightarrow$S/C2$\rightarrow$S/C3$\rightarrow$S/C1,

Path 2: S/C1$\rightarrow$S/C3$\rightarrow$S/C2$\rightarrow$S/C1.

\begin{figure}[H]
  \centering
  \includegraphics[scale=1]{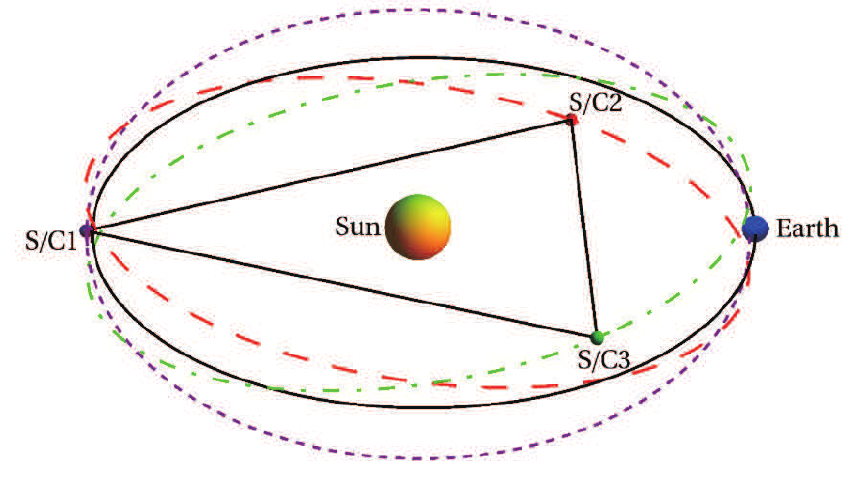}
  \caption{(Color online) Schematic of ASTROD-GW orbit configuration with inclination.}\label{Fig1}
\end{figure}

\begin{figure}[H]
 \centering
  \includegraphics[scale=1]{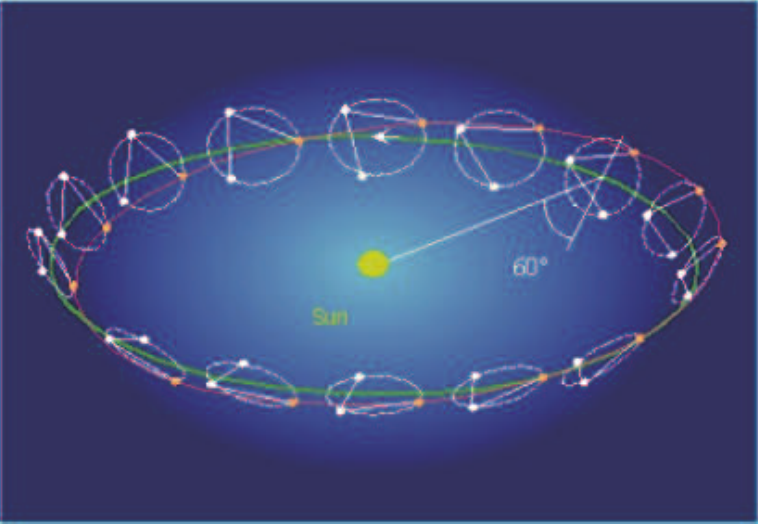}\vspace*{-2mm}
  \caption{(Color online) Schematic of LISA-type orbit configuration.\hspace*{6mm}}\label{Fig2}
\end{figure}

\begin{figure}[H]
\centering
\includegraphics[scale=1]{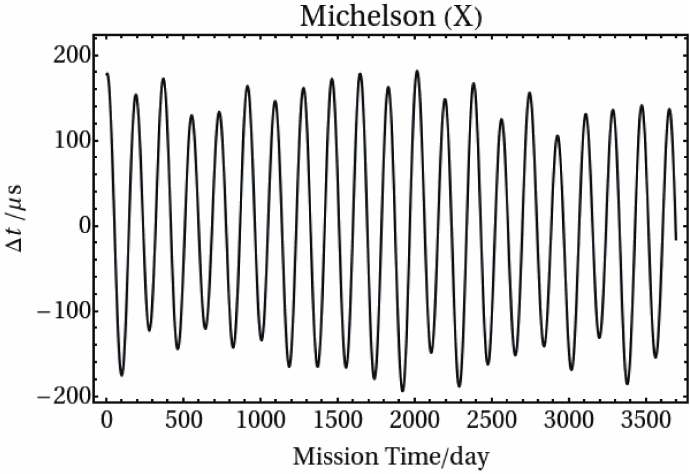}\vspace*{-2mm}
\caption{(Color online) Path length differences between two optical paths of the Unequal-arm Michelson TDI configuration (X-configuration) for ASTROD-GW orbit formation with $1^{\circ}$ inclination.}
\label{Fig3}
\end{figure}

Since then we have worked out the same things numerically for LISA\cite{10Dhurandhar2013}, eLISA/NGO\cite{11Wang2013}, NGO-LISA-type with $2\times10^6$ km arm length\,\cite{11Wang2013}, ASTROD-GW with no inclination\,\cite{12Wang2012,13Wang2013b}, and ASTROD-GW with inclination\,\cite{14Wang2015}. Schematic orbit configuration of ASTROD-GW mission design\,\cite{14Wang2015} and LISA-type mission design (http://sci.esa.int/lisa/35596-schematic-of-lisa-orbit/) are shown in Figures~\ref{Fig1} and ~\ref{Fig2}. For the numerical evaluation, we take a common receiving time epoch for both beams; the results would be very close to each other numerically if we take the same start time epoch and calculate the path differences. We refer to the path S/C1$\rightarrow$S/C2$\rightarrow$S/C1 as $a$(path) and the path S/C1$\rightarrow$S/C3$\rightarrow$S/C1 as $b$(path). Hence the difference $\Delta L$ between Path 1 and Path 2 for the unequal-arm Michelson can be denoted as $ab-ba=[a,b]$. Here $ab$ means $a$ path followed by $b$ path. The unequal-arm Michelson is now commonly called the X-configuration\,\cite{16Amstrong1999,17Tinto2014}. The result of this TDI calculation for ASTROD-GW orbit with 1$^{\circ}$inclination is shown in Figure~\ref{Fig3}.

\begin{table*}
\caption{Comparison of the resulting differences for second generation TDIs ($n=1$ and $n=2$) due to arm length variations for various mission proposals -- eLISA/NGO, an NGO-LISA-type mission with a nominal arm length of $2\times10^6$ km, LISA and ASTROD-GW}\label{tab2:compare}
\begin{center}\vspace{-2mm}\footnotesize \doublerulesep 0.2pt \tabcolsep 11.5pt
\begin{tabular*}{\textwidth}{cccccc}
\toprule
&& \multicolumn{4}{c}{TDI path difference $\Delta L$}\\
\cline{3-6\ }
\multicolumn{2}{c}{2*TDI configuration} & eLISA/NGO\cite{11Wang2013} & NGO-LISA-type with & LISA\cite{10Dhurandhar2013} & ASTROD-GW \\
&&&2$\times$10$^6$ km arm length\cite{11Wang2013} && (1$^{\circ}$ inclination)\cite{14Wang2015} \\\hline
\multicolumn{2}{c}{Duration} & 1000 d & 1000 d & 1000 d & 10 years \\\hline
$n=1$ & [$ab, ba$] & $-$1.5 to +1.5 ps & $-$11 to +12 ps & $-$70 to +80 ps & $-$228 to +228 ns \\ \hline
${3\times n=2}$ & [$a^2b^2, b^2a^2$] & $-$11 to +12 ps & $-$90 to +100 ps & $-$600 to +650 ps & $-$1813 to +1813 ns \\
\cline{2-6\ }
& [$abab, baba$] & $-$6 to +6 ps & $-$45 to +50 ps & $-$300 to +340 ps & $-$907 to +907 ns \\
\cline{2-6\ }
& [$ab^2a, ba^2b$] & $-$0.0032 to +0.0034 ps & $-$0.0036 to +0.004 ps & $-$0.015 to +0.013 ps & $-$0.66 to +0.66 ns \\\hline
\multicolumn{2}{c}{Nominal arm length} & 1 Gm (1 Mkm) & 2 Gm & 5 Gm & 260 Gm \\ \hline
\multicolumn{2}{c}{Requirement on $\Delta L$} & 10 m (30 ns) & 20 m (60 ns) & 50 m (150 ns) & 500 m (1500 ns) \\
\bottomrule
\end{tabular*}
\end{center}
\end{table*}

\begin{table*}[t]
\caption{Frequency classification of gravitational waves \cite{2Ni2010,317} \hspace*{112mm}}\label{tabA}
\footnotesize \begin{center}\vspace{-2mm}
\begin{tabular*}{\textwidth}
{p{7cm}p{9cm}}\toprule[0.65pt]  
Frequency band&Detection method\\\hline
Ultra high frequency band: above 1 THz
&Terahertz resonators, optical resonators, and magnetic conversion detectors\\

Very high frequency band:
100 kHz--1 THz
&Microwave resonator/wave guide detectors, laser interferometers and Gaussian beam detectors\\

High frequency band (audio band)$^{\rm a)}$: 10 Hz--100 kHz&Low-temperature resonators and ground-based laser-interferometric detectors\\

Middle frequency band:
0.1 Hz--10 Hz
&Space laser-interferometric detectors of arm length 1000 km--60000 km\\

Low frequency band (milli-Hz band)$^{\rm b)}$: 100 nHz--0.1 Hz&Space laser-interferometric detectors of arm length longer than 60000 km\\

Very low frequency band (nano-Hz band): 300 pHz--100 nHz&Pulsar timing arrays (PTAs)\\

Ultra low frequency band:
10 fHz--300 pHz
&Astrometry of quasar proper motions\\

Extremely low (Hubble) frequency band (cosmological band):
1 aHz--10 fHz
&Cosmic microwave background experiments\\

Beyond Hubble-frequency band: below 1 aHz&Through the verifications of inflationary/primordial cosmological models\\\bottomrule
\multicolumn{2}{l}{a) The range of audio band (also called LIGO band) normally goes only to 10 kHz}\\
b) The range of the milli-Hz band is 0.1 mHz to 100 mHz\\
\end{tabular*}
\end{center}
\end{table*}

Time delay interferometry has been investigated for LISA much more thoroughly since 1999\,\cite{16Amstrong1999,17Tinto2014}.  First-generation and second-generation TDIs have been proposed. In the first generation TDIs, static situations are considered. In the second generation TDIs, motions are compensated to a certain degree. The X-configurations considered above belongs to the first generation TDI configurations. We shall not review more about these developments here, but refer the readers to the excellent
review article by Tinto and Dhurandhar\,\cite{17Tinto2014} for a comprehensive treatment.

We compile for comparison the resulting differences for
to \linebreak
\vspace*{-3mm}

\noindent arm second generation two-arm TDIs with $n=1$ and $n=2$ ($n$ is the degree of polynominal in $ab$, $ba$, $a^2$ and $b^2$) due
length variations for various mission proposals -- eLISA/NGO, an NGO-LISA-type mission with a nominal arm length of $2\times 10^6$ km, LISA and ASTROD-GW.

\noindent
We note that:

i. All the second-generation TDIs considered for the one-detector case for eLISA/NGO, for NGO-LISA-type with $2\times 10^6$ km arm length, for LISA and for ASTROD-GW with 1$^{\circ}$ inclination basically satisfy this requirement on $\Delta L$ (Table~8).

ii. The requirement for unequal arm Michelson (X-configuration) TDI of ASTROD-GW needs to be relaxed by about 2 orders (Figure~\ref{Fig3}).

iii. In view of the possibility of a GW mission in Earth orbit, numerical TDI study for GW missions in Earth orbit are desired.

iv. Experimental demonstration of TDI in laboratory for LISA has been implemented in 2010\,\cite{18Vine2010}. eLISA and the original ASTROD-GW TDI requirement are based on LISA requirement, and hence also demonstrated. With the present pace of development in laser technology, the laser frequency noise requirement is expected to be able to compensate for 2--3 order of TDI requirement relaxation in 20 years.

v. X-configuration TDI sensitivity for GW sources has been studied extensively for eLISA  \cite{NGO_1}. It satisfies the present technological requirements well. With enhanced laser technology expected, it would also be good for studying the ASTROD-GW and
various GW missions in Earth orbit. The study for GW sensitivity and GW sources for other first-generation and second-generation TDIs would also be encouraged.


\subsection{Frequency classification of gravitational waves}\label{appendix}

In electromagnetism, the frequency spectrum of electromagnetic waves are classified into radio waves, millimeter waves, infrared, optical, ultraviolet, X-ray and $\gamma$-ray etc. For GWs, it is also desirable and useful to give a frequency classification. We extend the common classification of Thorne \cite{1Thorne1995} to reach a complete classification \cite{2Ni2010} shown in Table~\ref{tabA}.

%
%
%
%
%
%
%

\Acknowledgements{\bahao The authors thank the Kavli Institute for Theoretical Physics, China for funding the Next Detectors for Gravitational Astronomy Program, and for their hospitality.  Sect. 2 is based on DR's talks given during the course of this program. DR and KA are supported by the US National Science Foundation (Grant No. PHY-0757058), FZ is supported by the National Natural Science Foundation of China (Grant Nos. 11443008 and 11503003), a Returned Overseas Chinese Scholars Foundation grant, and Fundamental Research Funds for the Central Universities (Grant No. 2015KJJCB06). XZ, LW and GH acknowledge funding support from the Australian Research Council. This work was  supported by the National Space Science Center, Chinese
Academy of Sciences (Grant Nos. XDA04070400 and XDA04077700). Prof. HU WenRui's  effort to promote GW detection in space in China motivated our study in the first place. We are also grateful to Prof. ZHANG ShuangNan for his support throughout the course of our work.  Professors YAU Shing-Tung  and YANG Lo  have
been very supportive and the Morningside Center of Mathematics provides a very conductive
research environment to carry out the study. Partial supports from the National Natural Science Foundation of China (Grant Nos.
11305255,  11171329 and 41404019).}

\end{multicols}

\end{document}